\documentclass[final,5p,times]{elsarticle} 
\usepackage{hyperref} 

\usepackage{graphicx} 
\usepackage{natbib} 
\usepackage{amsmath} 
\usepackage{geometry}
\usepackage{pdfpages}
\usepackage{expl3}
\usepackage{float}
\usepackage[all]{hypcap}
\usepackage{amsfonts}
\usepackage{booktabs}
\usepackage{soul}
\usepackage{algorithm,algpseudocode}
\usepackage{siunitx} 

\graphicspath{ {SDSS_paper/} }

\journal{Astronomy and Computing}

\biboptions{authoryear}

\newcommand{\dgm}[1]{\text{Dgm}}

\usepackage{caption}
\usepackage{subcaption}

\begin{document}

\begin{frontmatter}

\title{Finding cosmic voids and filament loops using topological data analysis}

\author[sds]{Xin Xu}
\author[sds]{Jessi Cisewski-Kehe\corref{mycorrespondingauthor}}
\cortext[mycorrespondingauthor]{Corresponding author}
\ead{jessica.cisewski@yale.edu}
\author[phys]{Sheridan B. Green}
\author[phys,astr,ycaa]{Daisuke Nagai}

\address[sds]{Department of Statistics and Data Science, Yale University, New Haven, CT 06520, USA}
\address[phys]{Department of Physics, Yale University, New Haven, CT 06520, USA}
\address[astr]{Department of Astronomy, Yale University, New Haven, CT 06520, USA}
\address[ycaa]{Yale Center for Astronomy and Astrophysics, Yale University, New Haven, CT 06520, USA}

\begin{abstract}
We present a method called Significant Cosmic Holes in Universe (SCHU) for identifying cosmic voids and loops of filaments in cosmological datasets and assigning their statistical significance using techniques from topological data analysis.
In particular, persistent homology is used to find different dimensional holes.  For dark matter halo catalogs and galaxy surveys, the 0-, 1-, and 2-dimensional holes can be identified with connected components (i.e. clusters), loops of filaments, and voids.
The procedure overlays dark matter halos/galaxies on a three-dimensional grid, and a distance-to-measure (DTM) function is calculated at each point of the grid.
A nested set of simplicial complexes (a filtration) is generated over the lower-level sets of the DTM across increasing threshold values.   
The filtered simplicial complex can then be used to summarize the birth and death times of the different dimension homology group generators (i.e., the holes).  Persistent homology summary diagrams, called persistence diagrams, are produced from the dimension, birth times, and death times of each homology group generator.  
Using the persistence diagrams and bootstrap sampling, we explain how $p$-values can be assigned to each homology group generator. The homology group generators on a persistence diagram are not, in general, uniquely located back in the original dataset volume so we propose a method for finding a representation of the homology group generators.
This method provides a novel, statistically rigorous approach for locating informative generators in cosmological datasets, which may be useful for providing complementary cosmological constraints on the effects of, for example, the sum of the neutrino masses.
The method is tested on a Voronoi foam simulation, and then subsequently applied to a subset of the SDSS galaxy survey and a cosmological simulation.
Lastly, we calculate Betti functions for two of the MassiveNuS simulations and discuss implications for using the persistent homology of the density field to help break degeneracy in the cosmological parameters.
\end{abstract}


\end{frontmatter}


\section{Introduction}\label{intro}

The large-scale distribution of matter in the Universe forms a connected network known as the cosmic web \citep{Klypin1993,Bond1996}. Anisotropic gravitational collapse of matter has resulted in a picture where galaxy clusters form the nodes of this web and are interconnected by filaments, which form at the intersections of walls. The remaining majority of space is filled by cosmic voids: vast underdense regions that have experienced minimal non-linear growth of structure. 

\begin{sloppypar}
In the current era of multiband, high-resolution large-scale structure (LSS) surveys, there has been prolific investigation of the large dark matter halos at the nodes of the cosmic web, including cosmological analysis via the abundance of galaxy clusters (e.g., \citealt{Vikhlinin2009,Mantz2015,Planck15clusters}), and analysis of the large-scale matter distribution in dark matter halos via cosmic shear studies (e.g., \citealt{Joudaki2018,Troxel2018}) and the clustering of galaxies (e.g., \citealt{Tinker2012,SDSS_clustering_constraints}). However, there has been a growing tension between cosmological constraints derived using cosmic microwave background (CMB) and LSS measurements \citep{Planck15clusters,Joudaki2018,Troxel2018}.

Recently, voids have begun to attract interest as a cosmological probe complementary to halos. 
Unlike halos, which are regions that have experienced high levels of growth and virialization that can partially destroy signatures of the primordial density field, voids have only evolved minimally. 
Information about the geometry of the initial density field present in the voids has the potential to help break degeneracies in the cosmological parameters and tighten their current constraints \citep{Hamaus2016}. 
Additionally, motivated by the growing tension between CMB and LSS measurements, void statistics may be able to provide a complementary probe of the growth of structure that is less sensitive to non-linear structure formation physics \citep{Lavaux2010,Lavaux2012}.
Furthermore, methods have been proposed for utilizing voids to constrain dark energy \citep{Pisani2015} and the sum of the neutrino masses $\sum m_\nu$ \citep{Kreisch2018,Massara2015}, both of which have large effects on large-scale, low-density regions.
\end{sloppypar}

Various observables can be used for constraining cosmology through voids, some of which include weak gravitational lensing \citep{Sanchez2017,Cai2017,Kaiser1993}, the integrated Sachs--Wolfe effect in the CMB \citep{Nadarthur2016}, redshift-space distortions \citep{Cai2016,Hamaus2015}, void ellipticities \citep{Lee2009}, among others. Current datasets that are available and have been used to characterize some of the aforementioned void observables include galaxy redshift surveys such as the Sloan Digital Sky Survey (SDSS; \citealt{SDSS14}) and the Dark Energy Survey (DES; \citealt{DES}) as well as CMB anisotropy maps from Planck \citep{Planck2015isw}. 
In the near future, next-generation galaxy surveys will go online, including the Large Synoptic Survey Telescope (LSST; \citealt{LSST}) and the Dark Energy Spectroscopic Instrument (DESI; \citealt{DESI}), and the secondary CMB anisotropies will be observed to greater precision than ever by the Simons Observatory (SO; \citealt{Simons}) and the CMB-S4 experiment \citep{CMBS4}. In anticipation of these large upcoming datasets, many theoretical and computational approaches for identifying voids from an input simulation or galaxy survey have been and are currently being developed in order to characterize the potential future constraining power of void clustering and abundance statistics \citep[and references therein]{libeskind2018tracing, sutter2012public, neyrinck2008zobov, platen2007cosmic, pranav2016topology}. The majority of these methods allow for the identification of the physical location of the voids in the matter field, enabling one to study clustering statistics such as the void two-point correlation function and abundance statistics such as the void mass and volume functions.

In this paper, we propose a method called Significant Cosmic Holes in Universe (SCHU)\footnote{The implementation code and illustration example of SCHU can be found and downloaded from: \href{https://github.com/xinxuyale/SCHU}{https://github.com/xinxuyale/SCHU}} for relating the cosmic matter distribution to topology using persistent homology.  Persistent homology quantifies and summarizes the shape of a dataset by its hole structure, and SCHU uses this information to assign a measure of statistical significance to the individual holes and records locations of the representations of these structures back in the data volume, which enables analysis of void clustering and abundance.
The different dimensional homology groups are associated with different cosmic environment types.
For example, connected components (0th-dimensional homology groups, $H_0$), loops (1st-dimensional homology groups, $H_1$), and low density 3D volumes (2nd-dimensional homology groups, $H_2$) are analogous to galaxy clusters, closed loops of filaments, and cosmic voids, respectively.
Thus, cosmic voids can be identified as representations of $H_2$ homology group generators and newly-proposed \emph{filament loops} can be identified as representations of $H_1$ homology group generators.

%

Topological methods have previously been employed in cosmology.
For example, the topological evolution of the matter distribution of the Universe was studied in \citet{van2011alpha} by analyzing the changing Betti numbers, which are ranks of different order homology groups (i.e., number of clusters, filament loops, and voids), across a filtration, which is an indexed sequence of nested sets, constructed using alpha shapes\footnote{An alpha shape is a generalization of the convex hull, and captures the shape of a point set \citep{edelsbrunner1983shape}.}; in particular, they demonstrated that the Betti numbers across the filtration can be used to distinguish the matter distribution resulting from different dark energy models. 
Additionally, \citet{pranav2016topology} introduced a multiscale topological measurement of the cosmic matter distribution and explored the analysis of Betti numbers and topological persistence of different cosmological models.
%
A scale-free and parameter-free method for identifying the cosmic environments (voids, walls, filaments, nodes) called Discrete Persistent Structures Extractor (DisPerSE) was proposed in \citet{sousbie2011persistent}. DisPerSE computes the discrete Morse--Smale complex of a spatial dataset using the Delaunay tessellation field estimator (DTFE) technique \citep{schaap2000continuous, weygaert2008cosmic}. The mathematical background and algorithm implementation is described in \citet{sousbie2011persistent} and applications to 3D simulation datasets and observed galaxy surveys are found in \citet{sousbie2011persistentII}.

As noted previously, persistent homology is a tool within topological data analysis (TDA) that finds different dimensional holes in data (e.g. connected components, loops, and voids) and summarizes the generators by their lifetime in a particular filtration. These persistence diagrams and their associated Betti numbers can then be used for various types of statistical inference or as inputs into machine learning algorithms \citep{reininghaus2015stable}.
Persistent homology has proven to be useful in a variety of applications, such as natural language processing \citep{zhu2013persistent}, computational biology \citep{xia2014persistent}, Lyman-alpha forest studies \citep{cisewski2014non}, angiography \citep{bendich2016persistent}, and dynamical systems \citep{emrani2014persistent}. 

\begin{figure}[!ht]
  \centering
    \includegraphics[width=0.45\textwidth]{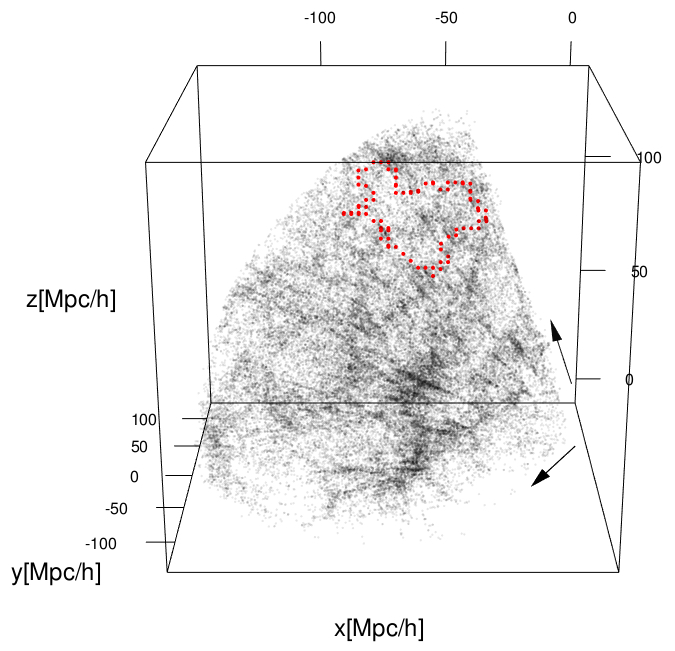}
\caption{The red points highlight an example of a \emph{filament void} identified using the proposed persistent homology method in the SDSS (Sloan Digital Sky Survey) dataset which is introduced in \S\ref{sec:sdss}.}\label{filament_example}
\end{figure}

Though useful for summarizing information for complex data, one shortcoming of persistent homology is that the homology group generators identified are not uniquely mapped back into the data volume. This is because the homology group generators displayed on the summary diagrams each represent an equivalence class of representations of that particular hole. 
SCHU uses the output of the persistent homology algorithm \citep{Edelsbrunner2002, zomorodian2005computing} in order to find a representation of the equivalence class back in the original data volume. 
SCHU detects and captures the locations of cosmic voids ($H_2$ generators) along with another cosmic structure that we call \emph{filament loops} ($H_1$ generators). Filament loops are formed when filaments are connected together in such a way that a loop forms, surrounding an empty or low density region, as shown in \autoref{filament_example}. 
Thus, SCHU, and the persistent homology underlying SCHU, enable analysis of the cosmological density field: the void and filament loop locations and sizes enable the standard clustering and abundance statistics, and the persistence diagrams and Betti numbers provide additional topological summary statistics of the density field that can be used to further discriminate between cosmological models.

This article is organized as follows. 
In \S\ref{sec:pers_hom}, we provide an overview of the formalism of persistent homology, describing filtrations, persistence diagrams, and bootstrap confidence bands. 
In \S\ref{sec:method}, we present SCHU for identifying statistically significant voids and filament loops in astronomical datasets. In \S\ref{sec:voronoi_foam}, we test the void identification capabilities of SCHU on Voronoi foam simulation data, which is generated such that the ground truth void locations are known. In \S\ref{sec:comparisons}, we apply SCHU to identify voids and filament loops in a subset of the SDSS galaxy survey dataset. Additionally, we identify voids and filament loops in the cosmological N-body simulation from \citet{libeskind2018tracing} and compare the void locations to those found by other methods. We then study the Betti numbers of two simulations from the MassiveNuS simulation suite \citep{MassiveNuS}. Finally, in \S\ref{sec:discussion}, we summarize our results and provide concluding remarks.

\section{Introduction to persistent homology}\label{sec:pers_hom}
\subsection{Background}
\begin{figure*}
 \centering
    \begin{subfigure}{0.45\textwidth}
        \centering
    \includegraphics[width=\textwidth]{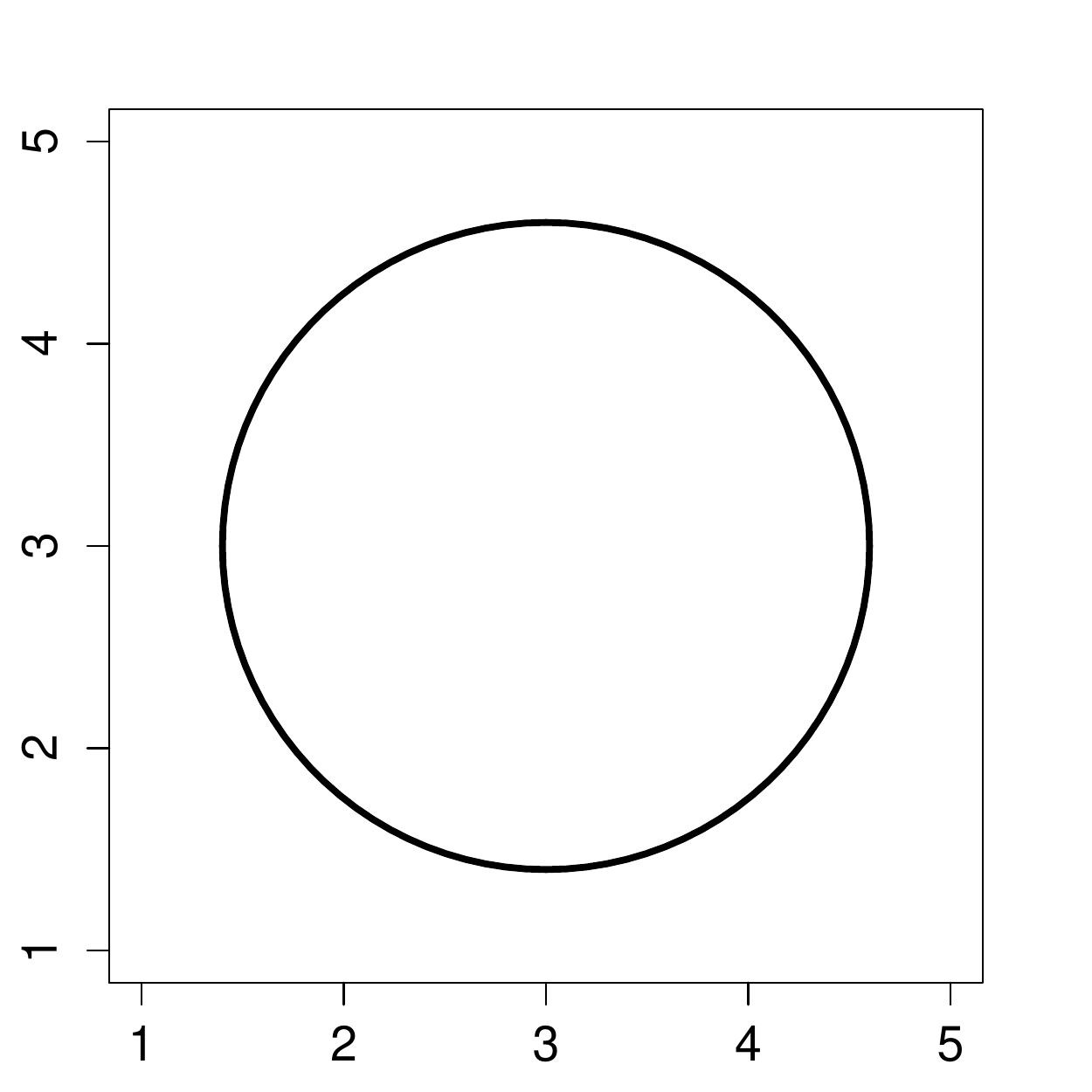}
        \caption{True manifold}\label{subfig:homo_true}
    \end{subfigure}
    \begin{subfigure}{0.45\textwidth}
        \centering
    \includegraphics[width=\textwidth]{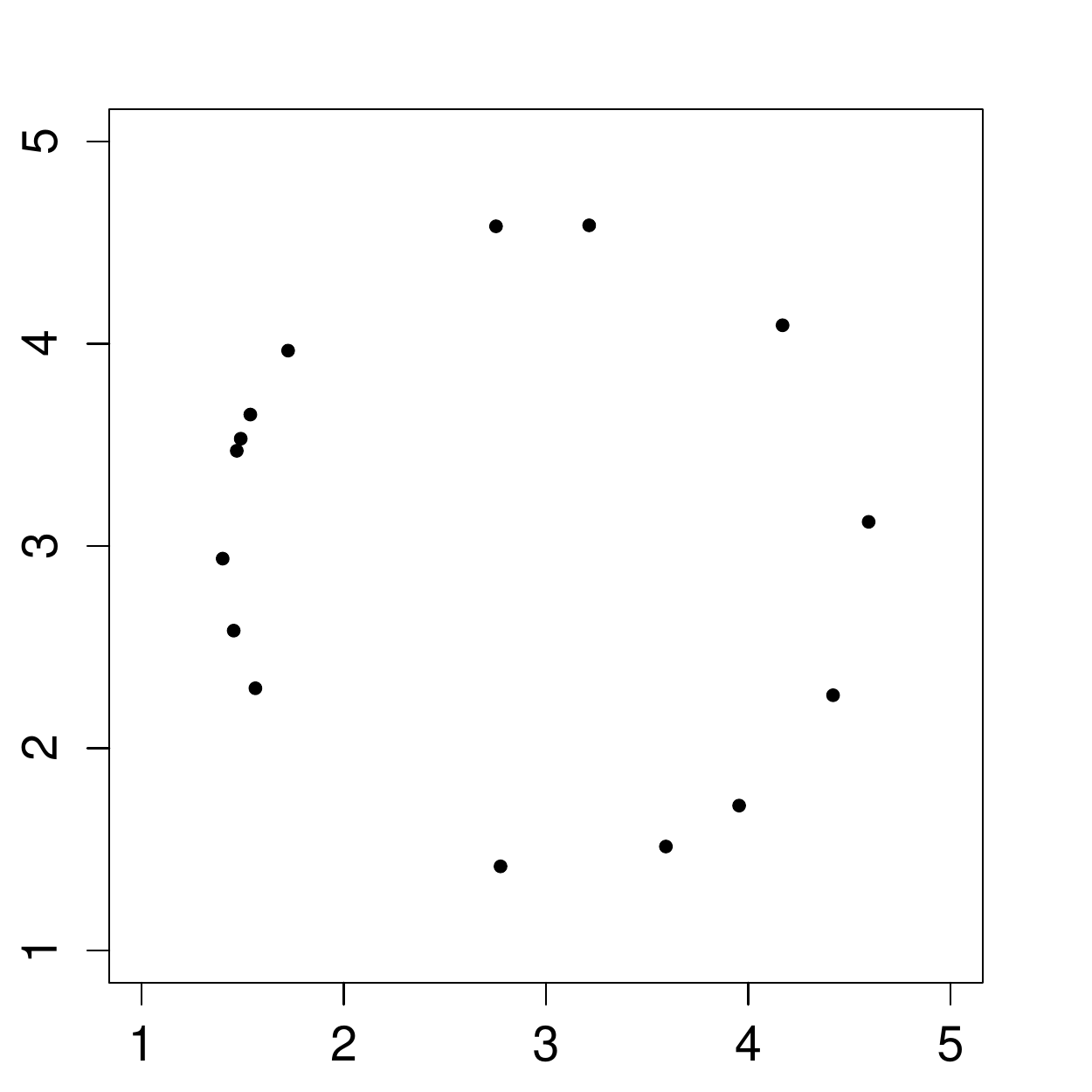}
        \caption{Sampled manifold}\label{subfig:homo_sampled}
    \end{subfigure}
\caption{(a) A circle, which has one connected component and one closed loop (one $H_0$ generator and one $H_1$ generator). (b) 15 points randomly sampled on a circle, which has 15 connected components and no closed loop (15 $H_0$ generators and 0 $H_1$ generators). 
} \label{homo}
\end{figure*}

Homology describes different dimensional holes of a manifold.
To be specific, the generators of $H_0$ describe connected components, the generators of $H_1$ describe closed loops, and the generators of $H_2$ describe voids (i.e. low-density or empty regions). 
Put into the context of cosmic web environments, the $H_0$ generators represent clusters of galaxies, the $H_1$ generators represent filaments that form loops, and the $H_2$ generators represent cosmic voids. \autoref{homo} illustrates an example of $H_0$ and $H_1$: Figure~\ref{subfig:homo_true} shows a circle, which forms one closed loop (one $H_1$ generator) and is one connected component (one $H_0$ generator). 
Figure~\ref{subfig:homo_sampled} shows 15 points randomly sampled on a circle, which is a dataset that has 15 connected components (15 $H_0$ generators) and no closed loop (zero $H_1$ generators).  While there is not a closed loop present, it is clear that the data were sampled on a circle; persistent homology provides a framework for identifying such a loop.
Our present interest is in the identification and analysis of voids and filament loops in cosmological datasets, and thus we will focus our attention in this work towards the $H_1$ and $H_2$ generators.

Persistent homology is a framework for computing the homology of real data across a particular filtration, and results are summarized in a persistence diagram. 
For example, in the popular ``friends of friends'' (FOF) clustering algorithm used for defining dark matter halos, if a sequence of increasing ``linking lengths'' are considered, the sets of halo clusters that form for the different linking lengths would form a sequence of nested sets (i.e., a filtration).  This is because the smaller halos that form with a small linking length would be contained in the larger halos that form with increasing linking lengths.
A persistence diagram summarizes the birth and death times of the homology group generators (i.e., the time when a homology group generator first appears in the filtration, and the time when it is no longer present due to merging with another homology group generator or the closing of the hole). In terms of the FOF algorithm, the death of a cluster (i.e., halo) can occur when it merges with another cluster at a larger linking length.
Regarding the homology group generators on a persistence diagram, an $H_2$ generator on the diagram indicates the presence of a cosmological void.  When the void generator lasts a long time in the filtration (i.e., it has a long lifetime), that suggests that the void may be larger and can be interpreted as being more statistically significant (see \S\ref{inference} for the notation of statistical significance).  Similarly, an $H_0$ generator indicates the presence of a cluster of halos and an $H_1$ generator indicates the presence of a filament loop.
For a detailed introduction of persistent homology, we refer the reader to \citet{edelsbrunner2010computational}. Several software and packages are available to compute persistent homology (e.g., \citet{GUDHI,Dionysus}). In this work, we use Dionysus \citep{Dionysus} via the R package TDA \citep{fasy2014Introduction} along with several of the other functions available in the R package.

\subsection{Filtration construction}\label{persistent_homology}
Given a point cloud of data, $S_n=\{X_1, X_2, \dots, X_n\}$ in $\mathbb{R}^d$, sampled from some true distribution $M$, one may be interested in the homology of $M$. In a cosmological context, this point cloud may be a galaxy survey dataset or dark matter halo catalog, which we assume traces the underlying dark matter distribution.
Persistent homology computes the homology of $S_n$ across different levels of a constructed filtration \citep{edelsbrunner2010computational}, where the levels of the filtration are based on a parameter analogous to the linking length in the FOF algorithm. 
We consider a function-based construction that relies on a robust version of a distance function.
A distance function, $d_{S_n}$, is a function that maps points from $\mathbb{R}^d$ to $\mathbb{R}$, defined as $d_{S_n}(x)=\inf_{y\in S_n}\|x-y\|_2$, where $\|\cdot\|_2$ is the $L^2$ norm. 
A filtration can then be constructed using the lower level sets of $d_{S_n}$, $L_{n,t}=\{x:d_{S_n}(x)\leq t\}$, where $t$ is the filtration parameter defining the threshold of the lower-level sets.
Notice that $L_{n,0}=S_n$ and $L_{n,\infty}=\mathbb{R}^d$. Persistent homology describes how the homology of $L_{n,t}$ changes as $t$ increases from 0 to $\infty$ (in practice, $t$ increases from the minimum distance to the maximum distance). 
Let $p$ be the homology group dimension, and then $H_p(L_{n,t})$ is the $p$th-dimensional homology group of $L_{n,t}$. 

In practice, we use a grid over the point cloud data and evaluate $d_{S_n}$ for each point on the grid. Each grid point is assigned a $d_{S_n}$ value based on its distance to the nearest observation in $S_n$; if an observation falls directly on the grid point, then $d_{S_n} = 0$.
Consider a sequence of $t$: $t_0, t_1, t_2, \dots, t_k, t_{k+1}$. Let $t_0$ be $0$ and $t_{k+1}$ be $\infty$, and then $L_{n,t_0}=S_n$ and $L_{n, t_{k+1}}=\mathbb{R}^d$. (In practice, this just goes to the size of the dataset box.) We can mathematically represent a filtration of homology groups as:
\begin{equation}\label{eq1}
H_p(L_{n,t_0}) \rightarrow H_p(L_{n, t_1}) \rightarrow H_p(L_{n,t_2}) \rightarrow \dots \rightarrow H_p( L_{n,t_{k+1}}).
\end{equation}

This filtration of homology groups is constructed by computing the homology of the lower-level sets at each threshold value $t$. For example, if the threshold is set to $t=0.5$, then all the grid points $x$ with $d_{S_n}(x)\leq 0.5$ would be contained in $L_{n,t}$.
Along the filtration, topological structures gradually change: homology groups appear at some point and disappear later during the filtration. 
The threshold value $t_k$ where $H_p(L_{n, t_k})$ contains the first appearance of a homology group along the filtration is called its \textit{birth time} and the threshold value $t_l$ where $H_p(L_{n, t_l})$ contains the last appearance of the homology group is called its \textit{death time}.

A more robust version of the distance function called the \emph{distance-to-measure} (DTM) function \citep{chazal2011geometric} is available and useful for TDA \citep{chazal2017robust}. Given a probability measure $P$, a DTM function with respect to a set $X$ is defined for each $y\in \mathbb{R}^d$ as
\begin{equation}
d_{m_0}(y)=\left(\frac{1}{m_0}\int_0^{m_0}(G_y^{-1}(u))^rdu\right)^{\frac{1}{r}},
\end{equation}
where $G_y(t)=P(\|X-y\|\leq t)$, and $m_0\in (0, 1)$ and $r\in [1, \infty)$ are tuning parameters. The $m_0$ can be seen as a smoothing parameter\footnote{The DTM function can be understood as a smoothed version of the distance function.}, and $r$ is commonly set to $2$ (using the $L^2$-norm to measure distance). Given an observed $S_n = \{x_1, \ldots, x_n\}$, the empirical version of the DTM function at some $y \in \mathbb{R}^d$ is
\begin{equation} \label{eq:dtm_obs}
\hat{d}_{m_0}(y)=\left(\frac{1}{k}\sum_{x_i\in N_k(y)}\|x_i-y\|^r\right)^{1/r},
\end{equation}
where $k=\lceil m_0n\rceil$ (the ceiling of $m_0n$) and $N_k(y)$ is the set containing the $k$ nearest neighbors of $y$ among $x_1, x_2, \dots, x_n$. The DTM provides a type of average distance to the $k$ nearest halos/galaxies at each grid point (see \citet{chazal2017robust} for additional properties of the TDM function).

\begin{figure*}
 \centering
    \begin{subfigure}{0.4\textwidth}
        \centering
    \includegraphics[width=\textwidth]{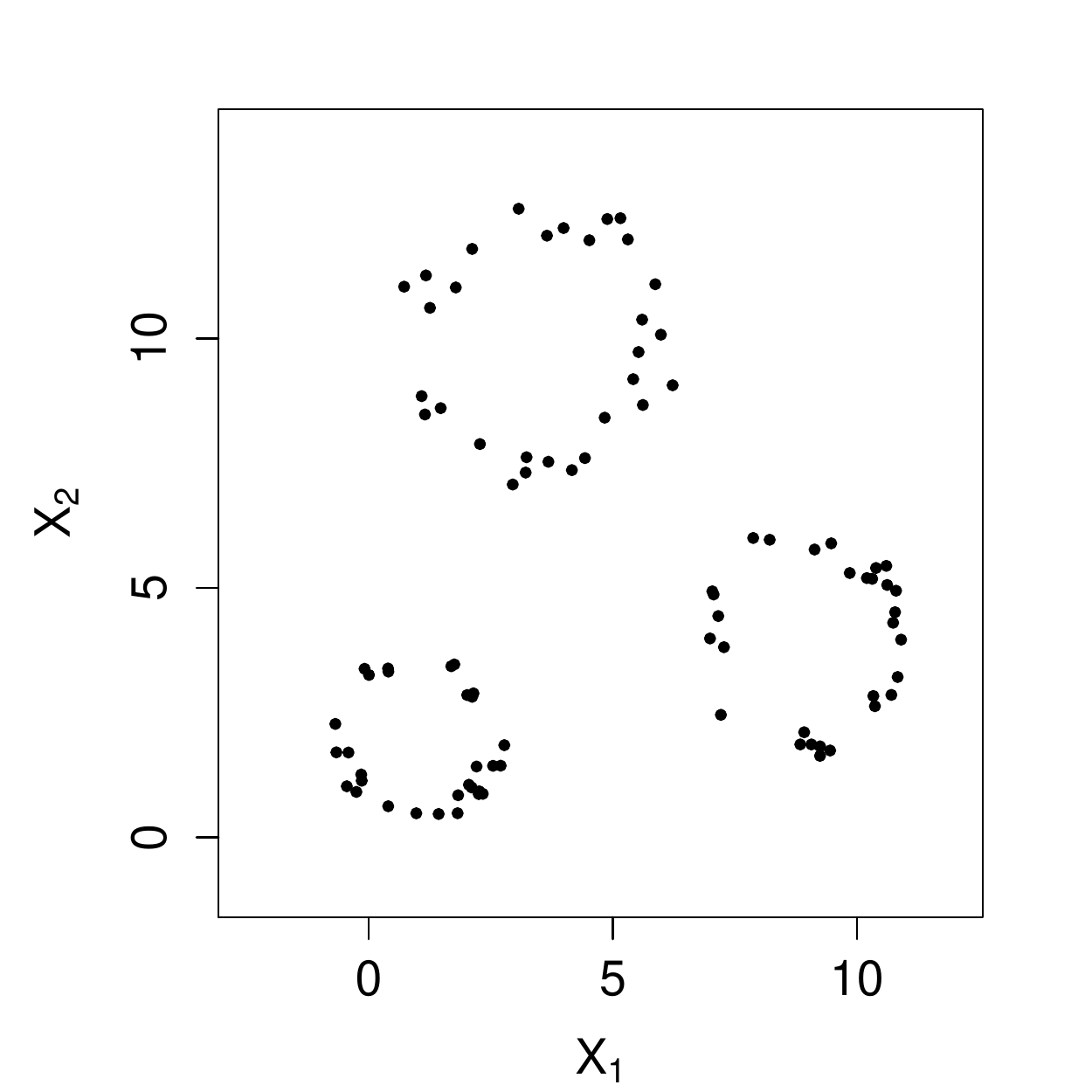}
        \caption{Point cloud}\label{subfig:dtm1}
    \end{subfigure}
    \begin{subfigure}{0.4\textwidth}
        \centering
    \includegraphics[width=\textwidth]{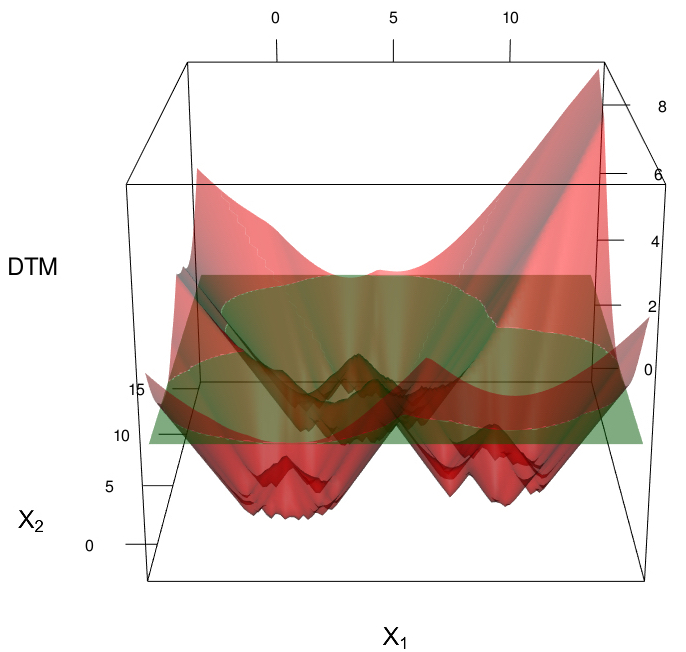}
        \caption{DTM}\label{subfig:dtm2}
    \end{subfigure}
        \begin{subfigure}{0.45\textwidth}
        \centering
    \includegraphics[width=\textwidth]{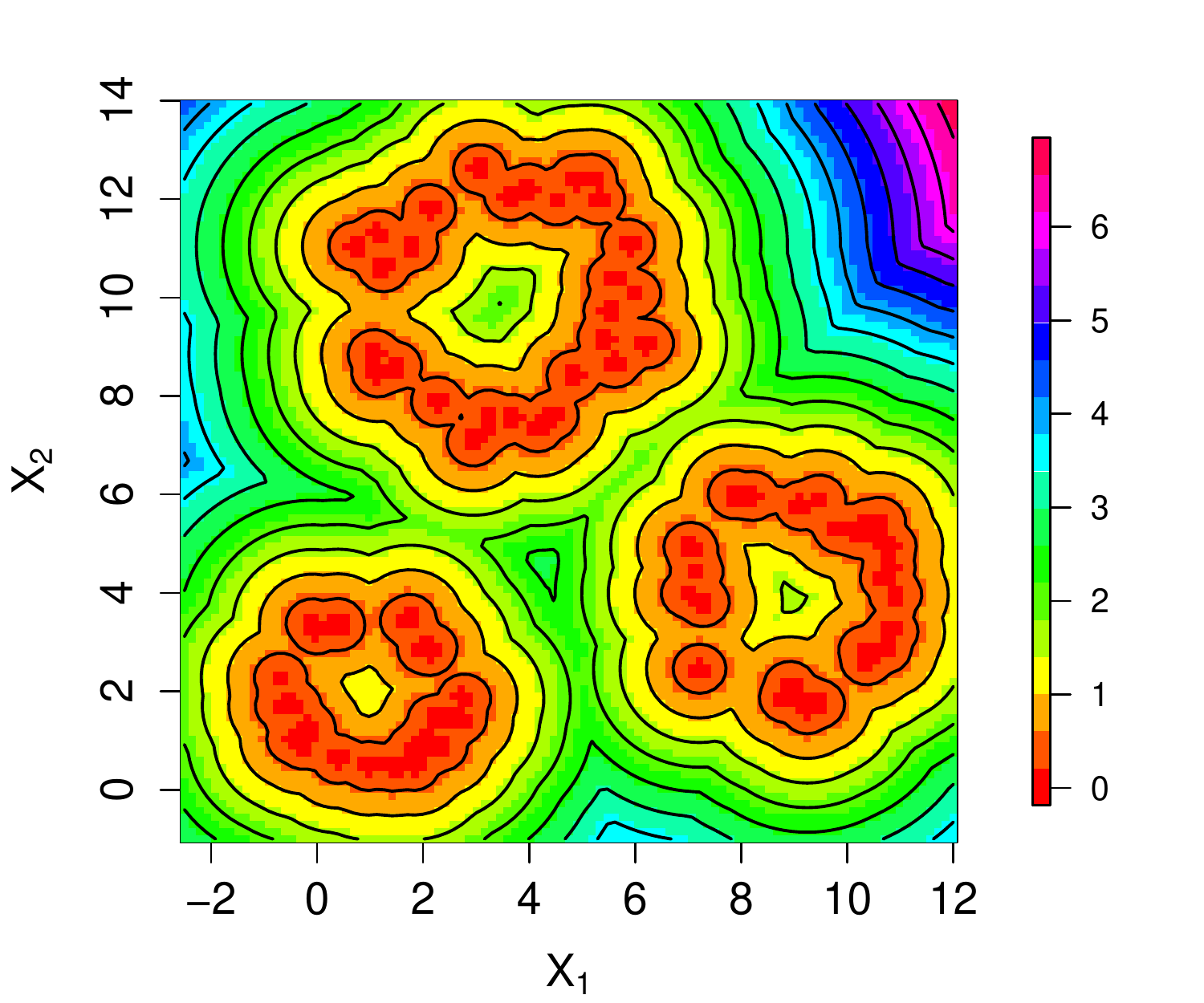}
        \caption{Contour map of DTM}\label{subfig:dtm3}
    \end{subfigure}
        \begin{subfigure}{0.35\textwidth}
        \centering
    \includegraphics[width=\textwidth]{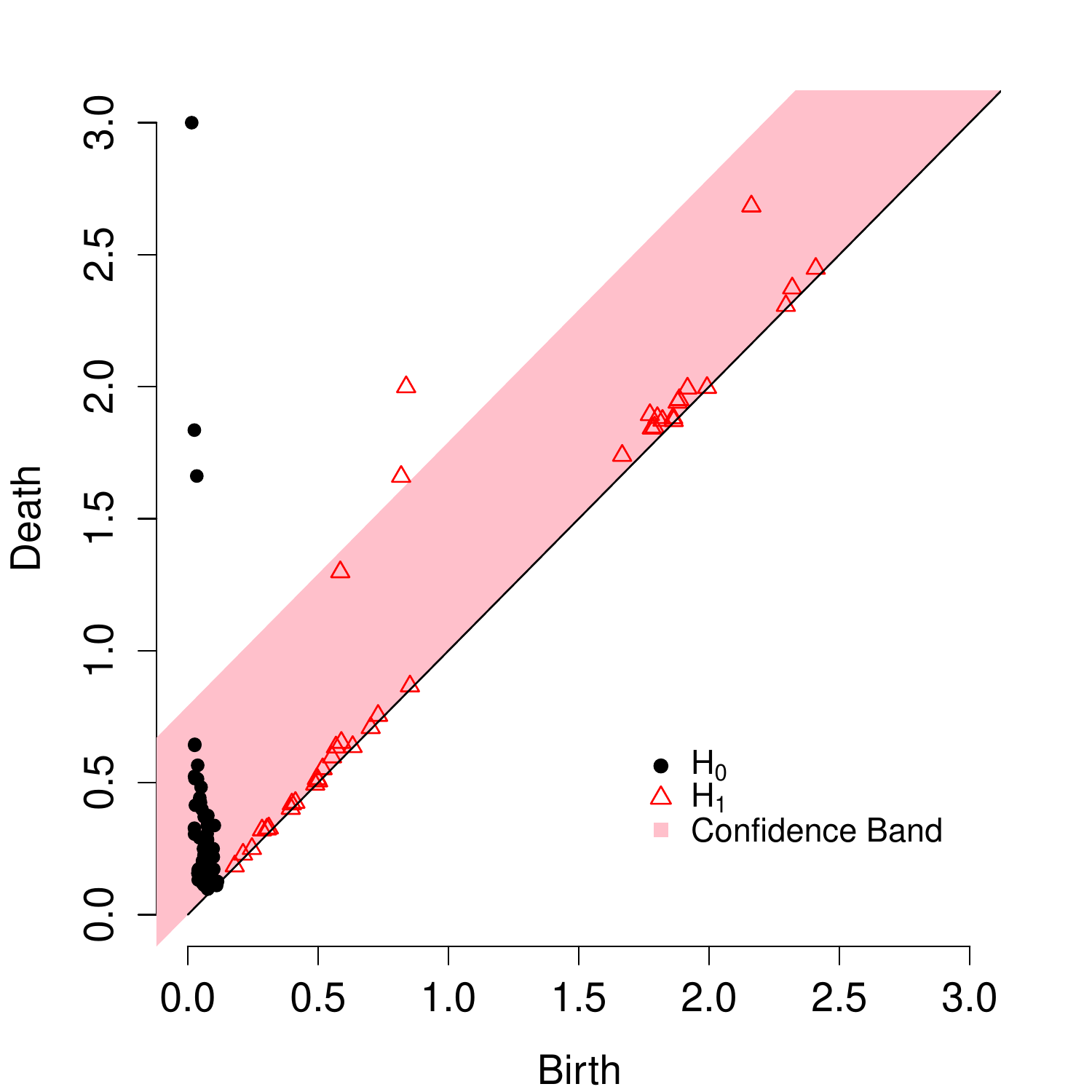}
        \caption{Persistence diagram}\label{subfig:dtm4}
    \end{subfigure}
\caption{(a) A simulated dataset, $S_n$. (b) The DTM function of $S_n$ in 3D over a grid of points. The green plane shows a threshold corresponding to a DTM value of 3. (c) The DTM function in colored contour over a grid of points. (d) The persistence diagram of $S_n$ using the lower level sets of the DTM. There are several black points and three red points obviously apart from the diagonal, which is consistent with the point cloud: three loops. The pink band is $90\%$ confidence band and is used as a method for distinguishing topological signal from noise, which is introduced in \S\ref{inference}.  
} \label{dtm}
\end{figure*}

Lower-level sets of $\hat{d}_{m_0}$ from Eq.~\eqref{eq:dtm_obs} can be used to construct the sequence of homology groups in \eqref{eq1}, where each homology generator has a birth time $b_i$ and a death time $d_i$, $i=1, \dots, m$. Then a persistence diagram is generated from the set of birth and death times $(b_1, d_1)$, $(b_2, d_2)$, $\dots$, $(b_m, d_m)$. 
For example, the minima of the $\hat{d}_{m_0}(y)$ (if present) are born early in the filtration as $H_0$ generators. 
As the filtration parameter $t$ increases, homology group generators merge together leading to the death of some of the $H_0$ generators, and eventually all merge into a single connected component. 
Additionally, some $H_0$ generators can merge in such a way that an $H_1$ generator, a closed loop, is born. 
As the filtration parameter, $t$, increases, the loop is filled in, indicating the death of the loop. 
\autoref{dtm} shows an example of a DTM function and its corresponding persistence diagram produced.
\autoref{subfig:dtm1} presents the point cloud we used to generate a persistence diagram, which contains three disconnected loops. \autoref{subfig:dtm2} and \autoref{subfig:dtm3} show the DTM function in 3D and as colored contours over a grid of points, respectively. The green plane in \autoref{subfig:dtm2} shows a threshold at the DTM value 3. The persistence diagram in \autoref{subfig:dtm4} has two types of points: the black dots are $H_0$ generators, and the red triangles are $H_1$ generators. The $x$-axis and $y$-axis represent birth time and death time, respectively. The longer the lifetime of the homology group generator, the further the feature is from the diagonal on the persistence diagram, the longer the homology group generator is present in the filtration. A long lifetime can also be interpreted as the feature being more significant.
There are three distinct $H_1$ generators far from the diagonal, which is consistent with the three loops in the point cloud data.

\subsection{Confidence band for persistence diagrams}\label{inference}
The lifetime of a homology group generator $i$ is called its \textit{persistence}, defined as $d_i-b_i$, where $b_i$ and $d_i$ are its birth time and death time, respectively. It is common to interpret more persistent generators (longer lifetimes) as topological signal and less persistent generators (shorter lifetimes) as topological noise. 
Thus, $H_2$ generators with longer lifetimes can be thought of as being more statistically significant than those with shorter lifetimes, and, in our context, are generally physically larger, cosmic voids.  This can be a helpful property since it will limit the number of spurious voids identified.
A method for defining confidence sets on a persistence diagram and discriminating topological signal from topological noise was developed in \citet{fasy2014confidence}.
To define these confidence sets, or to carry out other types of statistical inference, on a persistence diagram, an appropriate metric on the space of persistence diagrams is needed.
The \textit{bottleneck distance}, $W_{\infty}(\cdot, \cdot)$, is a popular metric for quantifying the distance between two persistence diagrams.
It is defined as $W_{\infty}(U, V)=\inf_{\eta:U\rightarrow V}\sup_{u\in U}\|u-\eta(u)\|_{\infty}$, where $\eta$ is a bijection between two persistence diagrams that matches each point on persistence diagram $U$ to a point on persistence diagram $V$. The diagonal (i.e. where $b_i=d_i$) is considered to have an infinite number of points so if there is an imbalance in the number of topological generators between the two diagrams, any excess of points can be mapped to the diagonal. 
The $\|u-v\|_{\infty} = \max\{|b_u-b_v|, |d_u-d_v|\}$, with $u=(b_u,d_u)$ and $v=(b_v, d_v)$, where $v = \eta(u)$. 
That is, the bottleneck distance considers the $L^{\infty}$ distance between every matching between two diagrams, and reports the minimum of all those distances.

Let $f$ be the DTM function of the true data distribution $M$ (i.e., $f$ can be interpreted as a function that encodes the topology of the underlying density field) and $\hat{f_n}$ be the empirical DTM function of the point cloud data $S_n$ (i.e., calculated using the observed galaxies/halos), with $Dgm_p(f)$ as the persistence diagram of homology dimension $p$ for $f$ and $Dgm_p(\hat{f_n})$ be the persistence diagram of dimension $p$ for $\hat{f_n}$.
Following the notation of \citet{fasy2014confidence}, for a given statistical significance level $\alpha \in (0, 1)$, we will find a constant $c_n=c_n(X_1, X_2, \dots, X_n)$ such that 
\begin{equation}\label{equation}
\limsup_{n\rightarrow \infty} \mathbb{P}(W_{\infty}(Dgm_p(\hat{f_n}), Dgm_p(f))>c_n)\leq \alpha.
\end{equation}
\begin{sloppypar}
\noindent Hence $C_n=[0, c_n]$ is an asymptotic $1-\alpha$ confidence set for $W_{\infty}(Dgm_p(\hat{f_n}), Dgm_p(f))$ \citep{fasy2014confidence}. 
\end{sloppypar} 
The bottleneck distance defines the shape of a confidence set as a square (since it is based on the $L^{\infty}$ distance). A higher confidence level would result in a larger square, and a lower confidence level would result in a smaller square; this is analogous to the interpretation of the confidence interval around an estimator.  A $1-\alpha$ confidence set on a persistence diagram specifies the region of the persistence diagram on which we are $1-\alpha$ confident that there is a \emph{real} homology group generator, and, hence, a real cluster, filament loop, or void.

Eq.~\eqref{equation} implies that a square $1-\alpha$ confidence set can be centered on each point of $Dgm_p(\hat{f_n})$ with a side of length $2c_n$.  If this box for some generator $i$ intersects the diagonal line on the persistence diagram (where the birth time = death time), then generator $i$ can be interpreted as topological noise at the significance level $\alpha$ (i.e., the $p$-value of generator $i$ would be greater than $\alpha$); if this box for some generator $i$ does not intersect the diagonal line, then generator $i$ can be interpreted as topological signal at the significance level $\alpha$.
Instead of considering the separate confidence sets around each generator, a \textit{confidence band} can be added to a persistence diagram using a band of width $\sqrt{2}c_n$ on the diagonal of $Dgm_p(\hat{f_n})$, where points in the band are considered indistinguishable from noise and points outside the band are considered significant topological generators (see Figure~\ref{subfig:dtm4} for an example of a confidence band for the $H_1$ group generators).

The following bootstrap procedure, as described in \citet{fasy2014confidence}, can be used as an option for computing the confidence band. First, randomly sample a new point cloud sample, $S_n^*$, from the original dataset $S_n$. 
Next, compute a persistence diagram $Dgm_p(\hat{f_n^*})$ for $S_n^*$, and calculate $$w = W_{\infty}\left(Dgm_p(\hat{f_n}), Dgm_p(\hat{f_n^*})\right).$$ 
Repeat the bootstrap sampling $N_{\textrm{boot}}$ times such that $N_{\textrm{boot}}$ persistence diagrams and bottleneck distances are obtained.

\begin{figure*}
 \centering
    \begin{subfigure}{0.47\textwidth}
        \centering
    \includegraphics[width=\textwidth]{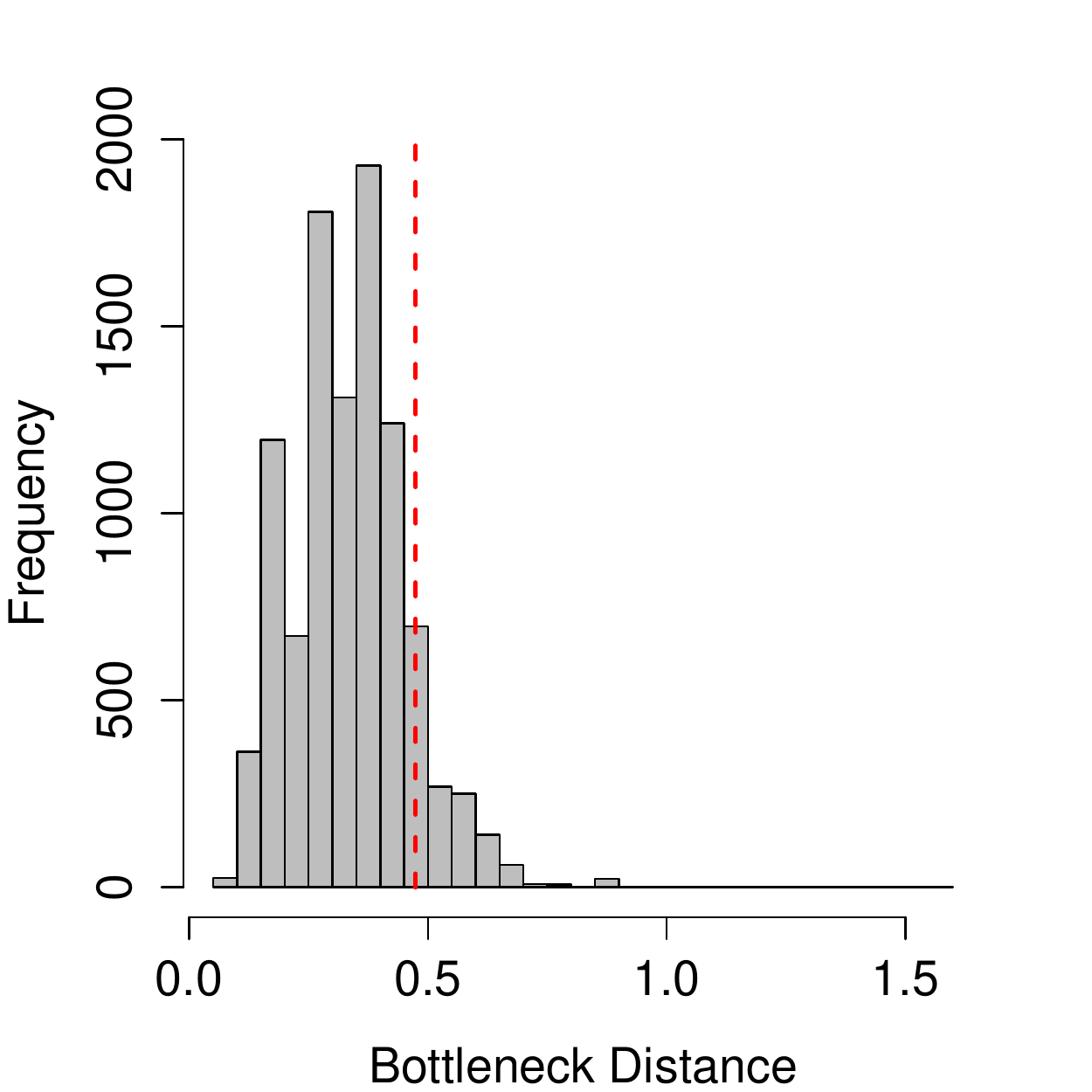}
        \caption{Distribution of $W_{\infty}(Dgm_0(\hat{f}_n), Dgm_0(\hat{f}_n^*))$}\label{subfig:dist1}
    \end{subfigure}
    \begin{subfigure}{0.47\textwidth}
        \centering
    \includegraphics[width=\textwidth]{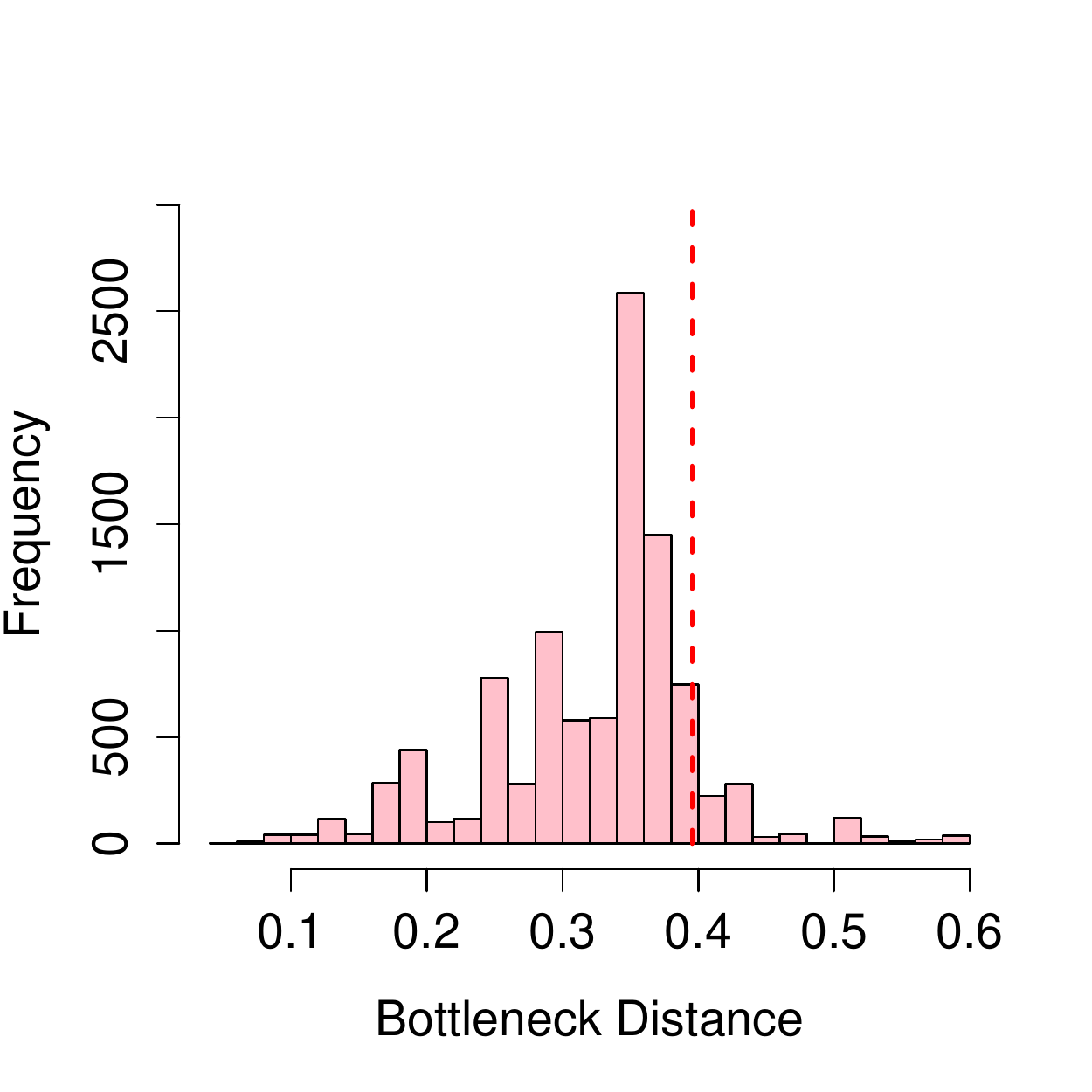}
        \caption{Distribution of $W_{\infty}(Dgm_1(\hat{f}_n), Dgm_1(\hat{f}_n^*))$}\label{subfig:dist2}
    \end{subfigure}
\caption{Empirical distribution of $W_{\infty}(Dgm_p(\hat{f_n}), Dgm_p(\hat{f_n^*}))$, the bottleneck distances between the persistence diagram of the data DTM and the persistence diagrams of the bootstrap realization DTMs. Red dashed lines are the 90\% quantiles. The number of bootstrap samples, $N_{\textrm{boot}}$, is $10^4$. The dataset used here is the same dataset shown in \autoref{subfig:dtm1}.}
 \label{dist}
\end{figure*}

The empirical distribution of $w$, denoted as $\hat{F}_{n,p}(w)$, can be used to approximate the distribution of $W_{\infty}(Dgm_p(\hat{f_n}), Dgm_p(f))$. 
Extracting the $1-\alpha$ quantile from $\hat{F}_{n,p}(w)$, we get the estimated value $\hat{c}_n$ from Eq.~\eqref{equation} for a $1-\alpha$ confidence band.  For example, \autoref{dist} displays the bootstrapped bottleneck distances from the dataset in \autoref{subfig:dtm1}, and the $\hat{c}_n$ for a 90\% confidence interval is displayed as a vertical, dashed, red line.

The persistence diagram in \autoref{subfig:dtm4} displays the confidence band for $H_1$ corresponding to $\alpha = 10\%$. On the diagram, the red triangles outside the pink band can be considered a statistically significant topological generator of $H_1$ at a significance level of 10\%.

This framework will be used to identify the statistically significant filament loops ($H_1$ generators) and cosmic voids ($H_2$ generators) on a persistence diagram.  Rather than using the confidence band construction, we will assign individual $p$-values to each topological generator on a persistence diagram.  This will be discussed in more detail in the next section.

\section{Method}\label{sec:method}

The SCHU code consists of four main steps described in Algorithm \ref{algo:SCHU}, and the persistent homology computation is performed using the TDA package \citep{fasy2014Introduction}. Below, we describe two key steps of SCHU in further detail: (i) computing $p$-values of the homology group generators of a dataset by adapting the framework from \S\ref{inference} and (ii) addressing how to find a representation (i.e. physical locations and boundaries) of the $H_1$ and $H_2$ homology group generators from the persistence diagram back in the physical volume. At the end, the SCHU code produces a catalog of cosmic voids and filament loops, each of which are assigned a $p$-value and a representation in the cosmological volume.



\begin{algorithm}
\caption{SCHU}\label{algo:SCHU}
\begin{algorithmic}
\State Step 1: Load in a galaxy or halo catalog dataset. 
\State Step 2: Perform the persistent homology computation using the DTM function.
\State Step 3: Compute $p$-value for each topological feature (i.e., filament loops and cosmic voids) using a bootstrap technique.
\Statex Step 4: Locate filament loops and cosmic voids in the cosmological volume.
\end{algorithmic}
\label{SCHU_algo}
\end{algorithm}

\subsection{\texorpdfstring{$p$}{p}-values for filament loops and cosmic voids}

\begin{figure}[!ht]
  \centering
    \includegraphics[width=0.5\textwidth]{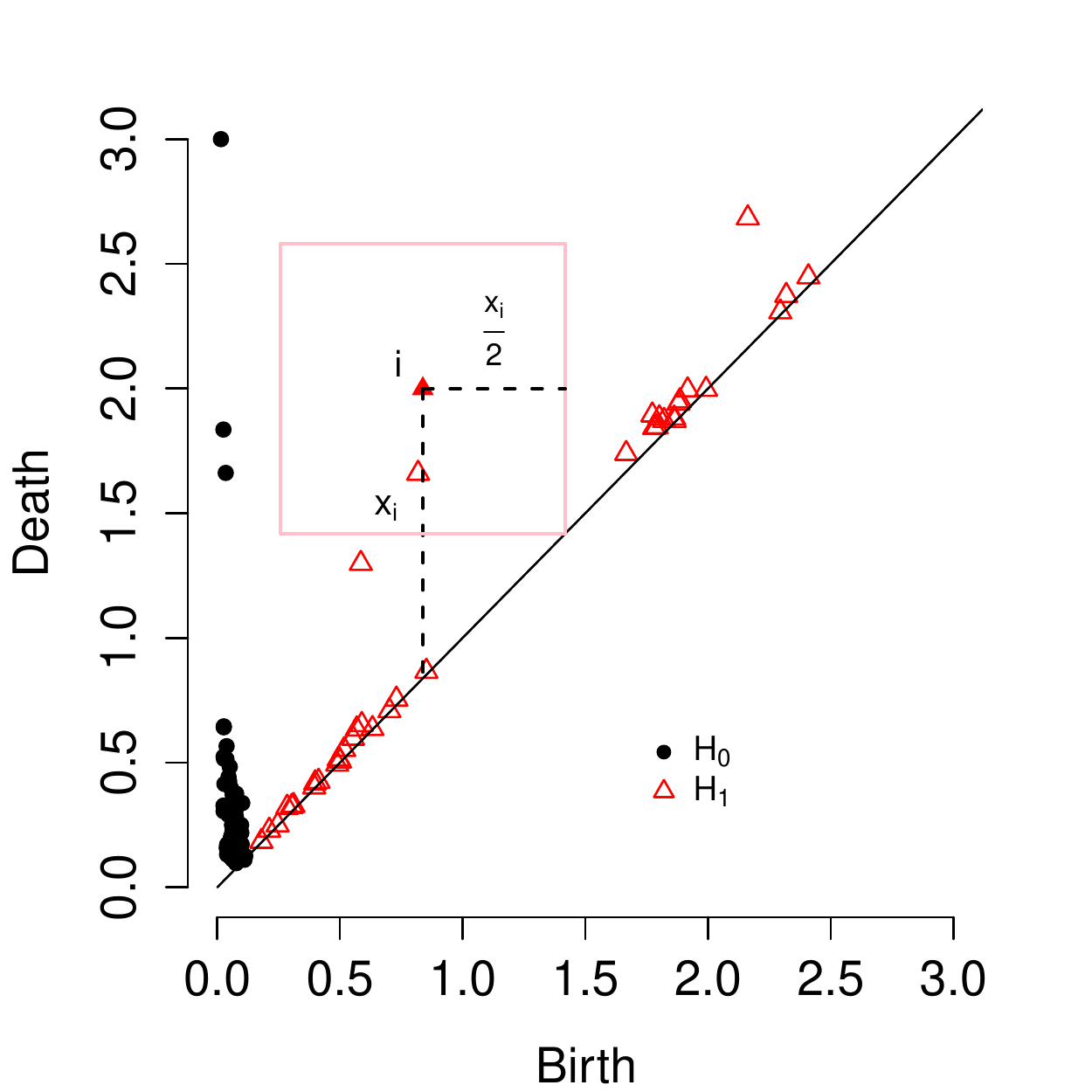}
\caption{A generator $i$ on $Dgm_p(\hat{f_n})$ with persistence $x_i$ (i.e., $x_i = d_i - b_i$). The shortest $L^2$ distance between the generator and the diagonal is $\frac{x_i}{\sqrt{2}}$ and the shortest $L^{\infty}$ distance between the generator and the diagonal is $\frac{x_i}{2}$. Any generator whose $L^{\infty}$ distance to $i$ is less than $\frac{x_i}{2}$ is located in the pink box.
}\label{pvalue}
\end{figure}

Rather than computing the confidence bands from \S\ref{inference}, $p$-values can be assigned for the individual homology group generators appearing on a persistence diagram.
%
%
By changing the value of $c_n$ from Eq.~\eqref{equation}, a corresponding probability $1-\hat{\alpha}=\hat{F}_{n,p}(c_n)$ can be obtained from $\hat{F}_{n,p}(w)$ ($\hat{\alpha}$ is the estimated probability that $W_{\infty}\left(Dgm_p(\hat{f_n}),Dgm_p(\hat{f_n^*})\right)$ is larger than $c_n$). 
For a generator $i$ on $Dgm_p(\hat{f_n})$ with persistence $x_i$ (i.e., $x_i = d_i - b_i$), the shortest distance between the generator and the diagonal is $\frac{x_i}{\sqrt{2}}$.  Noting that the bottleneck distance only considers vertical and horizontal distances (i.e. distances in the death time and birth time directions, respectively, because the $L^{\infty}$ distance is used in its definition), the lowest significance level where generator $i$ would not be considered topological noise (i.e. the generator's $p$-value) corresponds to a particular value of $\hat{F}_{n,p}(w)$ at $\frac{x_i}{2}$.
For example, in \autoref{pvalue}, generator $i$ is the solid red triangle with persistence $x_i$. The $L^{\infty}$ distance from $(b_i, d_i)$ to the diagonal is $\frac{x_i}{2}$. 
Any point with $L^{\infty}$ distance less than $\frac{x_i}{2}$ to $(b_i, d_i)$ will be located within the pink box as shown in \autoref{pvalue}. Thus, the lowest significance level where generator $i$ would not be rejected as noise corresponds to a bottleneck distance of $\frac{x_i}{2}$.

Instead of using a specific significance level $\alpha$ as a cutoff as in \S\ref{inference}, the individual $p$-values can be assigned as follows.  Let $x_i$ be the persistence of generator $i$ on the observed persistence diagram $Dgm_p(\hat{f_n})$.  Then the $p$-value assigned to generator $i$ is
\begin{equation}\label{eq:pvalue}
\text{$p$-value}(i)  = 1 - \hat{F}_{n,p}\left(\frac{x_i}{2}\right),
\end{equation}
where a longer lifetime (i.e., a larger $x_i$) results in a smaller $p$-value.
Generators within the $1 - \alpha$ confidence band discussed in  \S\ref{inference} suggests that the $p$-values are greater than $\alpha$ (i.e., the generators are considered indistinguishable from topological noise), and generators on or outside the confidence band have $p$-values less than or equal to $\alpha$ (i.e. it is considered topological signal). Points on a persistence diagram outside the confidence band with confidence level 90\% have $p$-values less than or equal to 10\%.
%
%

\subsection{Locating filament loops and cosmic voids in the cosmological volume}\label{finding}

Once a persistence diagram is obtained from a dataset with $p$-values assigned to the topological generators, a natural question is where these generators (i.e. the holes) are located in the original data volume.  
For example, if statistically significant filament loops and cosmic voids exist in a given dataset, locating a representation of these voids back into the dataset could be helpful for studying properties of the voids \citep{bos2012darkness}. 

Recall that the points on a persistence diagram represent homology group generators.  Because it is a group generator, multiple representations of the homology group exist that can be used in the original cosmological volume.  On a practical level, suppose some collection of halos make up the boundary of a void.  Among the collection of halos that make up the boundary of the void, various subsets of these halos could also be selected as representations of the void, each of which may have different physical characteristics, e.g., volume and enclosed mass.
%
%

\begin{figure*}
 \centering
    \begin{subfigure}{0.35\textwidth}
        \centering
    \includegraphics[width=\textwidth]{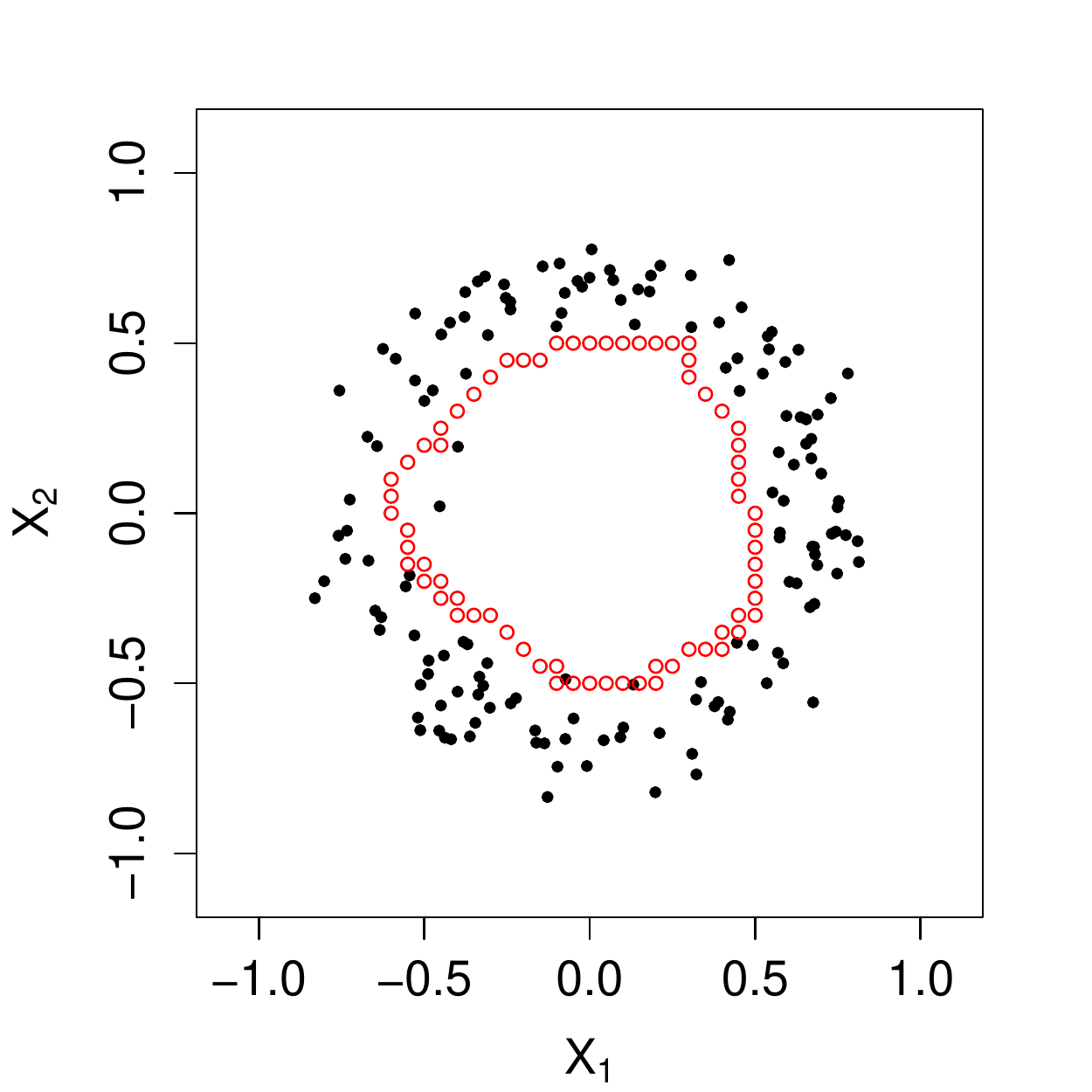}
        \caption{Representation: $grid~size=0.05$, $m_0=0.05$}\label{subfig:rep1}
    \end{subfigure}
        \begin{subfigure}{0.4\textwidth}
        \centering
    \includegraphics[width=\textwidth]{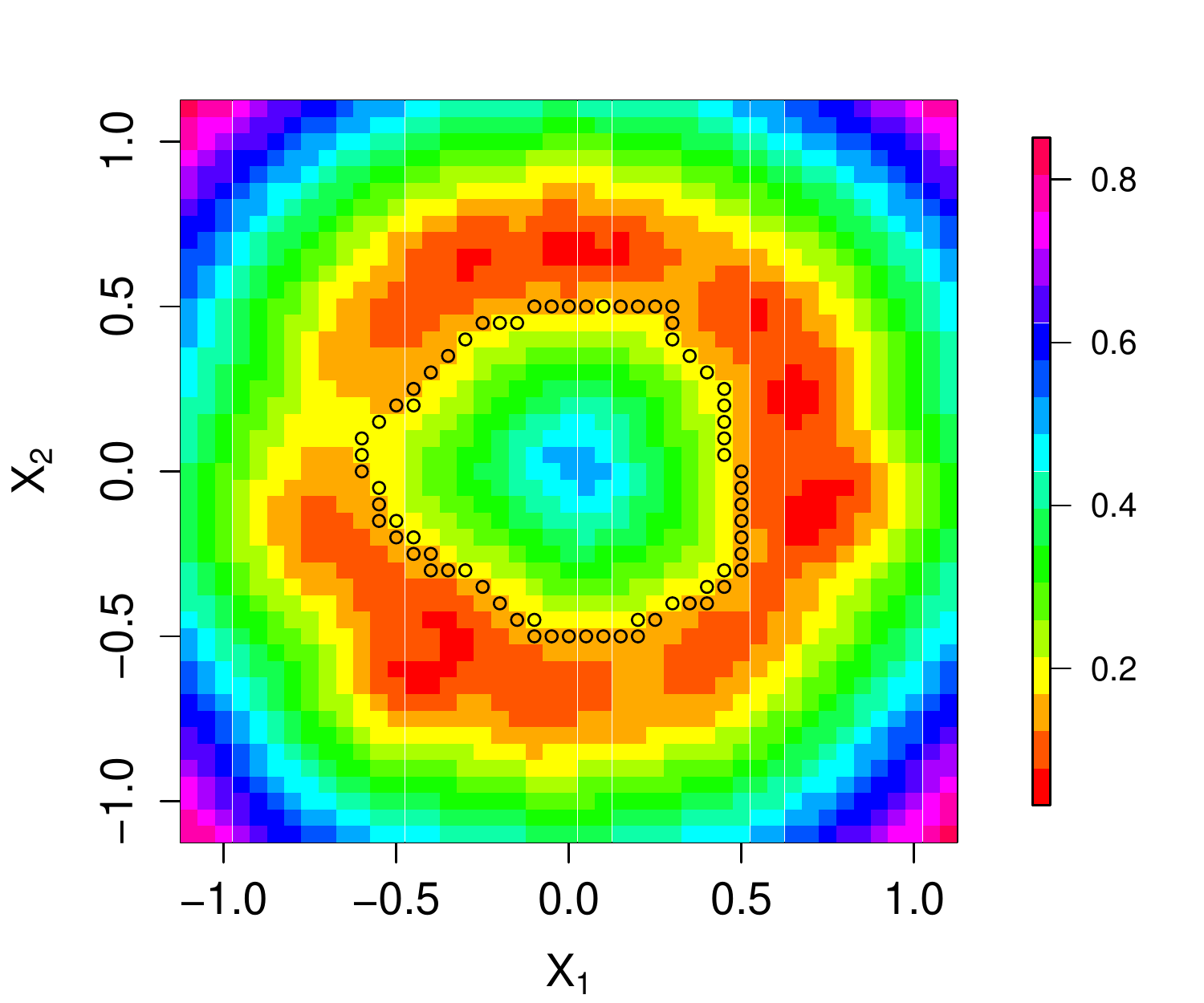}
        \caption{Contour: $grid~size=0.05$, $m_0=0.05$}\label{subfig:rep1_contour}
    \end{subfigure}
    \begin{subfigure}{0.35\textwidth}
        \centering
    \includegraphics[width=\textwidth]{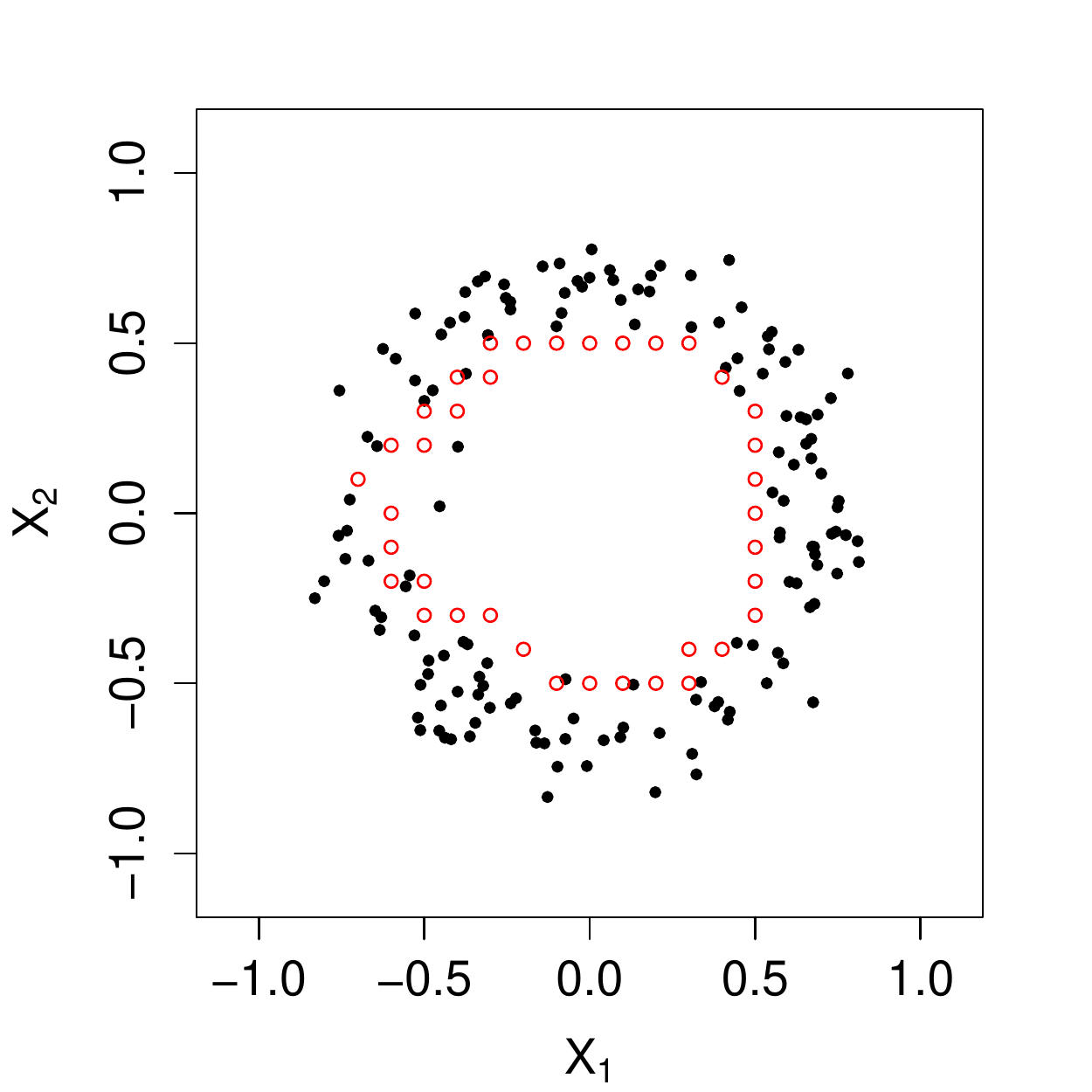}
        \caption{Representation: $grid~size=0.1$, $m_0=0.05$}\label{subfig:rep2}
    \end{subfigure}
        \begin{subfigure}{0.4\textwidth}
        \centering
    \includegraphics[width=\textwidth]{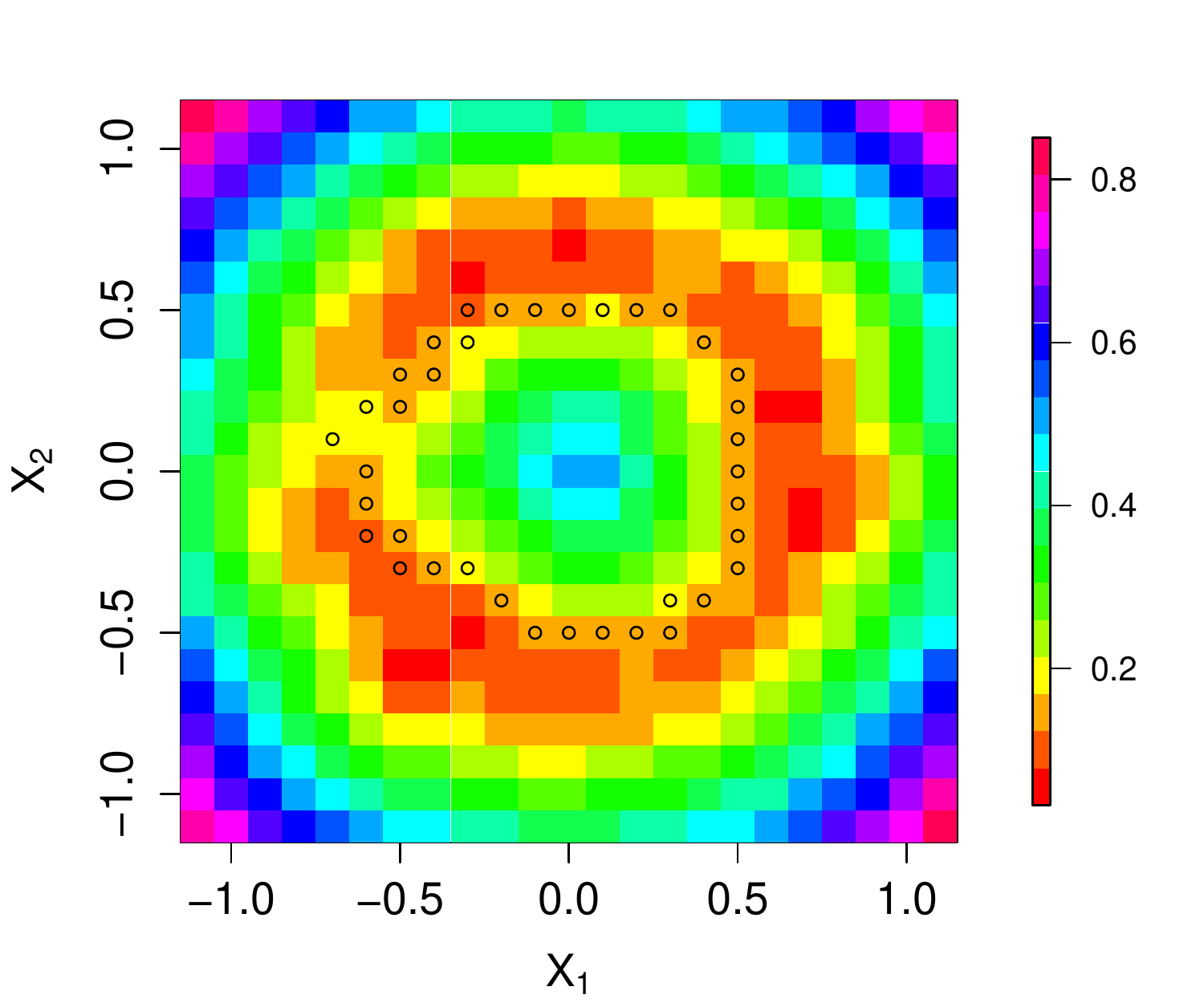}
        \caption{Contour: $grid~size=0.1$, $m_0=0.05$}\label{subfig:rep2_contour}
    \end{subfigure}
        \begin{subfigure}{0.35\textwidth}
        \centering
    \includegraphics[width=\textwidth]{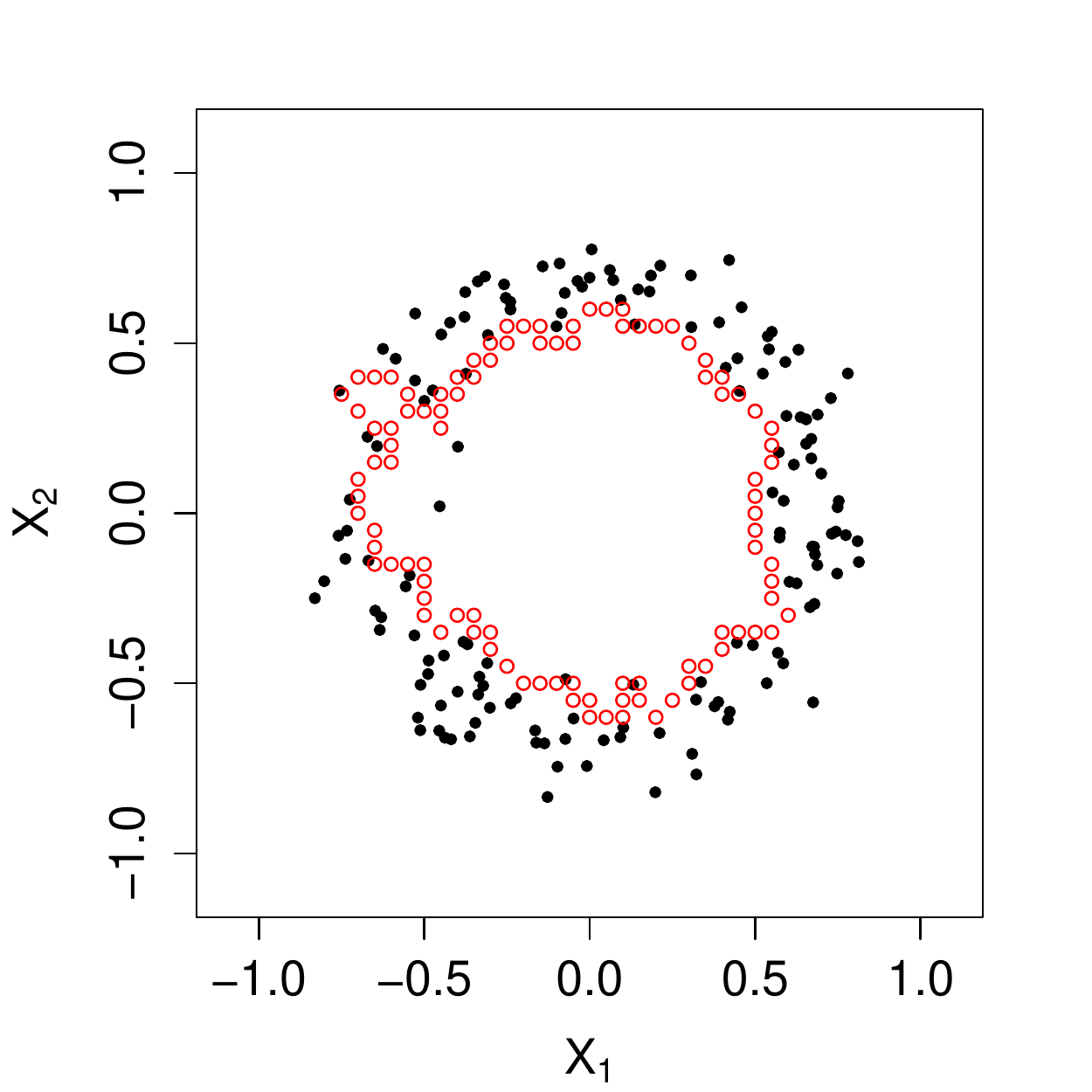}
        \caption{Representation: $grid~size=0.05$, $m_0=0.01$}\label{subfig:rep3}
    \end{subfigure}
        \begin{subfigure}{0.4\textwidth}
        \centering
    \includegraphics[width=\textwidth]{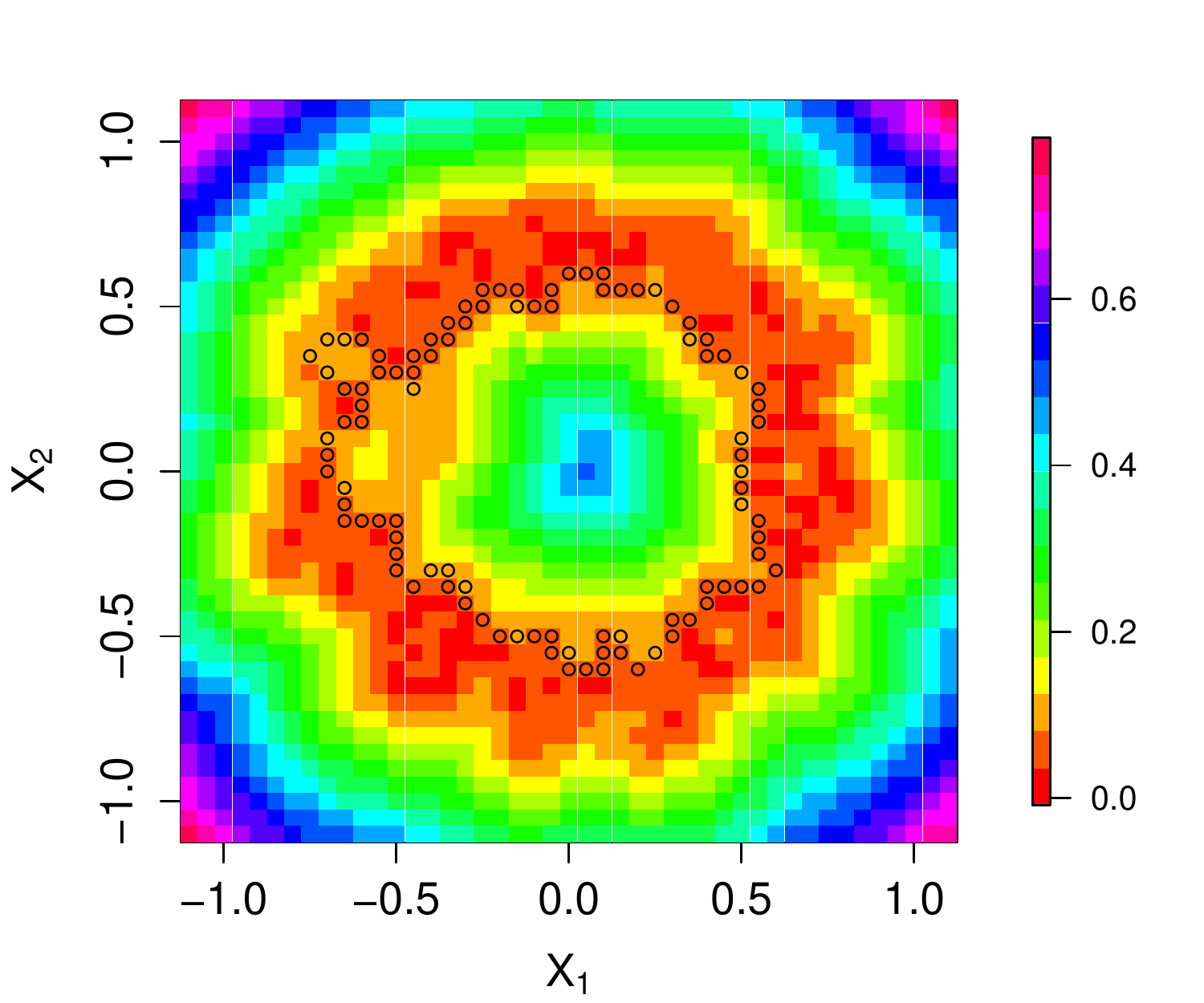}
        \caption{Contour: $grid~size=0.05$, $m_0=0.01$}\label{subfig:rep3_contour}
    \end{subfigure}
\caption{To compute a DTM function, a grid is generated with a specified grid size. The grid size and $m_0$ are two key parameters that influence resulting representations. (a), (c), and (e): Three representations of the same $H_1$ generator. (b), (d), and (f): Three contour maps of DTM functions with three representations marked in black circles.
} \label{rep_illustration}
\end{figure*}

Additionally, the DTM affects the representations of the homology group generators through the resolution of the grid and the selected $m_0$, (i.e., the proportion of nearest-neighbor halos/galaxies used in the DTM calculation).
In \autoref{subfig:rep1}, \autoref{subfig:rep2}, and \autoref{subfig:rep3}, there are three different representations of the $H_1$ generator shown in red circles. The three different representations correspond to contours of three lower level sets of three different DTM functions. 
In \autoref{subfig:rep1_contour}, \autoref{subfig:rep2_contour} and \autoref{subfig:rep3_contour}, the contour maps of DTM are shown and the three representations are marked in black circles. The representation reported from the persistent homology algorithm implemented in the {\tt R TDA} package is the inner contour of the smallest lower level set that forms a $H_1$ generator (which will be explained later). Thus, the representation is determined by the DTM function constructed over the dataset.

\begin{figure*}
 \centering
    \begin{subfigure}{0.40\textwidth}
        \centering
    \includegraphics[width=\textwidth]{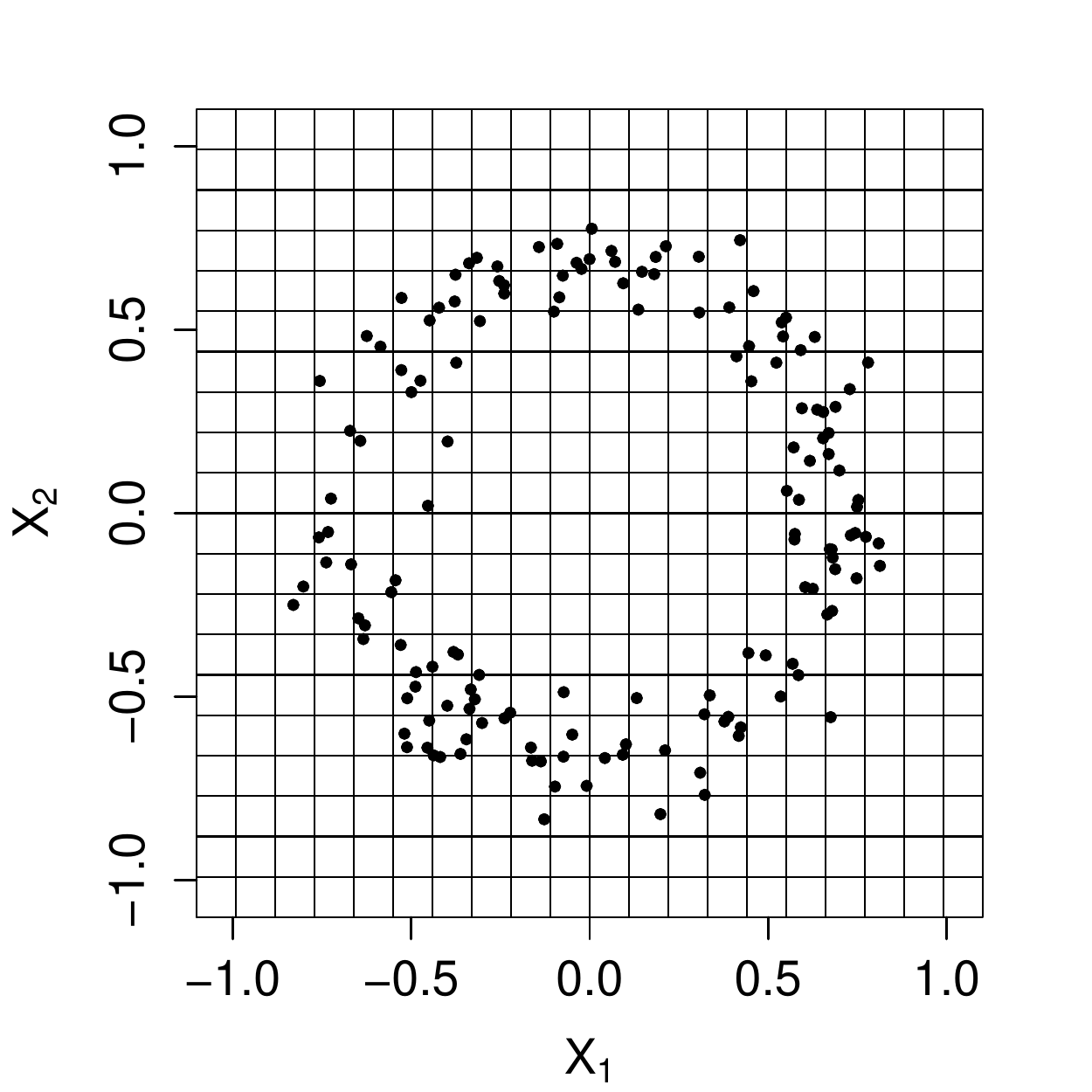}
        \caption{Constructing a grid}\label{subfig:filtra1}
    \end{subfigure}
        \begin{subfigure}{0.40\textwidth}
        \centering
    \includegraphics[width=\textwidth]{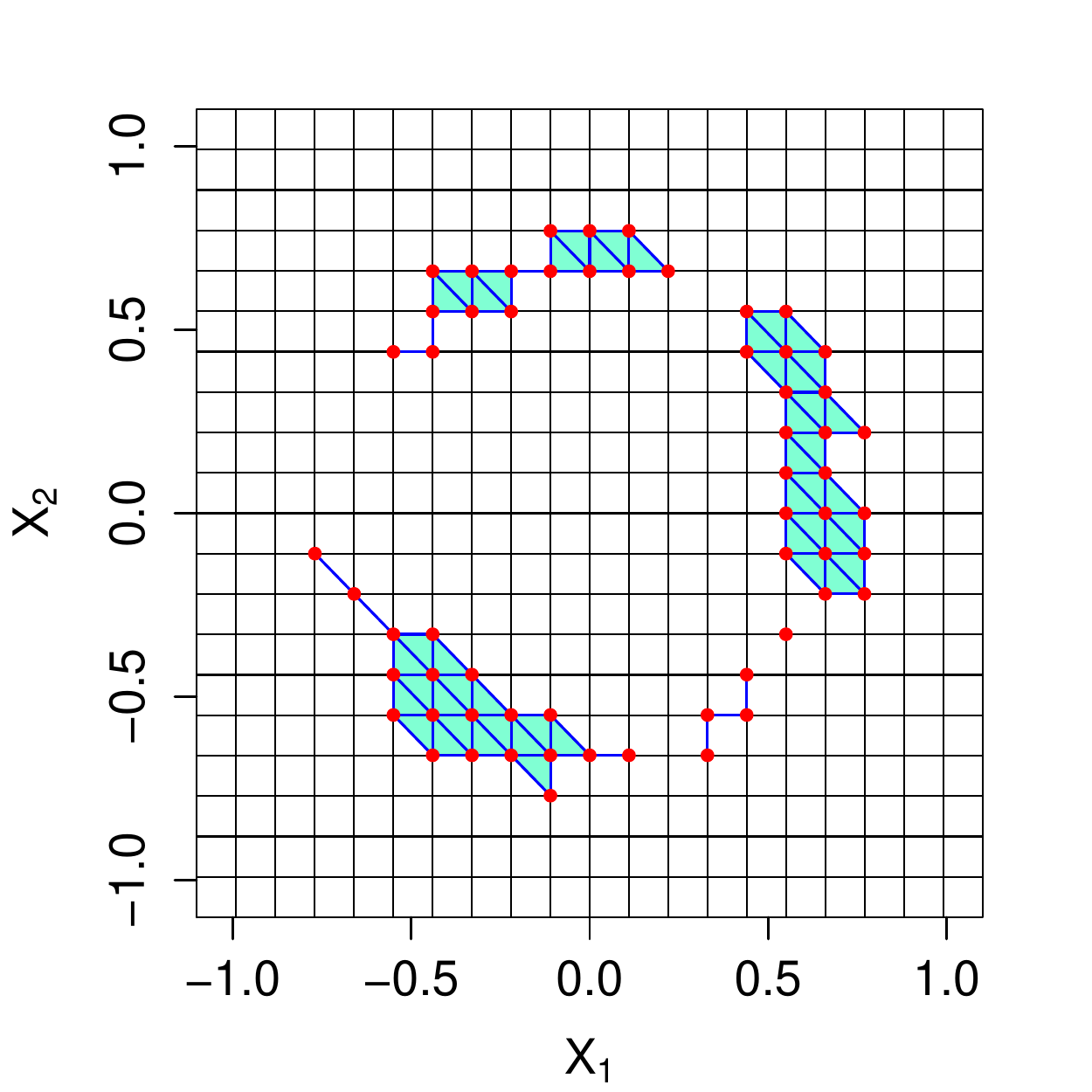}
        \caption{A simplicial complex on grid}\label{subfig:filtra2}
    \end{subfigure}
\caption{The same dataset as in \autoref{rep_illustration}. (a) A dataset with an overlaying grid for the DTM estimation. (b) The red points highlight an arbitrary lower level set on the grid.  The simplicial complex from this lower level set includes the red points, the blue segments, and the cyan triangles. 
} \label{filtration}
\end{figure*}

In order to compute a DTM function, a grid needs to be defined with a specified grid size: for example, see the grid displayed in \autoref{subfig:filtra1}.  Then, the DTM is computed for each point on the grid using Eq.~\eqref{eq:dtm_obs}. The grid size and $m_0$ in Eq.~\eqref{eq:dtm_obs} are two key parameters that influence resulting representations. The grid size determines the resolution of the DTM (and thus, the resolution of the void/filament loop representations reported by SCHU) and $m_0$ determines the spatial smoothness of the DTM.

\autoref{subfig:rep1_contour} and \autoref{subfig:rep2_contour} have the same $m_0$ (smoothness) but different grid sizes. \autoref{subfig:rep1_contour} and \autoref{subfig:rep3_contour} have the same grid size but different $m_0$.

In the persistent homology algorithm used in this work \citep{Dionysus}, there is a way to record homology group generator locations during the computation of a persistence diagram \citep{Edelsbrunner2002, zomorodian2005computing, de2011dualities}.
The method for finding generators from persistence diagrams back in the cosmological volume uses the output of the computation of the persistence diagram. To understand the persistent homology computation, we first provide more details on the filtration construction on a grid. 

As mentioned above, a grid needs to be defined to compute a DTM function, as displayed in \autoref{subfig:filtra1}.
%
After the DTM is computed using Eq.~\eqref{eq:dtm_obs}, a sequence of lower level sets  $L_t=\{x:\hat{d}_{m_0}(x)\leq t\}$ are determined (see \S\ref{persistent_homology}). 
See Figure~\ref{subfig:filtra2} for an example of a lower level set on a grid using an arbitrary threshold $t$, which consists of a subset of grid vertices drawn as red points.

A simplicial complex \citep{edelsbrunner2010computational, zhu2013persistent} is constructed using the lower level sets at threshold $t$ (i.e., the vertices of the grid points with corresponding DTM values less than the threshold, such as the red points displayed in Figure~\ref{subfig:filtra2}). 
A simplicial complex $K$ is a set of (possibly different order) simplices such that (i) any face of a simplex of $K$ is also a simplex in $K$, and (ii) the intersection of any two simplices in $K$ is a face of both simplices or empty.\footnote{A $0$-simplex is a vertex, a $1$-simplex is a segment, and a $2$-simplex is a triangle, etc. A face of a simplex $K$ is any simplex with order lower than $K$ that is part of $K$. For example, if $K$ is a triangle, one of the three vertices is a face of $K$, and one of the three edges is a face of $\sigma$. }
%
%
 In \autoref{subfig:filtra2}, we show the simplicial complex from the arbitrary lower level set displayed. 
 The simplicial complex includes the red points (0-simplices), the blue segments (1-simplices), and the cyan triangles (2-simplices).  
 Note that for the arbitrary threshold selected in the figure, the red points connected with blue segments represented a single connected component (i.e., an $H_0$ generator).  As the lower level set threshold increases, eventually the red points would be connected in such a way that a loop forms (i.e. an $H_1$ generator).


\begin{sloppypar}
The boundary of a $p$-simplex is the set of $(p-1)$-simplex faces \citep{zhu2013persistent}.\footnote{The boundary of an edge is the two vertices. The boundary of a triangle is the three edges \citep{zhu2013persistent}.}
A boundary matrix $D$ records the boundary information of the simplicial complex across the filtration.  As the DTM threshold increases, simplices get added to $D$. 
Each row and each column represents a simplex: $D_{i, j}=1$ if the $i$-th simplex is a boundary of the $j$-th simplex (e.g., if the $j$-th simplex is a segment and the $i$-th simplex is a point on its boundary) and $D_{i, j}=0$ if not. 
The details of the persistent homology computations (which take $D$ to a reduced form of $D$, $R$) are beyond the scope of this paper, but an interested reader can see  \citet{Edelsbrunner2002, zomorodian2005computing, de2011dualities, cohen2006vines} for details.  The reduced boundary matrix $R$ contains the information about the various generators of a persistence diagram.
If the $j$-th column of $R$ representing an $H_{p-1}$ generator (since the $j$-th simplex is a $p$-simplex) contains at least one non-zero element, the non-zero element(s) indicate which $(p-1)$-simplicies formed the generator.  The generator disappears at the time in the filtration when the $j$-th simplex appears \citep{de2011dualities}. 
The non-zero columns of matrix $R$ are used as representations of the homology group generators. 
For example, suppose column $j$ of $R$ represents an $H_1$ generator with only three non-zero elements at positions $i$, $i+1$, and $i+2$.  Then simplex $i$, $i+1$, and $i+2$ are $1$-simplices (i.e., segments) that can be used as a representation for the $H_1$ generator (i.e., the loop) of column $j$ of $R$. 
In practice, the vertices (i.e., the grid points) of the $p$-simplices can be used as the representation of an $H_p$ generator.
\end{sloppypar}

During the computation, lower level sets are constructed using an increasing threshold $t$. 
The lower level sets are in an increasing order, since a later set contains an earlier set. 
For a $H_1$ or $H_2$ generator, there exists a lower level set that first forms the $H_1$ or $H_2$ generator along the filtration, which is the smallest lower level set that forms the $H_1$ or $H_2$ generator.
This smallest lower level set is generally what appears in the boundary matrix for representing that homology group generator.
In general, the representation is the inner contour of the smallest lower level set that forms the $H_1$ or $H_2$ generator.

\begin{figure*}[!ht]
  \centering   
  \begin{subfigure}{0.42\textwidth}
    \includegraphics[width=1\textwidth]{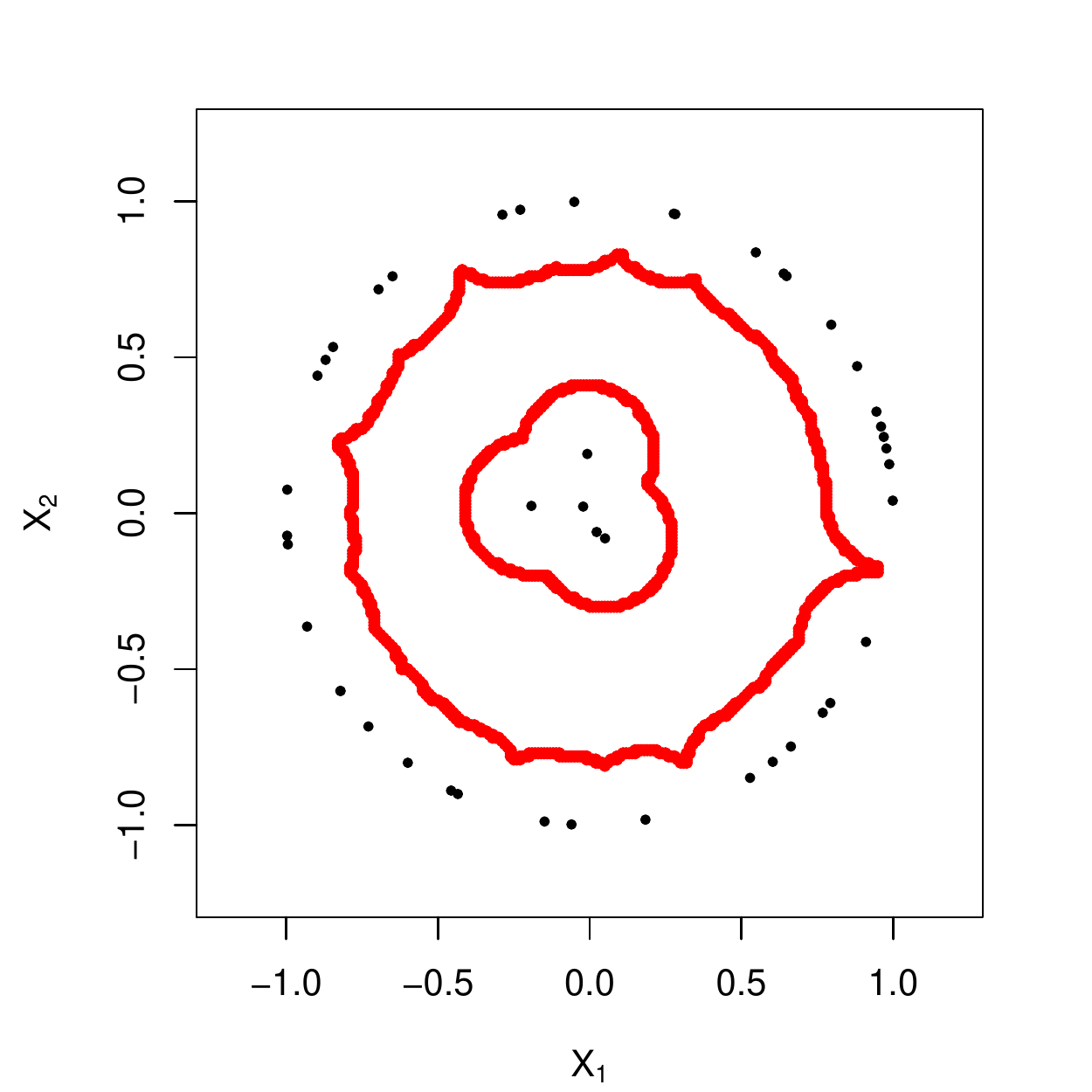}
\caption{A loop with several points inside.}\label{modification1}
    \end{subfigure}
      \centering   
       \begin{subfigure}{0.38\textwidth}
    \includegraphics[width=1\textwidth]{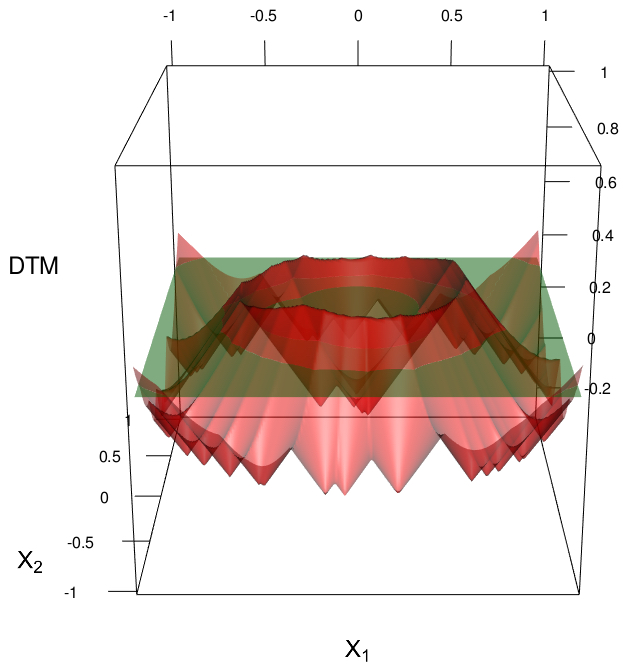}
\caption{Corresponding DTM with a threshold hyperplane. }\label{modification2}
    \end{subfigure}
    \caption{(a) The red lines are the representation returned by the persistent homology algorithm. There are two separate loops in this representation, but there is only one meaningful $H_1$ generator representation (the larger red loop). (b) The red surface shows the DTM function for the dataset. The green surface is the threshold for the lower level set that corresponds to the representation in (a). The intersection of the red and green surfaces has two separate closed loops, which are returned as the representation in (a). }\label{modification}
\end{figure*}

\textit{Remark 1}:  For $H_1$ generators, when there are points inside the loop, the representation returned by the algorithm can include an extra loop around the inner points. \autoref{modification} illustrates this special case. 
The representation is shown in red and includes two closed loops. 
However, only the larger loop is relevant as the inner points do not contribute to the $H_1$ generator and is an artifact of the algorithm. 
As discussed in \S\ref{finding}, the homology group generator representation corresponds to a lower level set of the DTM function. 
\autoref{modification2} shows the DTM function of the data in \autoref{modification1}. The green surface is the threshold for the lower level set that corresponds to the $H_1$ generator represented in red in \autoref{modification1}. 
Because of the inner points, the lower level set is empty around the inner points producing two closed loops.  SCHU includes a step to remove the artificial inner loop from the representation.

\textit{Remark 2}:  In order to define a DTM for a given dataset, the grid resolution and $m_0$ need to be selected. The grid resolution should be high enough to pick out features at a meaningful physical scale; generally the higher the resolution the better. The thickness of filaments and walls lies in the range of $1-5 h^{-1}$ Mpc \citep{Cautun2014}; therefore, a grid resolution at or below the Mpc scale should be sufficient for identifying distinct voids separated by such walls and filaments.
A statistically rigorous method for selecting $m_0$ for persistent homology is still an open question. However, \citet{chazal2017robust} provide two suggestions for selecting $m_0$ using computationally intensive approaches (e.g. one approach selects the $m_0$ that maximizes the number of statistically significant homology group generators).  Since there is not a statistically rigorous or computationally feasible approach for selecting $m_0$, we selected an $m_0$ so that the number of nearest neighbors was around 50.
If $m_0$ is selected to be too small then the resulting DTM may be too noisy to isolate important homological signals; if $m_0$ is selected to be too large, the resulting DTM may be too smooth and important homological signals may get washed out.

\section{Application to Voronoi Foam Data}\label{sec:voronoi_foam}
In order to demonstrate the performance of SCHU for finding statistically significant generators on a persistence diagram and then locating those generators in the original data, we consider a simulation study using data that mimic the large-scale structure of the Universe and focus on locating cosmic voids using $H_2$ generator representations. In the simulation study, we know the ground truth of where the voids are located and so can test its ability to find the true voids.  

\begin{figure*}
 \centering
    \begin{subfigure}{0.4\textwidth}
        \centering
    \includegraphics[width=\textwidth]{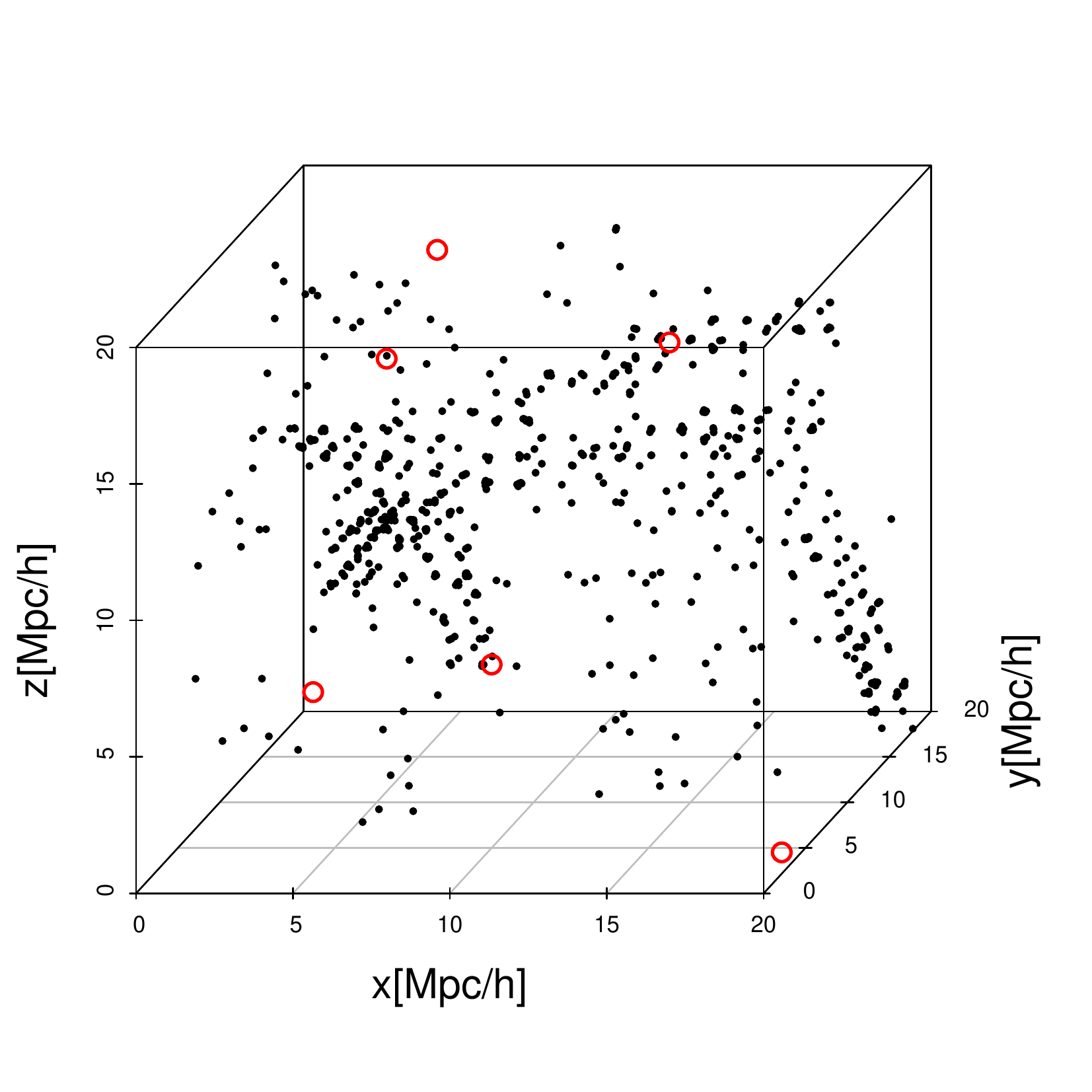}
        \caption{A 6-cell Voronoi foam dataset.}\label{subfig:voro_illustrate1}
    \end{subfigure}
        \begin{subfigure}{0.4\textwidth}
        \centering
    \includegraphics[width=\textwidth]{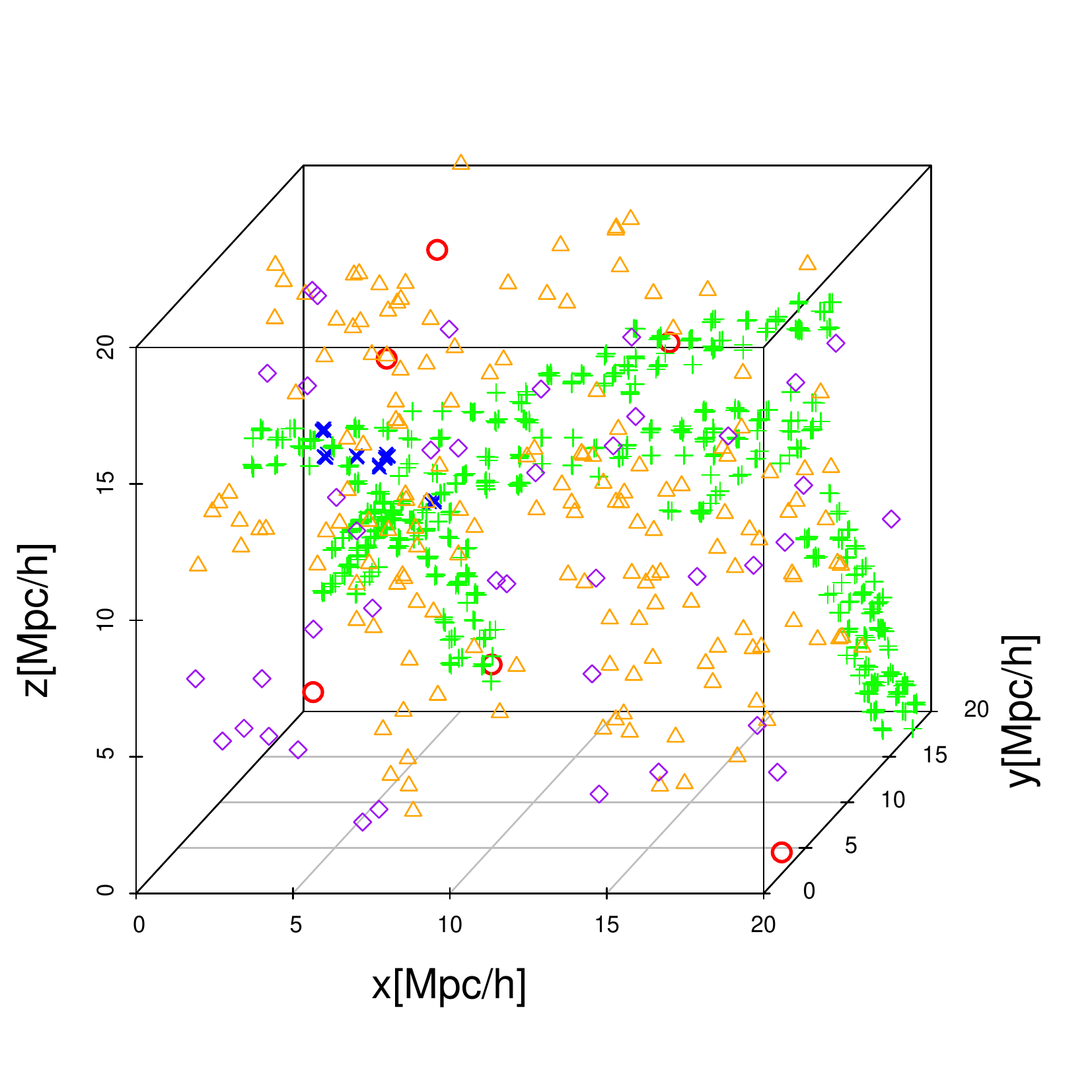}
        \caption{Different points in different colors.}\label{subfig:voro_illustrate2}
    \end{subfigure}
        \begin{subfigure}{0.13\textwidth}
        \centering
    \includegraphics[width=\textwidth]{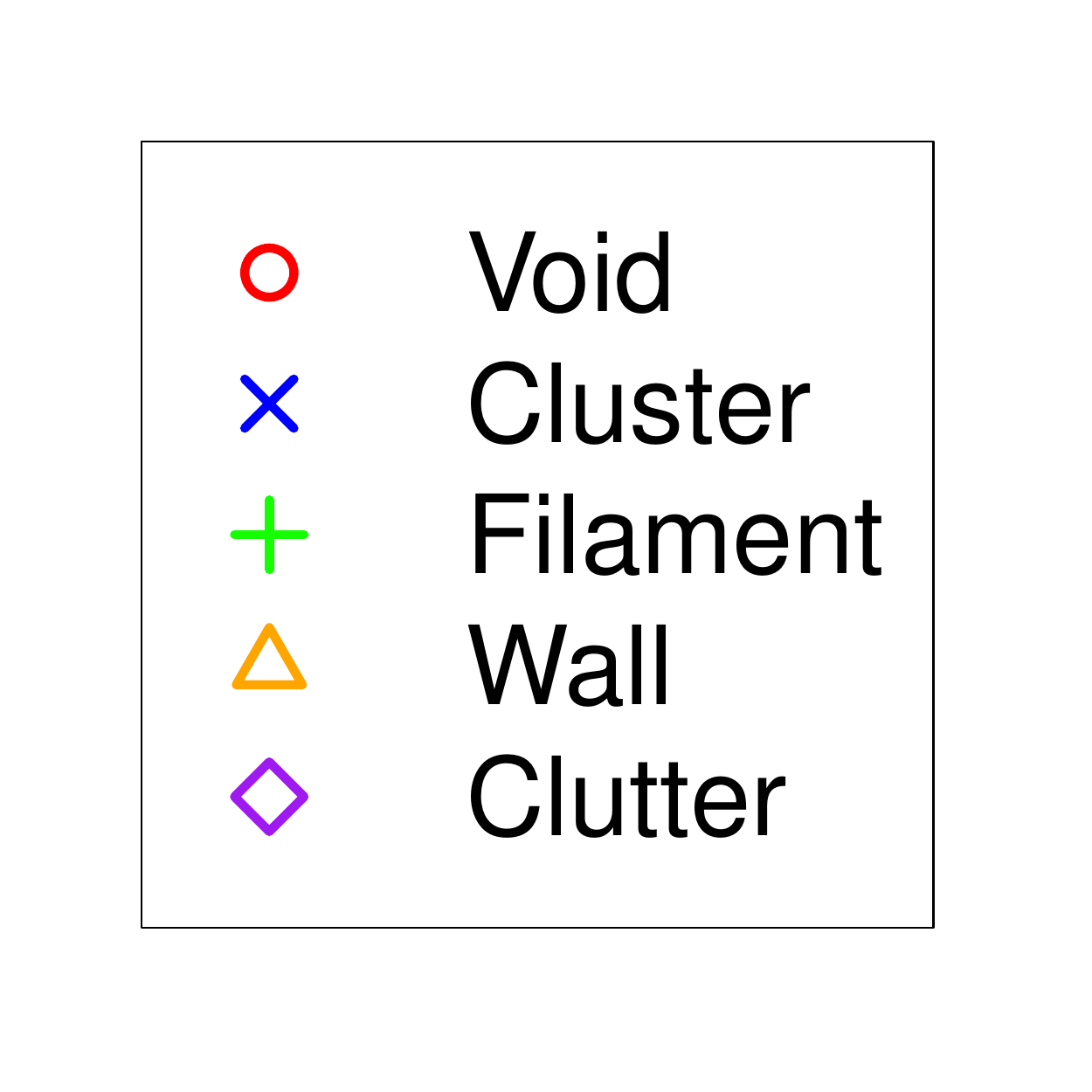}

    \end{subfigure}
\caption{An illustration example of a 6-cell Voronoi foam dataset. (a) Void points are in red and the generated Voronoi foam dataset are in black. (b) Perturbed cluster points, filament points, and wall points are drawn in blue, green, and orange. Clutter (noise) points are in purple, which were sampled uniformly at random in the simulation cube. 
} \label{voro_illustrate}
\end{figure*}

The generated data used in this study come from an approximation to the \emph{Voronoi foam} model \citep{icke1987fragmenting, weygaert1989fragmenting, weygaert1994fragmenting}, which can be considered an approximation of the large-scale structure of the Universe \citep{icke1991galaxy}. 
Our implementation of the Voronoi foam model begins with a Voronoi tesselation\footnote{A Voronoi tessellation is a type of partition that divides a space into several regions based on distance to a set of points called seeds. The set of seeds is specified beforehand and each seed is associated with a region or cell where all points are closer to this seed than to any other seed.} 
on a random set of points, or seeds, in the 3D simulation cube volume.  These seeds are referred to as \emph{void points}
%
because their associated Voronoi cells become the cosmic voids. 
An illustration example of a 6-cell Voronoi foam simulation dataset is shown in \autoref{subfig:voro_illustrate1}, where void points are drawn in red and the generated Voronoi foam data are drawn in black.
A grid is generated over the 3D simulation cube at a specified resolution. 
Each grid point is assigned a void label indicating the nearest void point (i.e., indicating in which Voronoi cell the point is). For each grid point, its neighboring grid points are checked to see how many different void labels are assigned to its neighbors. 
If there are more than three different labels, the grid point is considered as a cluster point.  
This is because this grid point is equal-distance to more than three void points and so it is at the intersection of more than three Voronoi cells. 
Similarly, the grid point is considered as a filament point if there are exactly three different labels, and is considered as a wall point if there are exactly two different labels. If there is just one label, the grid point is considered within a void region.
Then we have a set of cluster points, a set of filament points and a set of wall points.
With specified proportions ($f_c$ for clusters, $f_f$ for filaments, and $f_w$ for walls), certain numbers of cluster points, filament points and wall points are sampled from these sets. 
In order for the points not to fall exactly on the grid, Gaussian noise with specified standard deviation, $\sigma_{\text{perturb}}$, is added to the sampled points. 
In \autoref{subfig:voro_illustrate2}, perturbed cluster points, filament points, and wall points are drawn as blue x's, green pluses, and orange triangles, respectively.
Also, a fraction, $f_n$, of clutter noise points are sampled uniformly over the simulation cube, shown in \autoref{subfig:voro_illustrate2} as purple diamonds. 
The perturbed cluster points, filament points and wall points, together with clutter noise points, form a Voronoi simulation dataset, with $f_c + f_f + f_w + f_n = 1$.
%
%
%

A persistence diagram is computed for a realization of the Voronoi foam simulation in order to find the cosmic voids (i.e., the $H_2$ generators). 
Since the positions of the seeds are known (i.e., the locations of cosmic voids), we are able to compare the locations of cosmic voids found using SCHU to the true void points to check how well we are able to locate a representation of the $H_2$ generators in the data. The $H_2$ representation identified by SCHU defines the boundary surface of the void, and thus the average of these boundary surface points is approximately the void's center of volume.

The Voronoi foam simulated cosmological volume is represented by a cubic lattice with $100h^{-1}$Mpc per dimension and a grid spacing of $1h^{-1}$Mpc. 
In order to initiate the foam simulation, we randomly draw 23 seeds in order to form 23 cosmic voids in the volume. 
This number of voids was motivated by past results suggesting that voids have characteristic length scales on order tens of $h^{-1}$Mpc \citep{Kreisch2018} and occupy roughly 75\% of the cosmic volume \citep{Cautun2014}. The environment sample fractions are chosen to in order to best visually reproduce the cosmic web by-eye with the Voronoi foam model: $f_c = 0.08$, $f_f = 0.2$, $f_w = 0.65$, and $f_n=0.07$, with added Gaussian noise of $\sigma_{\textrm{perturb}} = 1 h^{-1}$Mpc applied to the sample\footnote{
We tried various environment fractions, including values based on Table 2 of \citet{libeskind2018tracing}, and found that the void/filament loop identification and significance levels are relatively insensitive to these ratios.}.
The total number of sampled data points was selected to match the mean halo density in the N-body simulation from \citet{libeskind2018tracing} that we use in Section~\ref{libeskind}. This consideration yields a total of $35,000$ Voronoi foam points plus $4,800$ boundary wall points.
To capture cosmic voids that may extend beyond the boundaries of the survey volume or simulation boundaries, boundary points are generated by uniform sampling on each of the six surfaces of the cube, while adding Gaussian noise with $\sigma=0.25h^{-1}$Mpc on the direction perpendicular to the surface.

\begin{figure*}
 \centering
    \begin{subfigure}{0.36\textwidth}
        \centering
    \includegraphics[width=\textwidth]{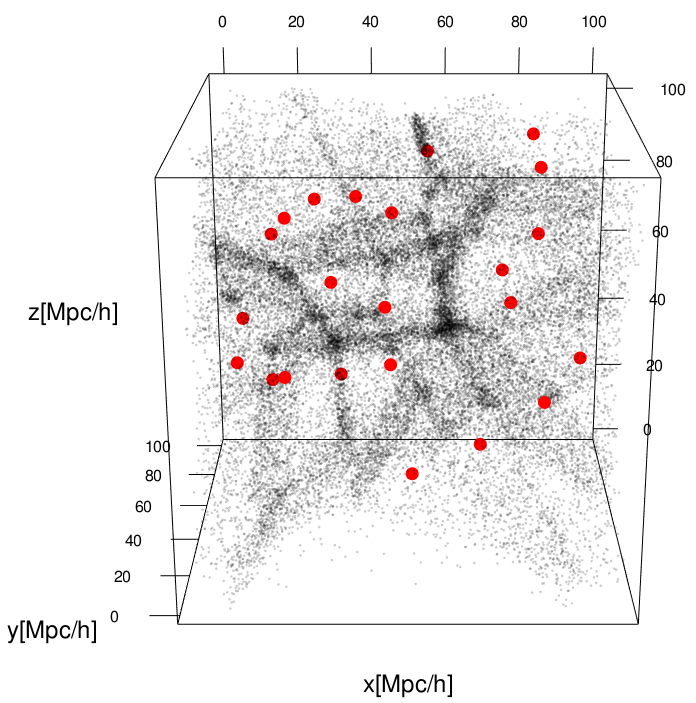}
        \caption{Voronoi foam data}\label{subfig:voro_1}
    \end{subfigure}
    \begin{subfigure}{0.36\textwidth}
        \centering
    \includegraphics[width=\textwidth]{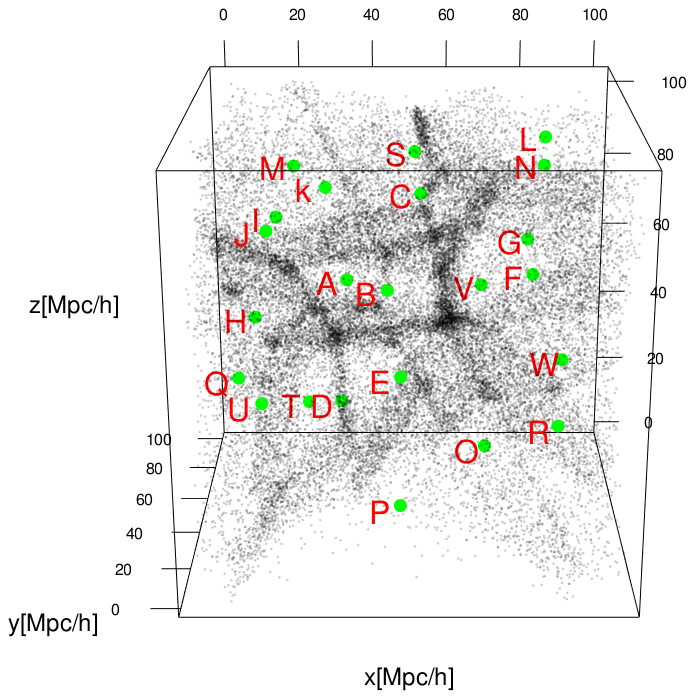}
        \caption{Volume centers}\label{subfig:voro_2}
    \end{subfigure}
    
        \begin{subfigure}{0.32\textwidth}
        \centering
    \includegraphics[width=\textwidth]{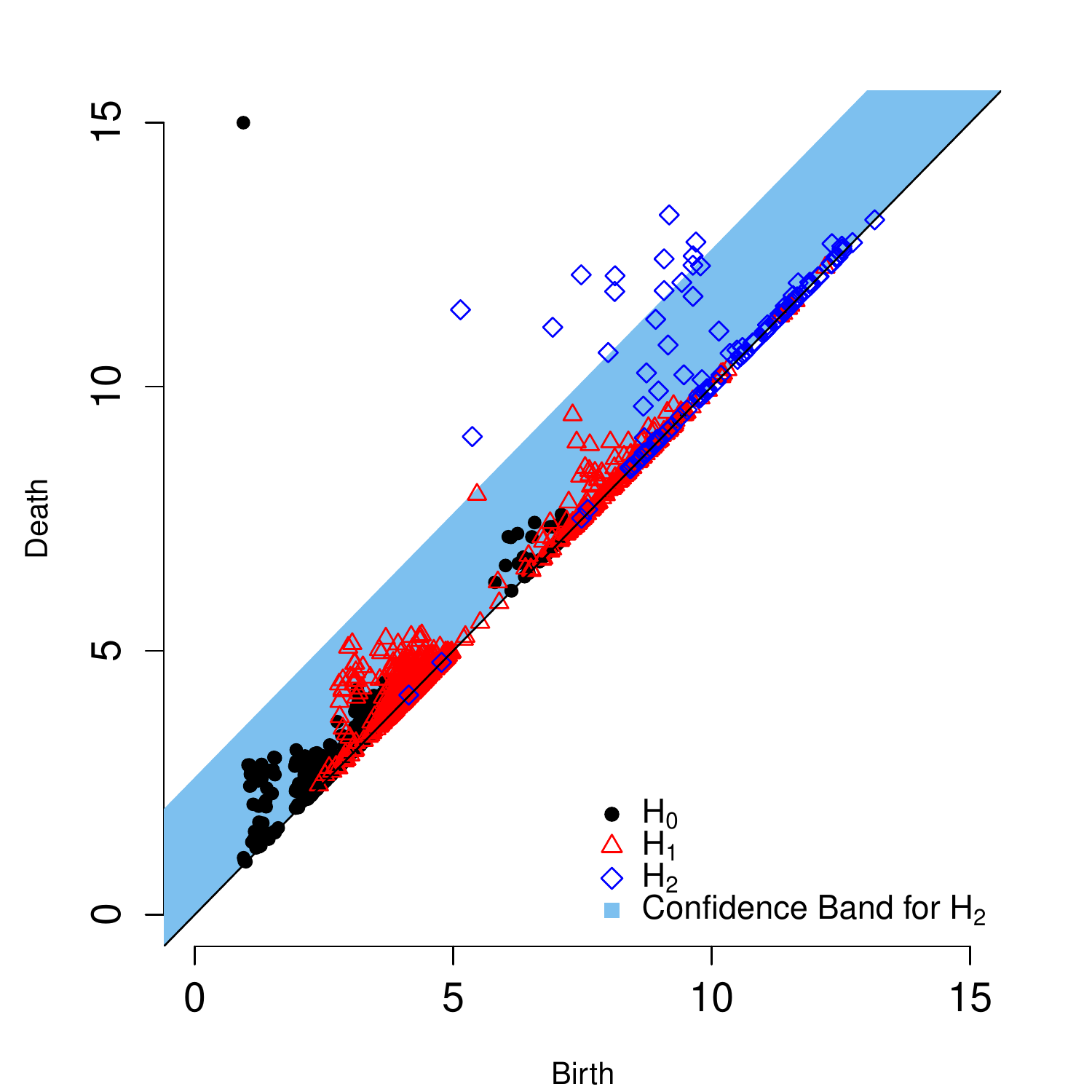}
        \caption{Persistence diagram}\label{subfig:voro_3}
    \end{subfigure}
        \begin{subfigure}{0.32\textwidth}
        \centering
    \includegraphics[width=\textwidth]{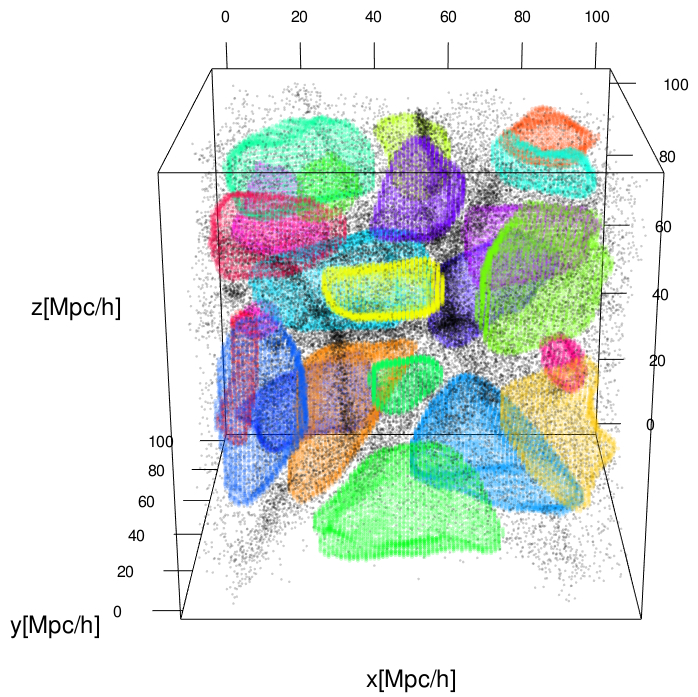}
        \caption{Example voids}\label{subfig:voro_4}
    \end{subfigure}
            \begin{subfigure}{0.32\textwidth}
        \centering
    \includegraphics[width=\textwidth]{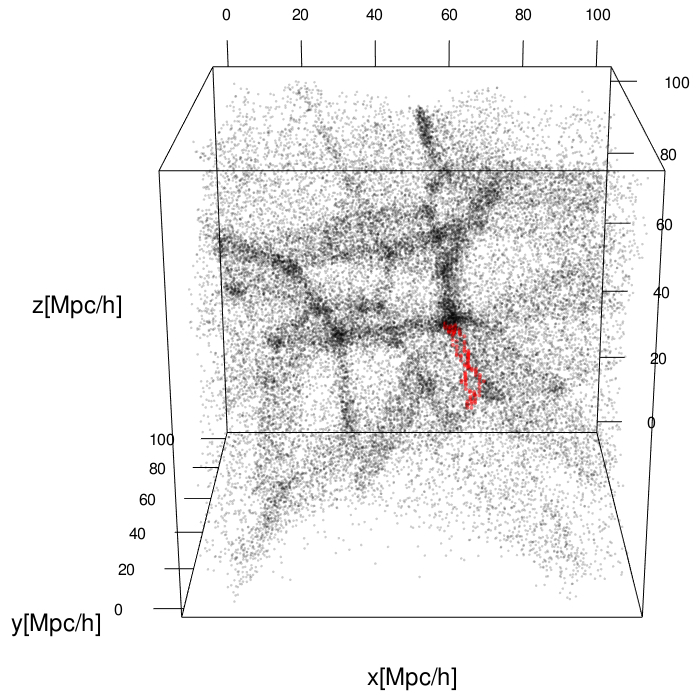}
        \caption{Example filament loop}\label{subfig:voro_5}
    \end{subfigure}
\caption{(a) Voronoi foam data where the red points are the void points used to generate the Voronoi tessellation. (b) Volume centers of the 23 most significant $H_2$ generators with labels (c) Persistence diagram of (a), where the blue band is 90\% confidence band for $H_2$. (d) Detailed shapes of the 23 voids found by SCHU. (e) For illustration, the red points show a representation of a $H_1$ generator with $p$-value 0.011.
} \label{voro}
\end{figure*}

\autoref{subfig:voro_1} shows the simulated Voronoi foam data, where the red points are seeds generating the Voronoi diagram. Using a $m_0$ of 0.001, which corresponds to using 40 nearest neighbors when calculating the DTM function at each grid point.  \autoref{subfig:voro_3} is the resulting persistence diagram from this dataset, where the blue band is the 90\% confidence band for $H_2$.
In \autoref{subfig:voro_3}, there are 23 $H_2$ generators that are noticeably separate from the diagonal, although some of them are not outside the 90\% confidence band. 
We map the center of each of the 23 most persistent $H_2$ generators, which is found by computing the volume center of the $H_2$ representation located in the data, to its nearest void seed point. 
By matching each of the 23 $H_2$ generator volume centers to its nearest void seed, each of the 23 seeds were uniquely matched with a corresponding $H_2$ generator suggesting that SCHU was able to accurately locate the cosmic voids. 
Using $N_{\textrm{boot}}=1,000$ bootstrap samples, a $p$-value for each $H_2$ generators was calculated. 
\autoref{subfig:voro_2} shows the volume centers of the 23 $H_2$ generators, labeled A through W, for each void, and \autoref{voro_table} displays the birth times, death times, and $p$-values of the corresponding generators.
\autoref{subfig:voro_4} highlights the shapes of the 23 $H_2$ homology group generators found by SCHU in different colors. Similarly, the $H_1$ generators could be analyzed, but a larger dataset would be necessary so there is room for the filaments to form more closed loops. For the current dataset, a $H_1$ generator with $p$-value 0.011 is shown as the red loop in \autoref{subfig:voro_5}.

%

\begin{table*}[ht]
\centering
\begin{tabular}{rrrrrrrrrrrrrrrrrrrrrrrr}
  \toprule
  Generator & A & B & C & D & E & F & G & H & I & J & K & L \\
  \midrule
  Birth & 9.16 & 5.37 & 5.14 & 7.48 & 8.12 & 9.64 & 9.70 & 9.64 & 8.74 & 8.92 & 9.46 & 9.64 \\ 
  Death & 10.79 & 9.06 & 11.46 & 12.12 & 11.80 & 11.71 & 12.74 & 12.48 & 10.26 & 11.28 & 10.22 & 12.30 \\ 
  $p$-value & 0.866 & 0.001 & $<$0.001 & $<$0.001 & 0.001 & 0.406 & 0.034 & 0.058 & 0.931 & 0.192 & $>$0.999 & 0.084 \\ 
       \bottomrule
     \\
     \toprule
  Generator & M & N & O & P & Q & R & S & T & U & V & W \\ 
    \midrule
  Birth & 8.13 & 9.42 & 9.08 & 8.97 & 9.08 & 8.68 & 9.18 & 10.14 & 6.93 & 9.79 & 8.00 \\ 
  Death & 12.10 & 11.97 & 11.82 & 9.92 & 12.42 & 9.63 & 13.25 & 11.05 & 11.12 & 12.29 & 10.65 \\ 
  $p$-value & 0.001 & 0.114 & 0.071 & $>$0.999 & 0.009 & $>$0.999 & 0.001 & $>$0.999 & 0.001 & 0.133 & 0.088 \\ 
   \bottomrule
\end{tabular}
\caption{$p$-values, birth and death times of all the 23 void generators found by SCHU, as shown in \autoref{subfig:voro_3}. 
}\label{voro_table}
\end{table*}

\section{Comparison Studies}\label{sec:comparisons}
In this section, SCHU is applied to galaxy survey and N-body simulation datasets in order to compare to several other void-finding techniques. We first study the results of SCHU as applied to a subset of the Sloan Digital Sky Survey (SDSS) galaxy catalog \citep{strauss2002spectroscopic} used in \citet{sutter2012public}. Next, we apply SCHU to the dark matter halo catalog from a cosmological simulation that is used in the cosmic web identification comparison study from \citet{libeskind2018tracing}. 
Finally, we present initial work focused on constraining the sum of the neutrino masses $\sum m_\nu$ by computing the persistence diagrams of two simulations from the MassiveNuS simulation suite \citep{MassiveNuS} and comparing their Betti functions.

\subsection{Cosmic voids and filament loops in SDSS} \label{sec:sdss}
\begin{figure}
 \centering
    \includegraphics[width=0.45\textwidth]{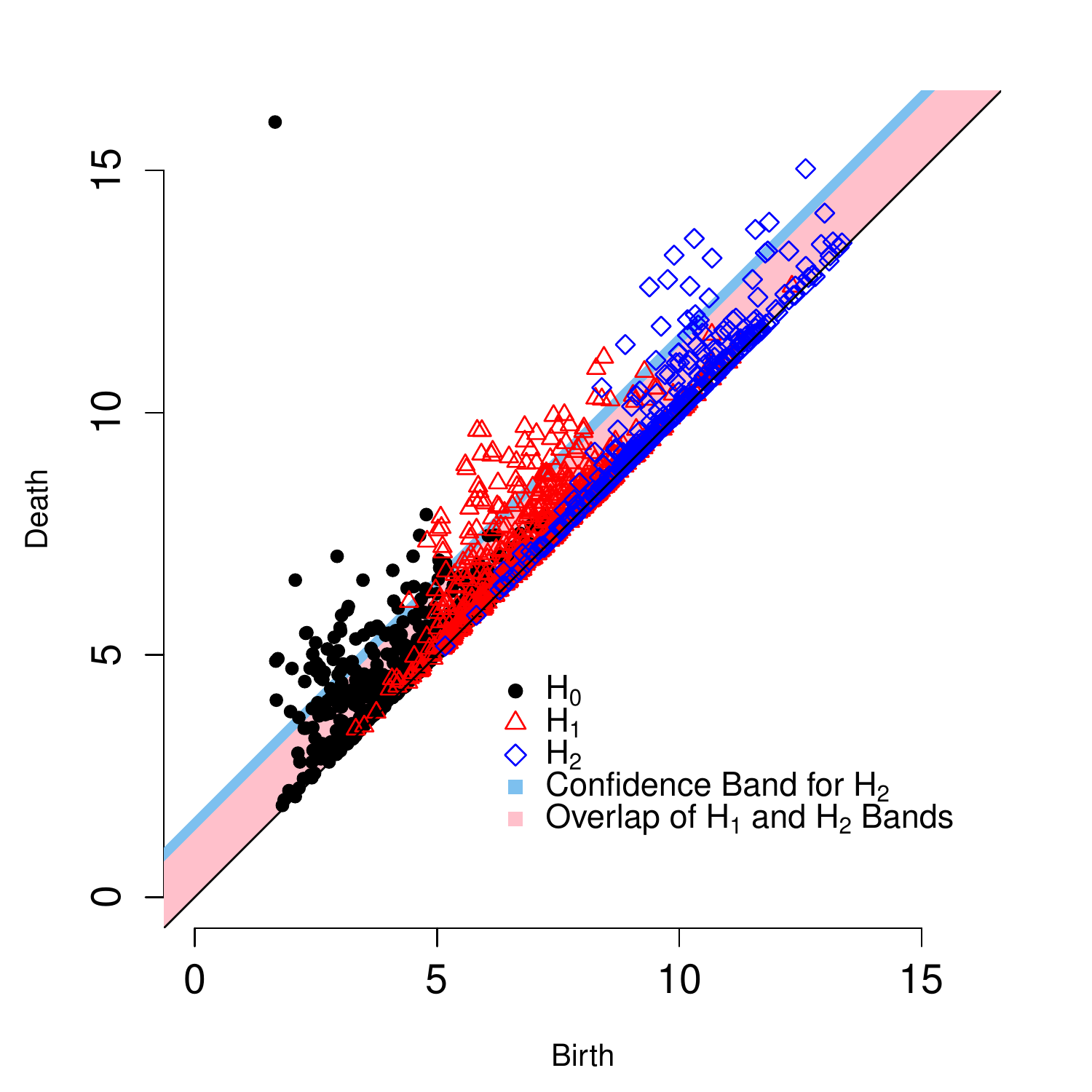}
\caption{Persistence Diagrams with 90\% confidence band for $H_1$ and $H_2$ of the SDSS dataset (with artificial boundary walls). A DTM function with $m_0=0.001$ is used to generate the persistence diagrams. The confidence band for $H_1$ is drawn in pink and the confidence band for $H_2$ is drawn in blue: the $H_1$ band is the overlap since it is shorter.} \label{sdss_pers}
\end{figure}

\begin{figure*}
 \centering
    \begin{subfigure}{0.45\textwidth}
        \centering
    \includegraphics[width=\textwidth]{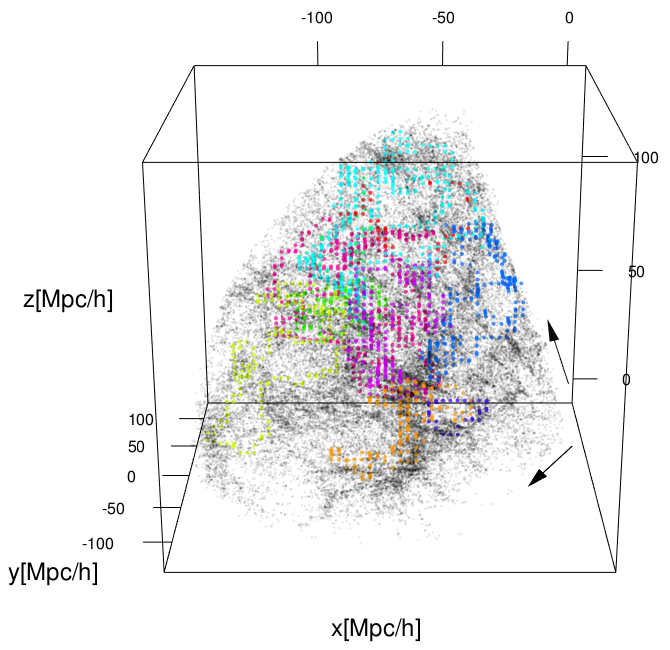}
        \caption{Filament loops, $H_1$}\label{subfig:sdss_loc1}
    \end{subfigure}
    \begin{subfigure}{0.45\textwidth}
        \centering
    \includegraphics[width=\textwidth]{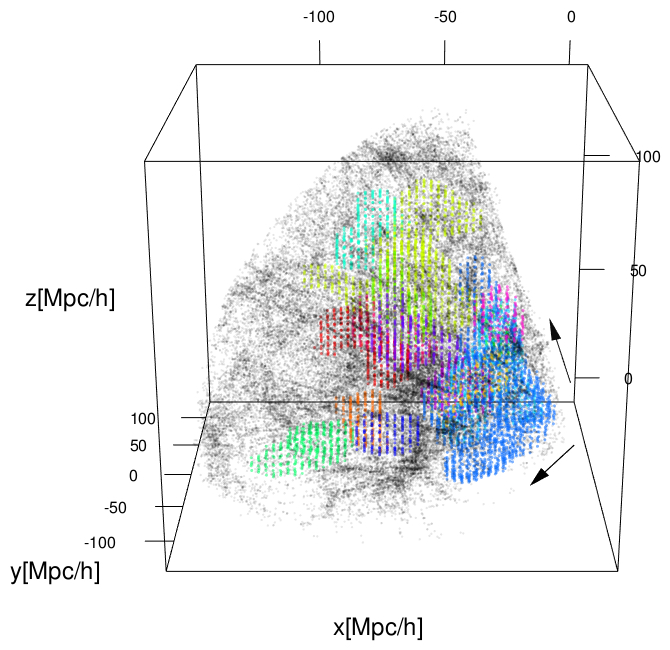}
        \caption{Cosmic voids, $H_2$}\label{subfig:sdss_loc2}
    \end{subfigure}
\caption{Filament loops (a) and voids (b) identified in the SDSS dataset using SCHU. The DTM function was constructed with $m_0=0.001$. The most significant 10 filament loops and the most significant 15 cosmic voids generators are shown in different colors. In the left figure, the 10 most significant filament loops are displayed with $p$-values$< 0.001$; in the right figure, the 15 most significant cosmic voids are displayed with $p$-values$< 0.09$.
} \label{sdss_loc}
\end{figure*}

In this section, we identify filaments and voids in a subset of the SDSS main galaxy redshift survey from the \emph{Public Cosmic Void Catalog}\footnote{\href{http://www.cosmicvoids.net/documents}{http://www.cosmicvoids.net/documents}}, and compare our cosmic voids to the cosmic voids identified in \citet{sutter2012public}.
The \textit{dim1} catalog subset contains 63,639 galaxies within the redshift range $0.0<z<0.05$. We only use the part of the catalog from the large contiguous region of the SDSS footprint in the northern part of the sky, which contains 57,795 galaxies within $109.81 \leq \text{RA} \leq 261.25$ and $-3.71 \leq \text{Dec} \leq 70.13$.
Artificial boundary walls are added on the 3D boundary surface of the grid in order to detect generators that may extend beyond the boundary. In the celestial coordinate system (RA, DEC, z), the dataset is like a cylinder with the redshift direction as its height. 
We impose a grid on the side surface of the cylinder and randomly sample points from the top surface (the bottom surface is no longer a boundary after the transformation). 
A total of 11,130 grid points along the cylinder wall plus 3,000 sampled points at the top surface form the artificial walls. 
Then, the dataset is transformed into the Cartesian coordinate system and a DTM function is computed with an $m_0$ of 0.001, 
which corresponds to using the 72 nearest neighbors when  calculating  the  DTM  function  at  each  grid  point.
The resulting persistence diagram is displayed in \autoref{sdss_pers}, and confidence bands are added as a reference (the 90\% confidence band for $H_1$ is drawn in pink, the 90\% confidence band for $H_2$ is drawn in blue, and the pink band shows the overlap). 
Using the bootstrap-derived empirical distribution of the bottleneck distances with $N_{\textrm{boot}} =1000$,  the $p$-values were computed for all the $H_1$ and $H_2$ generators. 
In \autoref{sdss_loc}, we highlight 10 of the most significant filament loops and 15 of the most significant cosmic voids found by SCHU. \autoref{subfig:sdss_loc1} shows the 10 most significant filament loops in different colors ($p$-values $<0.001$) and \autoref{subfig:sdss_loc2} shows the 15 most significant cosmic voids ($p$-values $<0.09$). 
As expected, the highlighted filament loops are surrounding sparse areas in \autoref{subfig:sdss_loc1}, and the highlighted cosmic voids in \autoref{subfig:sdss_loc2} are located in areas with low galaxy density; this indicates that SCHU is indeed capable of identifying reasonable filament loops and cosmic voids based on our visual comparison with the data. 
We note that while most cosmic voids are clustered around the bottom right corner of the Cartesian box (i.e. the area with lower redshift values), the identified filament loops span the entire redshift range.

\begin{figure*}[!ht]
  \centering
    \includegraphics[width=0.8\textwidth]{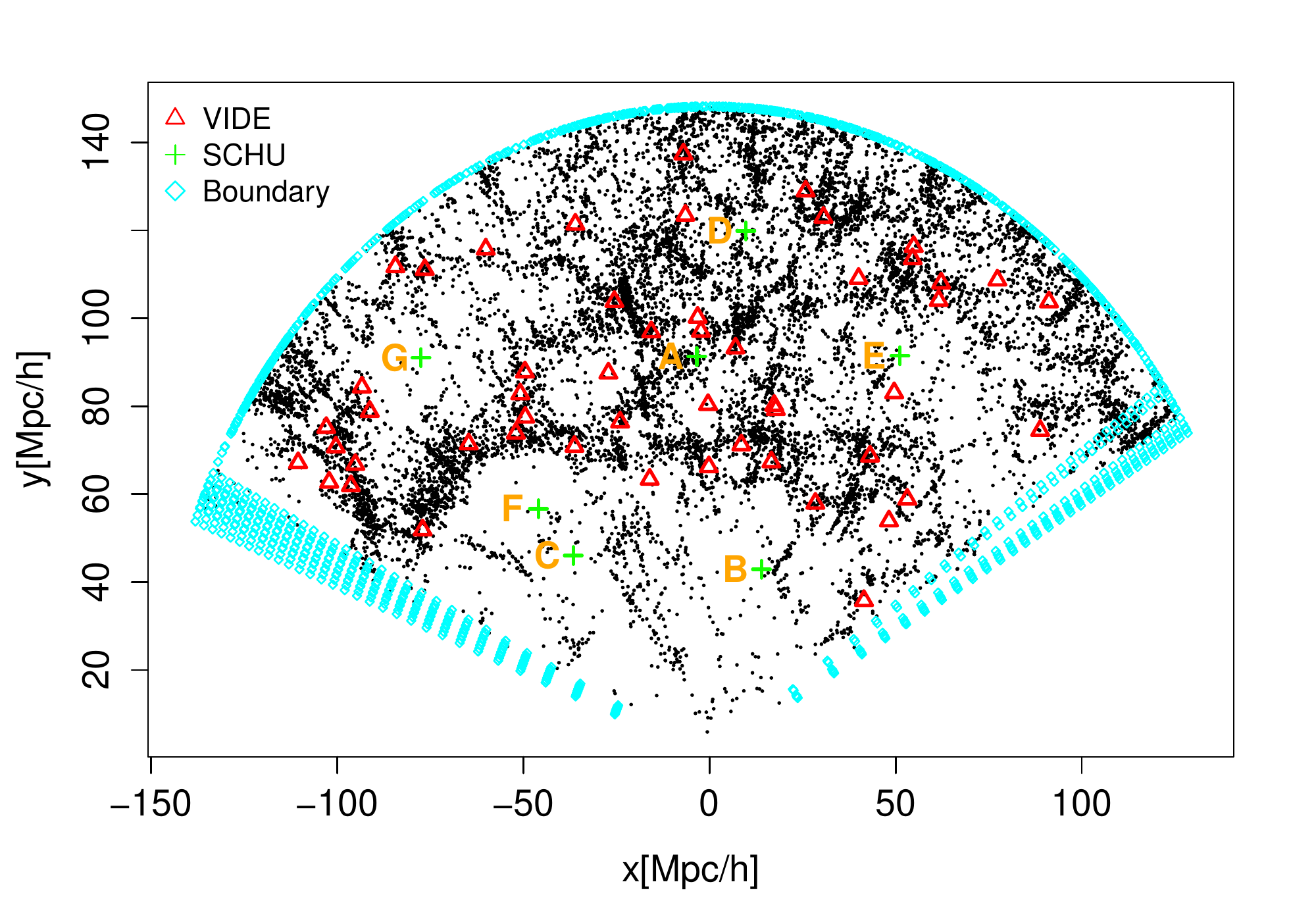}
\caption{Comparison between VIDE and SCHU. A slice of the SDSS dataset within a 20 degree opening angle after the rotation projection is shown. The red triangles are barycenters of voids found by VIDE; the green pluses are cosmic voids with $p<0.1$ and the corresponding $p$-values can be found in \autoref{pvalues_sdss}.
}\label{comp}
\end{figure*}

We compare the results from SCHU with the identified cosmic voids from VIDE \citep{sutter2012public}, which is based on the ZOBOV void-finding method \citep{neyrinck2008zobov}. 
The VIDE algorithm begins by constructing a Delaunay tessellation based on the galaxy positions and assigns a density to each galaxy based on the volume of its corresponding Voronoi cell. 
Then, a watershed procedure \citep{platen2007cosmic} is carried out to combine some neighboring cells and ultimately assign certain regions as cosmic voids.
To aid in visualizing the cosmic void locations and to easily compare the TDA-identified voids with those from VIDE, we use a similar rotation and subset as in \citet{sutter2012public} (i.e., the galaxy and void positions are rotated about the $y$-axis so that they lie on the $x-y$ plane, and only those galaxies and voids within a 20 degree opening angle are plotted). 
The volume centers of the voids found by VIDE are plotted as red triangles in \autoref{comp}.

For SCHU, the same DTM function as described above is used to generate the filtration with $m_0 = 0.001$ (i.e., using the 72 nearest galaxies).
Since VIDE locates cosmic voids, we only compare cosmic voids ($H_2$ generators) here. 
After bootstrapping with $N_{\textrm{boot}}=1000$, $p$-values for $H_2$ generators are obtained. 
In the whole dataset, SCHU identifies 311 cosmic voids, with 15 of them statistically significant ($p$-values $< 0.1$), while VIDE identifies 209 cosmic voids. In the subset visualized below, SCHU identifies 97 cosmic voids, with 7 of them statistically significant ($p$-values $< 0.1$), while VIDE identifies 51 cosmic voids.
The volume centers of the seven statistically significant voids found using SCHU ($p$-values $<0.1$) are plotted as green pluses in \autoref{comp} with labels A through G; the corresponding $p$-value can be found in \autoref{pvalues_sdss}.
Although SCHU identifies a smaller total number of statistically significant voids (with $p$-values $<0.1$) than the total number of voids identified by VIDE, the voids from SCHU are better-centered on locations where the eye would identify voids in the dataset.
Some of the voids found by VIDE are located in non-empty regions; however, these may be the smaller voids ($5-15 h^{-1}$ Mpc) that \citet{sutter2012public} note are located at the edges of filaments and walls (non-empty regions). 
The cosmic voids found using SCHU are concentrated more in empty regions. 
VIDE does not find the voids in the bottom region of \autoref{comp}, while SCHU classifies those empty regions as cosmic voids. 
The orange pluses in \autoref{comp} are the artificial boundary points added to detect boundary voids; the side walls appear thick because of the projection.

\begin{table}[ht]
\centering

\begin{tabular}{
  r
  S[table-format = <1.3]}
\toprule
{Label} & {$p$-value}\\
\midrule
   A & <0.001 \\
   B & <0.001 \\
   C & <0.001 \\
   D & .064 \\
   E & <0.001 \\
   F & <0.001 \\
   G & .006 \\
\bottomrule
\end{tabular}
\caption{$p$-values for corresponding cosmic voids in \autoref{comp}.}\label{pvalues_sdss}
\end{table}

\subsection{Cosmic web identification method comparison simulation} \label{libeskind}
In this section, we compare SCHU with other void-finding methods by applying SCHU to the cosmological simulation used in the \citet{libeskind2018tracing} comparison study, created via the GADGET-2 dark-matter only N-body simulation code with $512^3$ particles \citep{springel2005cosmological}.
\citet{libeskind2018tracing} used this dataset in order to compare several cosmic environment classification methods; the data and the results for the void-finding methods considered are available online\footnote{\href{http://data.aip.de/tracingthecosmicweb/}{http://data.aip.de/tracingthecosmicweb/}}.

The simulation cube is $200h^{-1}$ Mpc on each side and was run using the 2015 Planck $\Lambda $CDM parameters \citep{planck2015params}\footnote{$h=0.68$, $\Omega_M=0.31$, $\Omega_{\Lambda}=0.69$, $n_s=0.96$ and $\sigma_8=0.82$}. 
Dark matter halos were identified in the $z=0$ snapshot using a friends-of-friends (FOF) algorithm \citep{huchra1982groups, press1982identify} with a linking length of $0.2$, and requiring a minimum of 20 particles per halo. There are a total of $281,465$ halos in the catalog. 
Applying SCHU to a halo catalog is similarly well-motivated to that of a galaxy survey catalog, as dark matter halos (and the observable galaxies that lie within them) are both assumed to be biased tracers of the underlying matter field. Thus, underdensities in the halo field should correspond to physical voids in the matter distribution.

\begin{figure}
 \centering
    \includegraphics[width=0.45\textwidth]{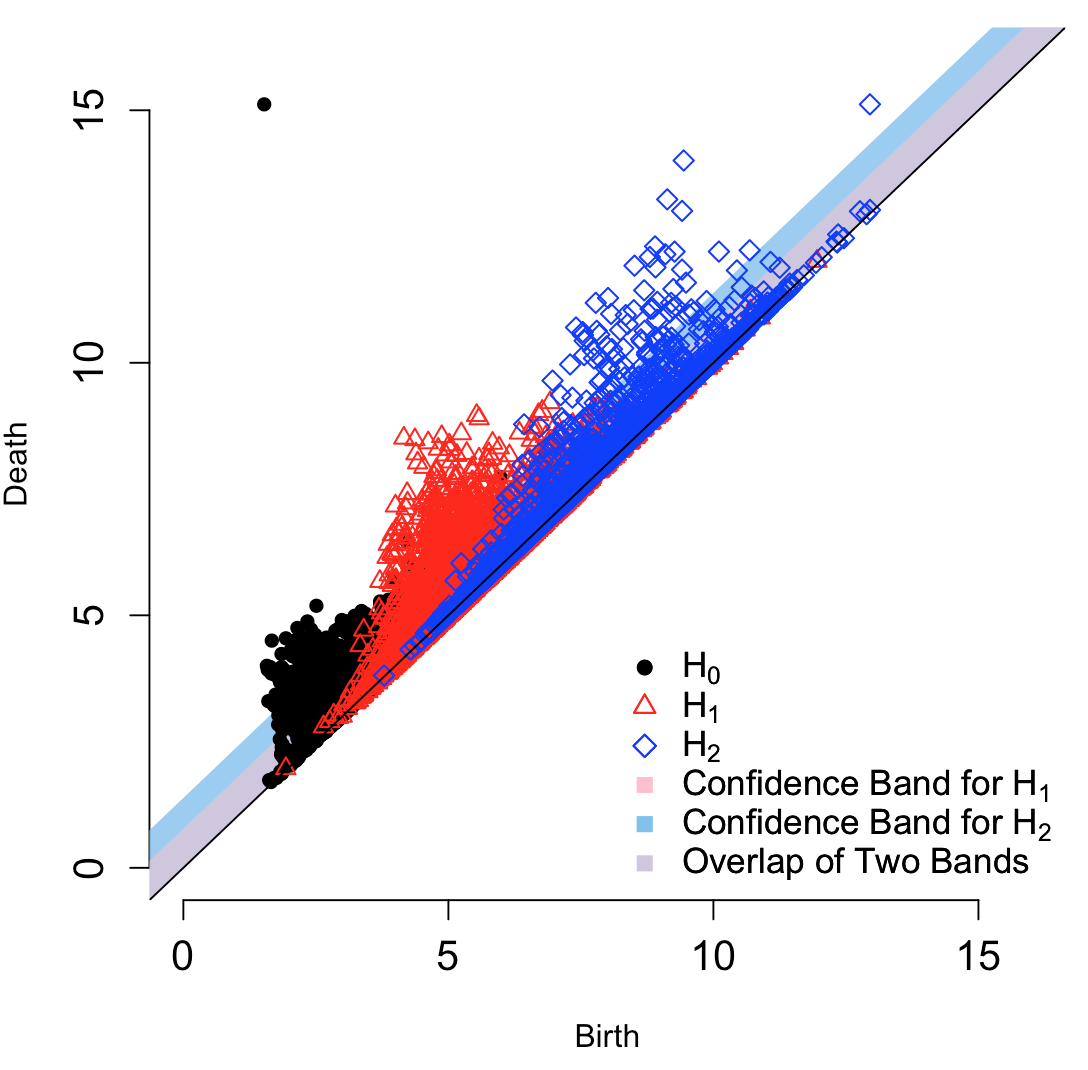}
\caption{Persistence Diagrams with 90\% confidence band for $H_1$ and $H_2$ from the \citet{libeskind2018tracing} dataset. A DTM function with $m_0=0.0002$ is used to generate the persistence diagrams. The confidence band for $H_1$ is drawn in pink and confidence band for $H_2$ is drawn in blue: they are overlapping so that the pink band, which is shorter, is the overlap. 
} \label{GADGET_pers}
\end{figure}

\begin{figure*}
 \centering
    \begin{subfigure}{0.45\textwidth}
        \centering
    \includegraphics[width=\textwidth]{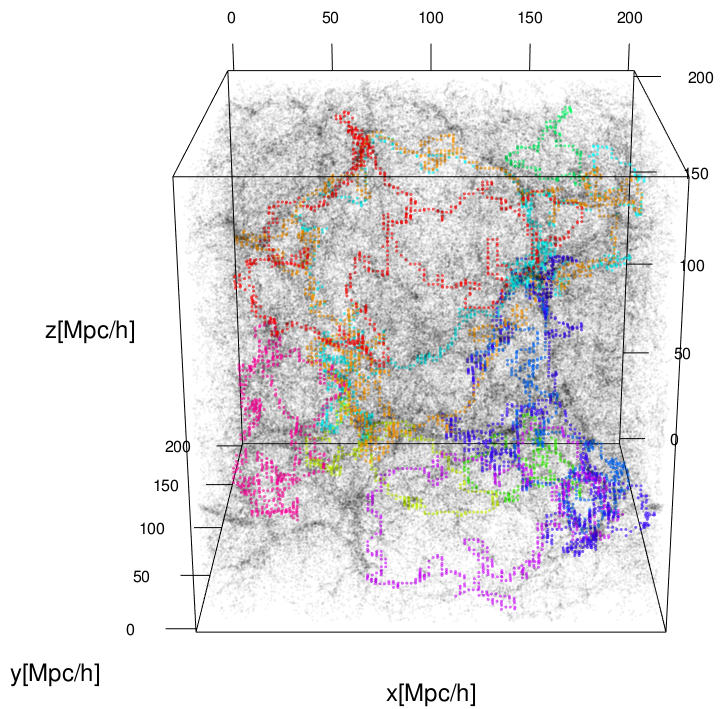}
        \caption{Filament loops ($H_1$)}\label{subfig:GADGET_filament}
    \end{subfigure}
    \begin{subfigure}{0.46\textwidth}
        \centering
    \includegraphics[width=\textwidth]{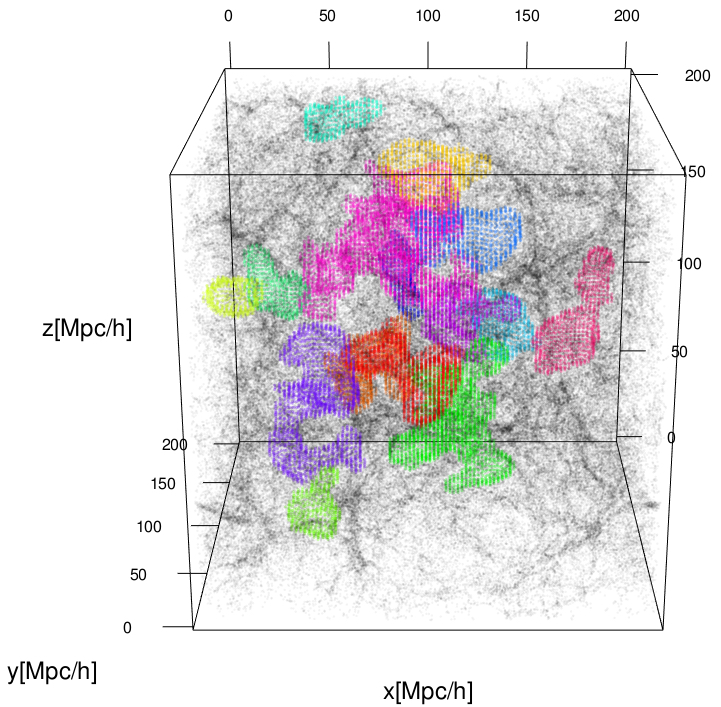}
        \caption{Cosmic voids ($H_2$)}\label{subfig:GADGET_void}
    \end{subfigure}
\caption{Filament loops (a) and voids (b) identified in the \citet{libeskind2018tracing} dataset using SCHU. The DTM function was constructed with $m_0=0.0002$. The most significant 10 filament loops (a) and the most significant 15 cosmic voids generators (b) are shown in different colors. All of their $p$-values are less than $0.001$.
} 
\end{figure*}

\citet{libeskind2018tracing} specified a $200\times200\times200$ grid for cosmic environment prediction in order to facilitate the comparison of the methods;  we also use this same grid for our predictions.
To detect boundary voids, we add artificial boundary walls by randomly sampling $1,000$ points on each of the six sides of the cube.
The $m_0$ is set to $0.0002$, which corresponds to using 58 nearest neighbors when  calculating  the  DTM  function  at  each  grid  point
\autoref{GADGET_pers} displays the resulting persistence diagram of the dataset along with the confidence bands. \autoref{subfig:GADGET_filament} shows the 10 most significant filament loops with $p$-value $<0.001$ in different colors and \autoref{subfig:GADGET_void} shows the 15 most significant cosmic voids with $p$-value $<0.001$ in different colors. 
Cosmic voids are more straightforward to visualize and we can see that they are located at low density areas, as expected. 

\begin{sloppypar}
We compare SCHU with the nine different methods used in \citet{libeskind2018tracing}: the Multiscale Morphology Filter-2 (MMF-2, \citet{aragon2014hierarchical}),
Multi-Stream Web Analysis (MSWA, \citealt{ramachandra2015multi}), 
CLASSIC \citep{kitaura2012linearization}, 
DisPerSE \citep{sousbie2011persistent}, 
NEXUS+ \citep{cautun2012nexus}, 
Spineweb \citep{aragon2010spine}, 
ORIGAMI \citep{falck2012origami, falck2015persistent}, 
the tidal shear tensor (T-web, \citealt{forero2009dynamical}), 
and the velocity shear tensor (V-web, \citealt{hoffman2012kinematic}). 
For a brief overview of these methods, we refer the reader to Table 1 of \citet{libeskind2018tracing}.  These methods are broadly classified as Hessian (CLASSIC, T-web, V-web), 
scale-space Hessian (MMF-2, NEXUS+), 
topological (DisPerSE, Spineweb), 
and phase-space (ORIGAMI,  MSWA) methods.
The \emph{Hessian} methods use the Hessian of the gravitational potential \citep{hahn2007properties} or velocity shear tensor \citep{hoffman2012kinematic} in order to classify the cosmic environment type by counting the number of eigenvalues of the resulting matrix at each point in space. 
The \emph{scale-space Hessian} methods are based on the Multiscale Morphology Filter (MMF) approach \citep{aragon2007multiscale}, which extends the Hessian method into a smoothing-scale-independent calculation by computing the optimum eigenvalue threshold after repeating the calculation across many different smoothing scales.
The \emph{topological} methods compute the topological structures of a reconstructed density field from the spatial datasets that can be constructed in various ways, such as using a Voronoi tessellation or a discrete Morse--Smale complex. SCHU falls into the topological category of cosmic environment classifiers.
Lastly, the \emph{phase-space} methods identify the cosmic environment based on the number of dimensions along which particles have shell-crossed. 
\end{sloppypar}

\begin{figure*}[htp!]
 \centering
      \begin{subfigure}{0.26\textwidth}
        \centering
    \includegraphics[width=\textwidth]{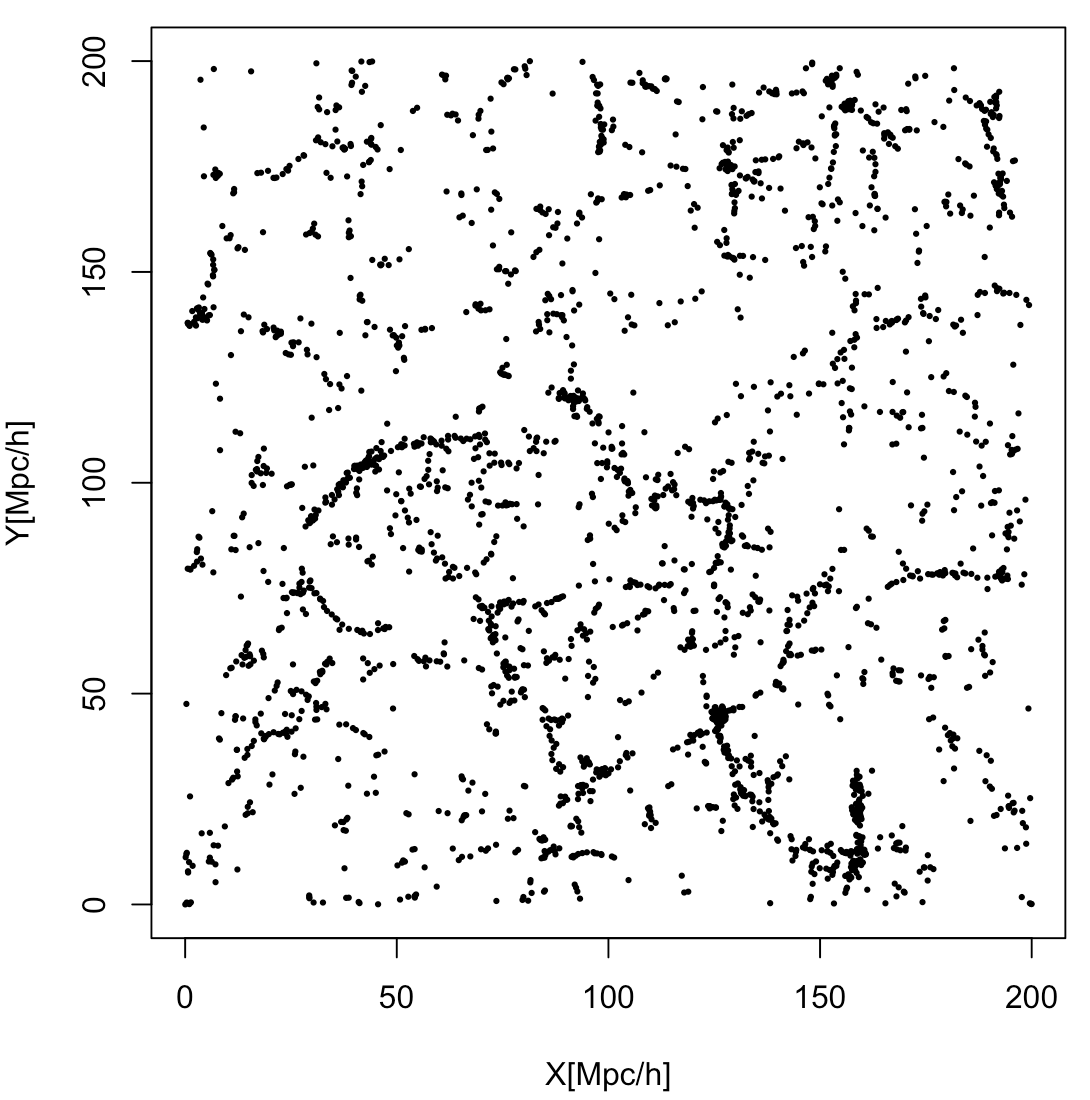}
        \caption{Halo locations in slice}\label{GADGET-data}
    \end{subfigure}
         \begin{subfigure}{0.26\textwidth}
        \centering
    \includegraphics[width=\textwidth]{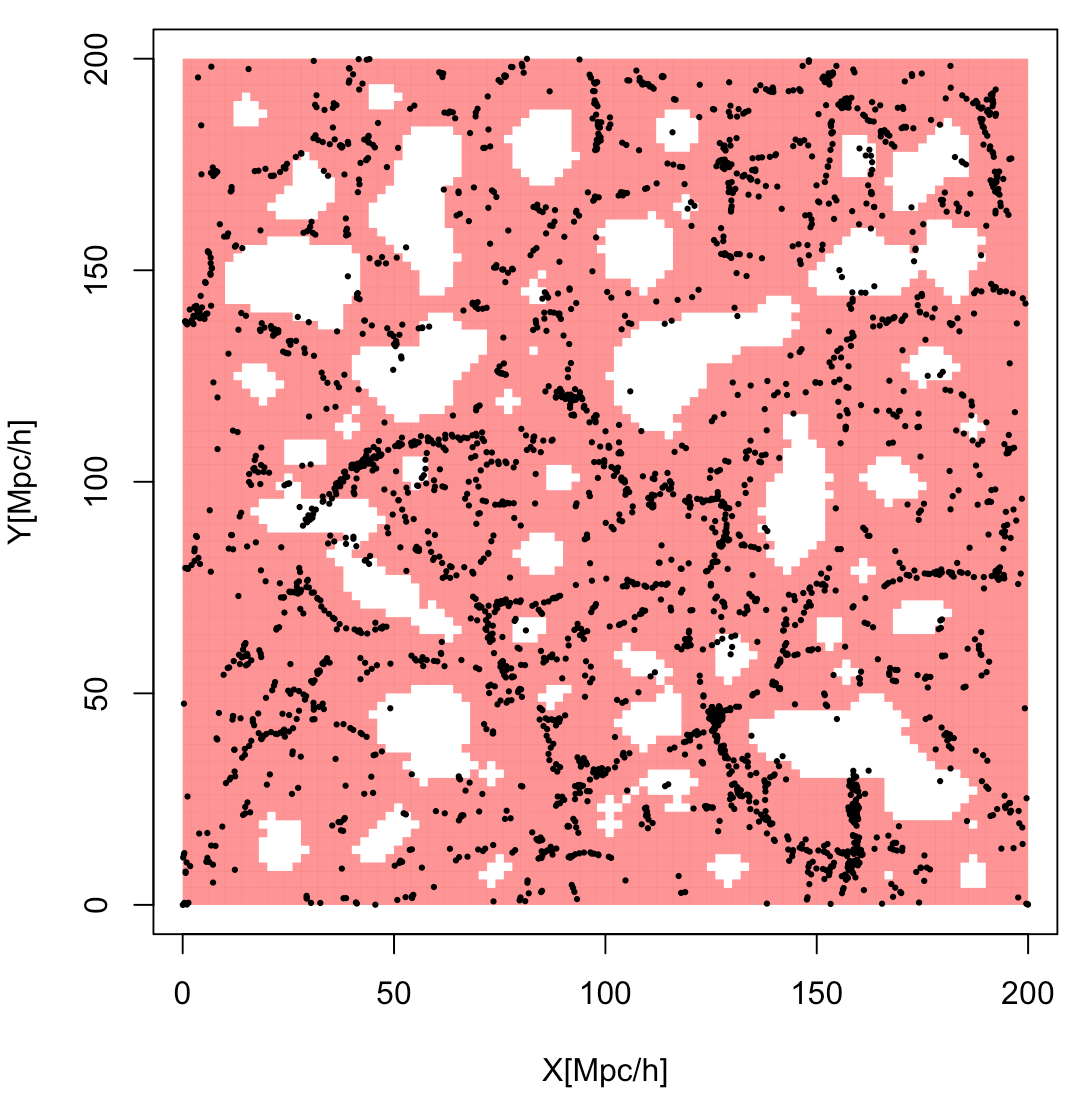}
        \caption{SCHU}\label{ours1}
    \end{subfigure}
     \begin{subfigure}{0.26\textwidth}
        \centering
    \includegraphics[width=\textwidth]{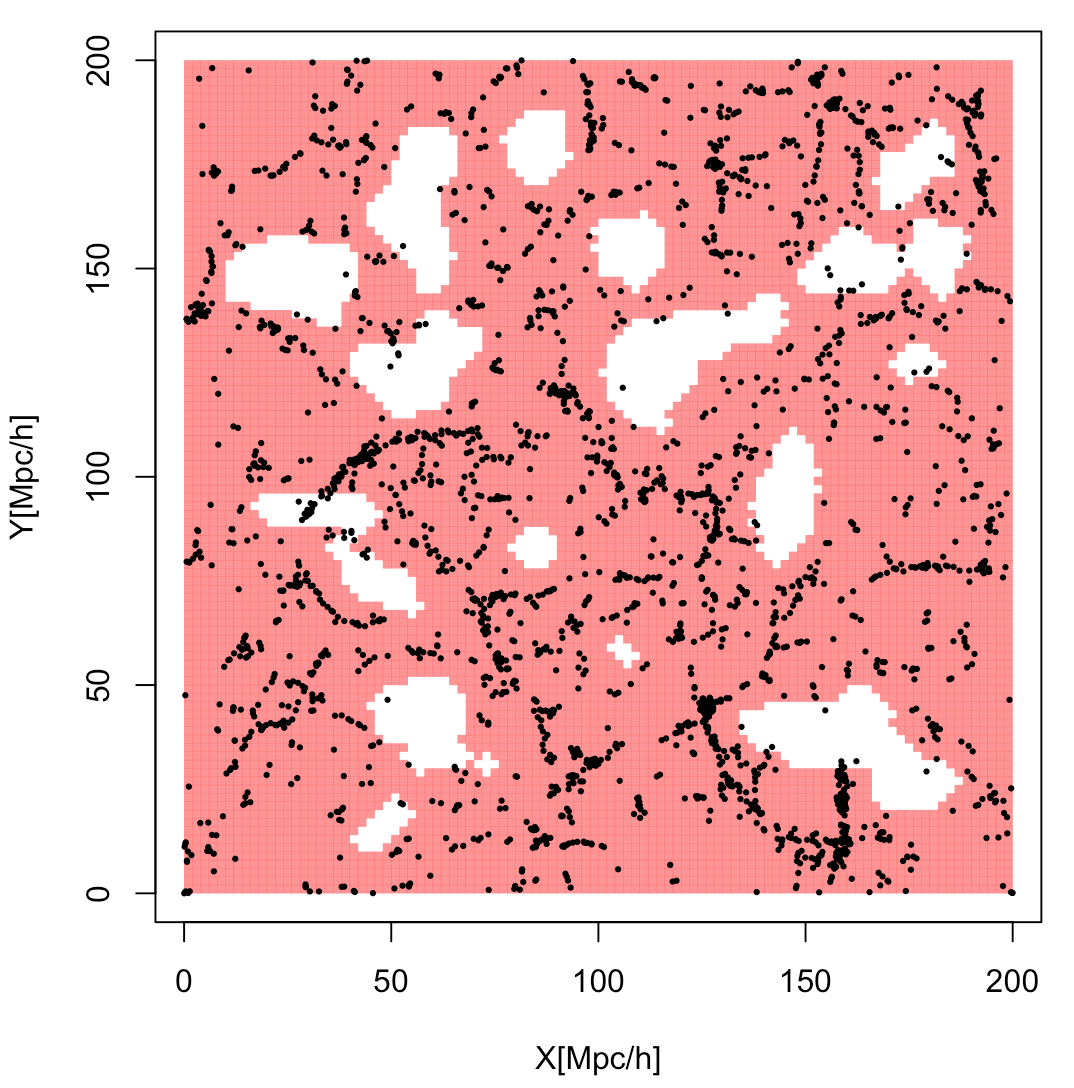}
        \caption{SCHU: $p$-value $<0.1$}\label{ours2}
    \end{subfigure}
    \begin{subfigure}{0.26\textwidth}
        \centering
    \includegraphics[width=\textwidth]{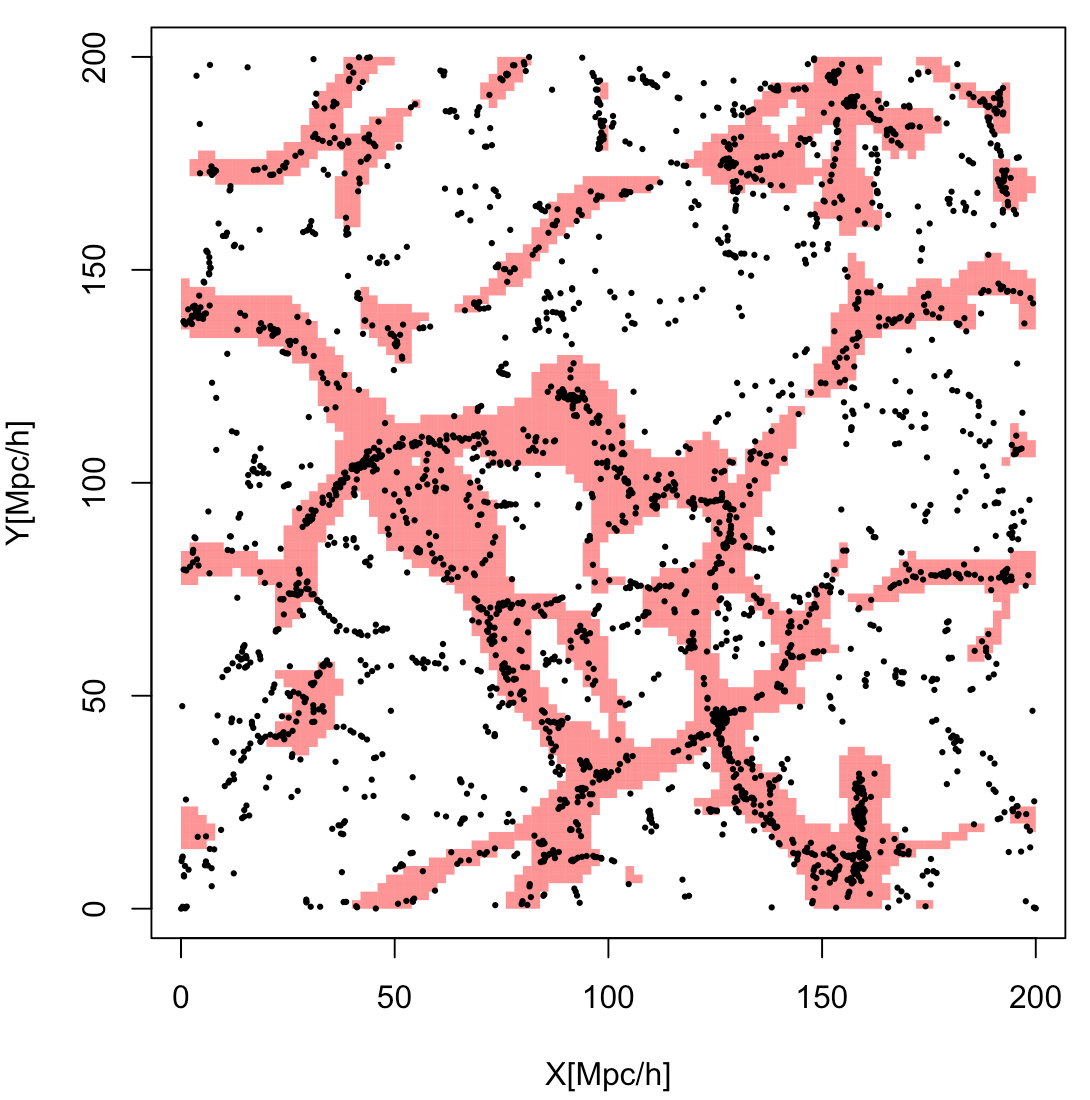} 
        \caption{CLASSIC} \label{classic}
    \end{subfigure}
    \begin{subfigure}{0.26\textwidth}
        \centering
    \includegraphics[width=\textwidth]{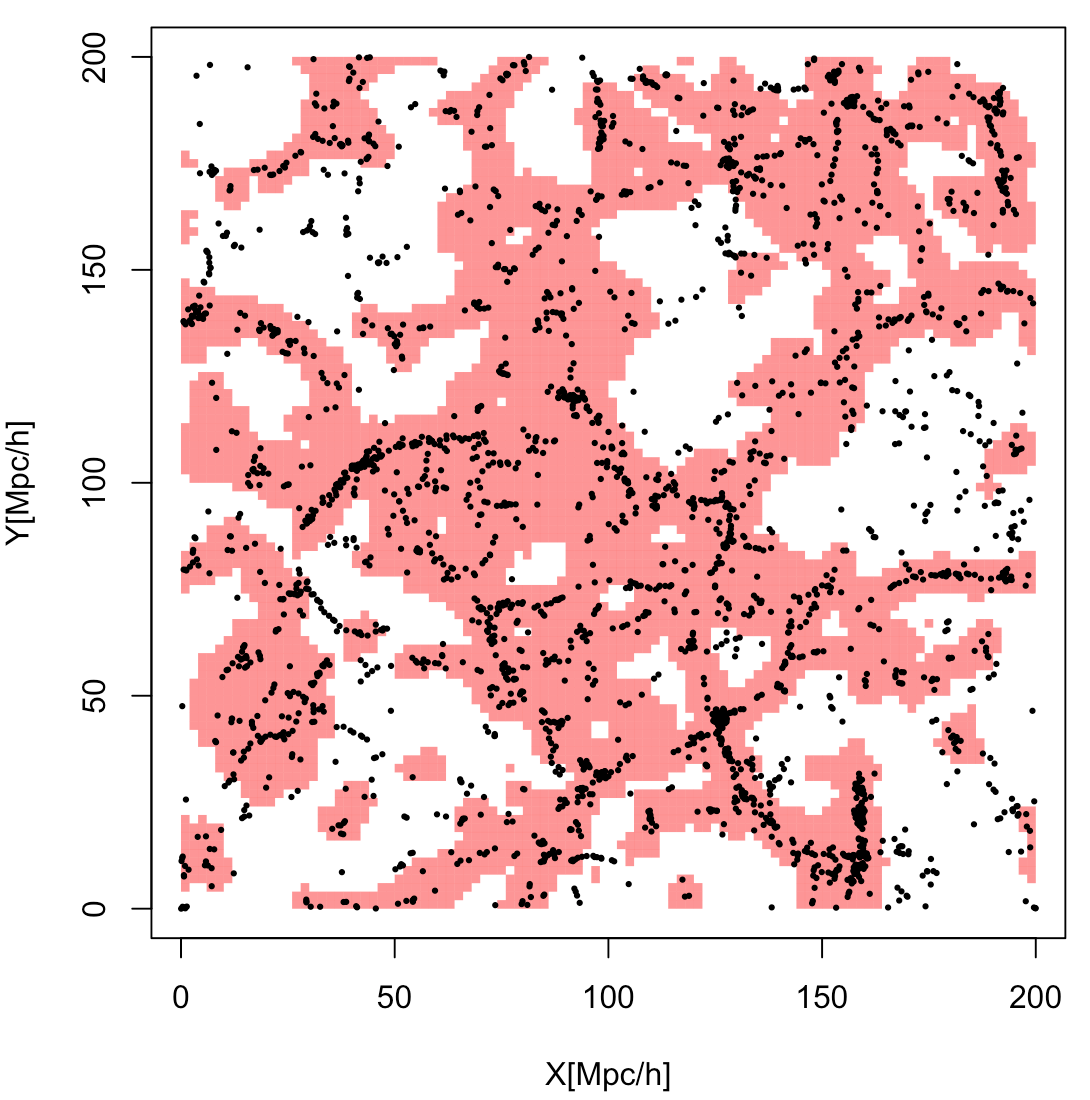}
        \caption{DisPerSE} \label{DisPerSE}
    \end{subfigure}
    \begin{subfigure}{0.26\textwidth}
        \centering
    \includegraphics[width=\textwidth]{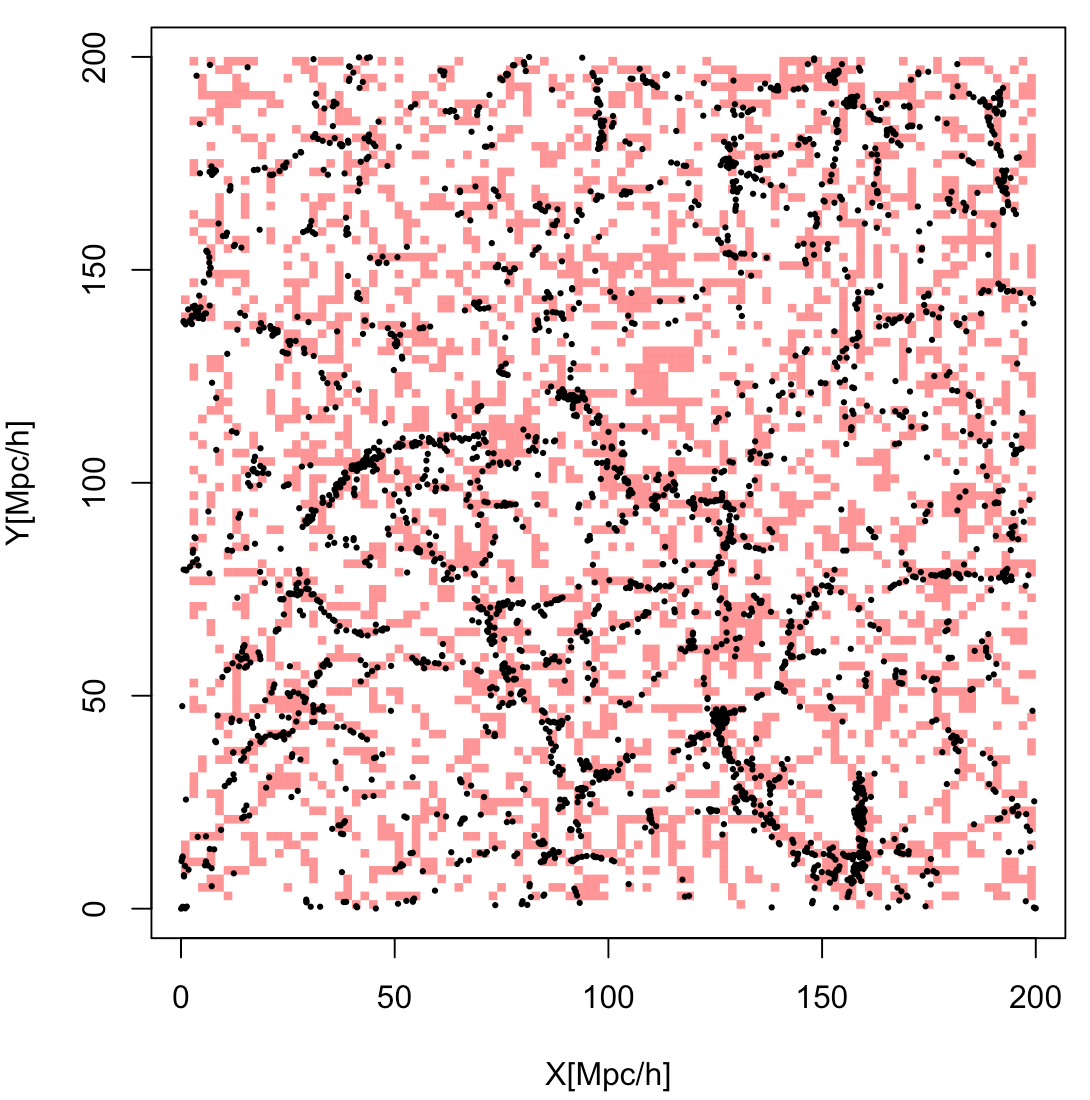}
        \caption{MMF-2} \label{mmf2}
    \end{subfigure}
        \begin{subfigure}{0.26\textwidth}
        \centering
    \includegraphics[width=\textwidth]{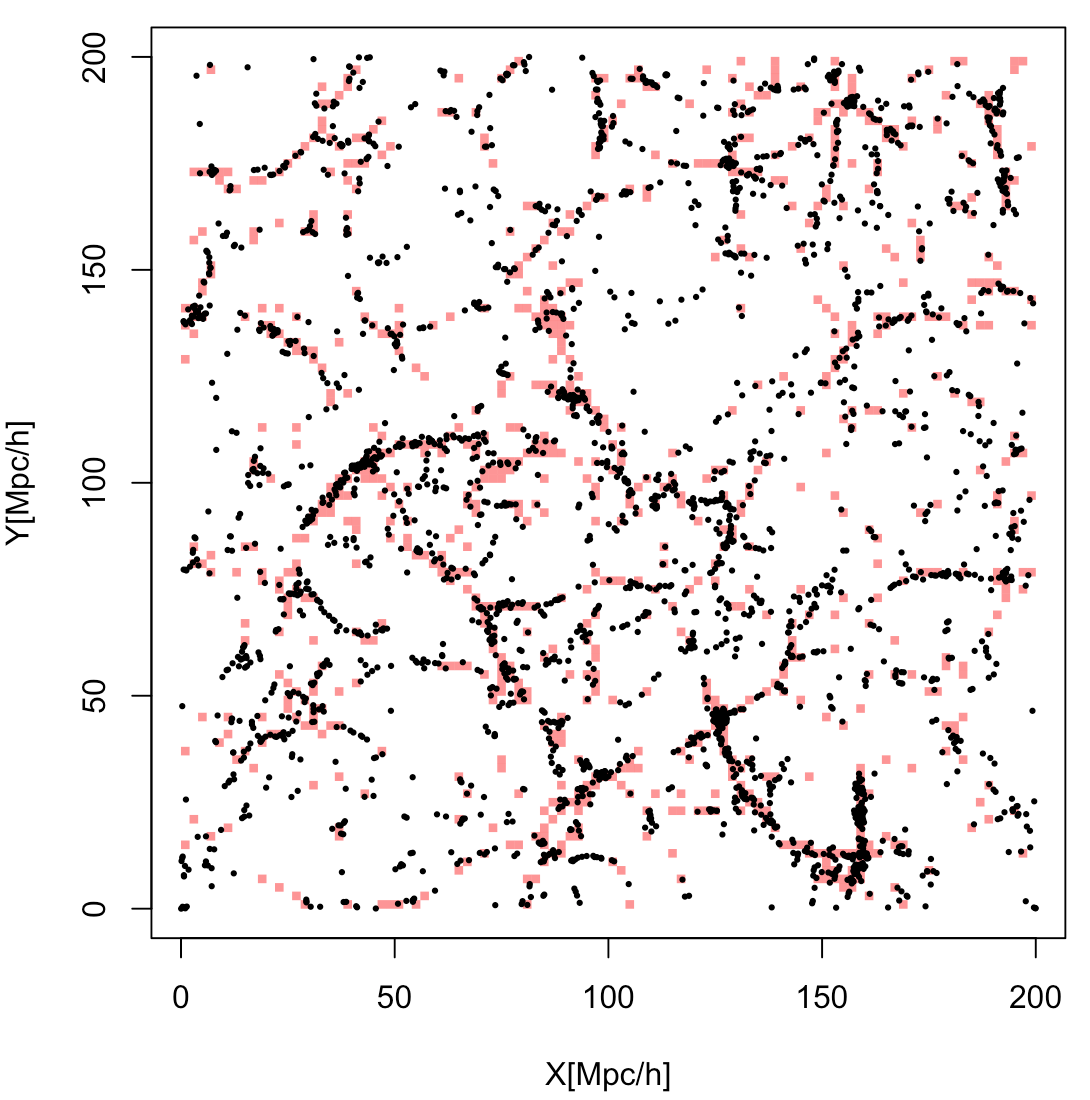}
        \caption{MSWA} \label{mswa}
    \end{subfigure}
        \begin{subfigure}{0.26\textwidth}
        \centering
    \includegraphics[width=\textwidth]{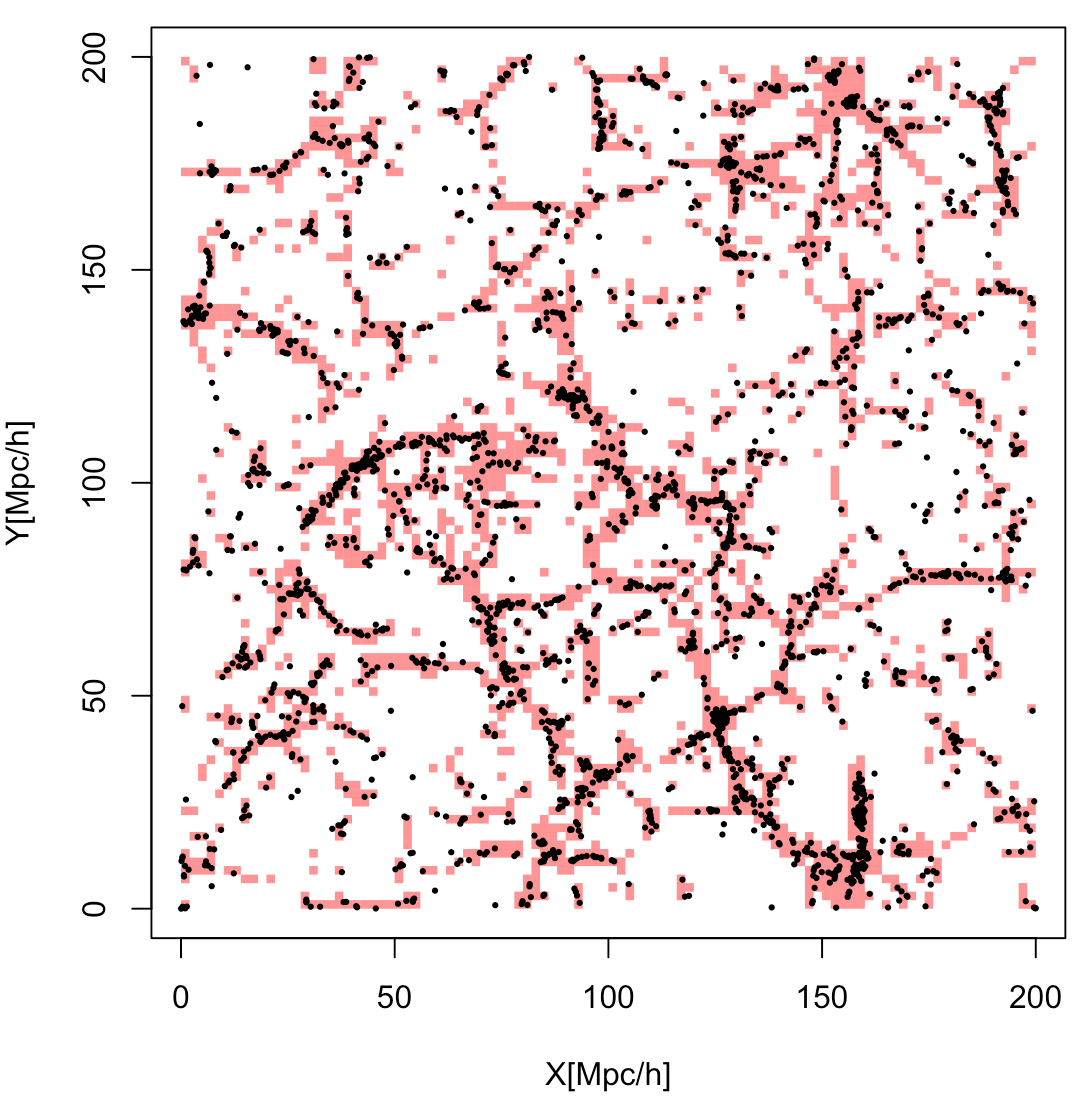}
        \caption{NEXUS+} \label{nexus}
    \end{subfigure}
        \begin{subfigure}{0.26\textwidth}
        \centering
    \includegraphics[width=\textwidth]{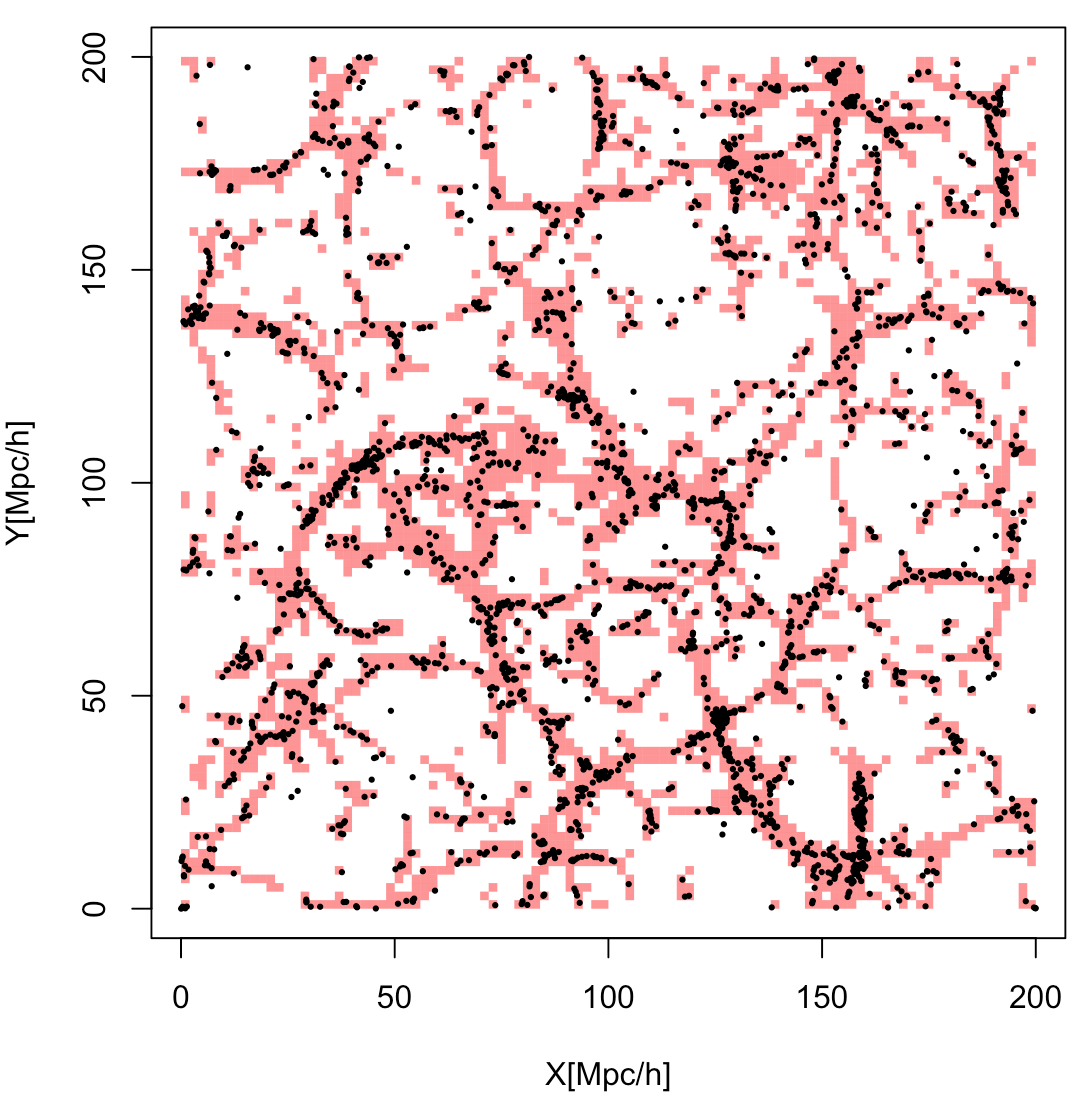}
        \caption{ORIGAMI} \label{origami}
    \end{subfigure}
        \begin{subfigure}{0.26\textwidth}
        \centering
    \includegraphics[width=\textwidth]{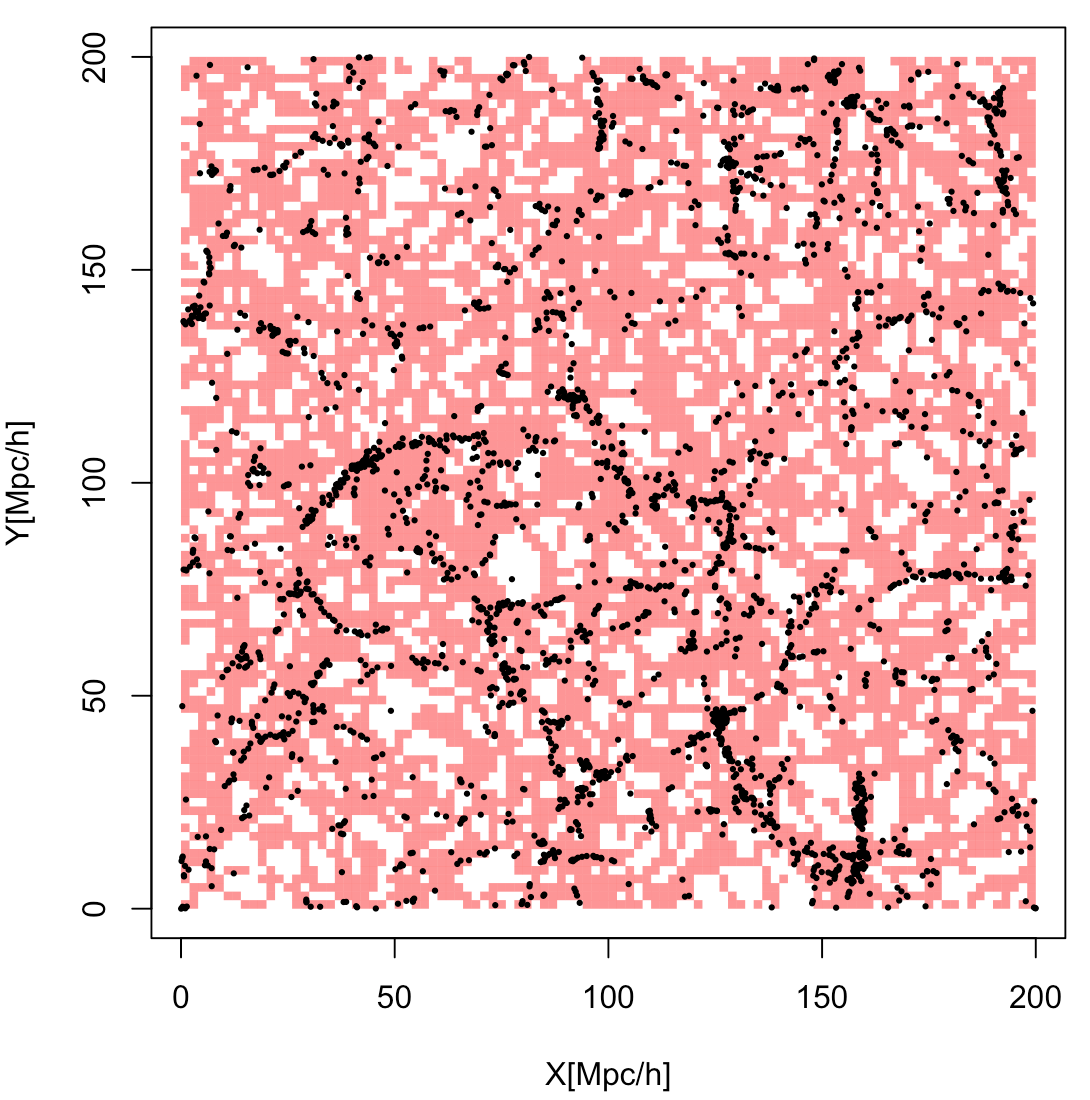}
        \caption{Spineweb} \label{spineweb}
    \end{subfigure}
        \begin{subfigure}{0.26\textwidth}
        \centering
    \includegraphics[width=\textwidth]{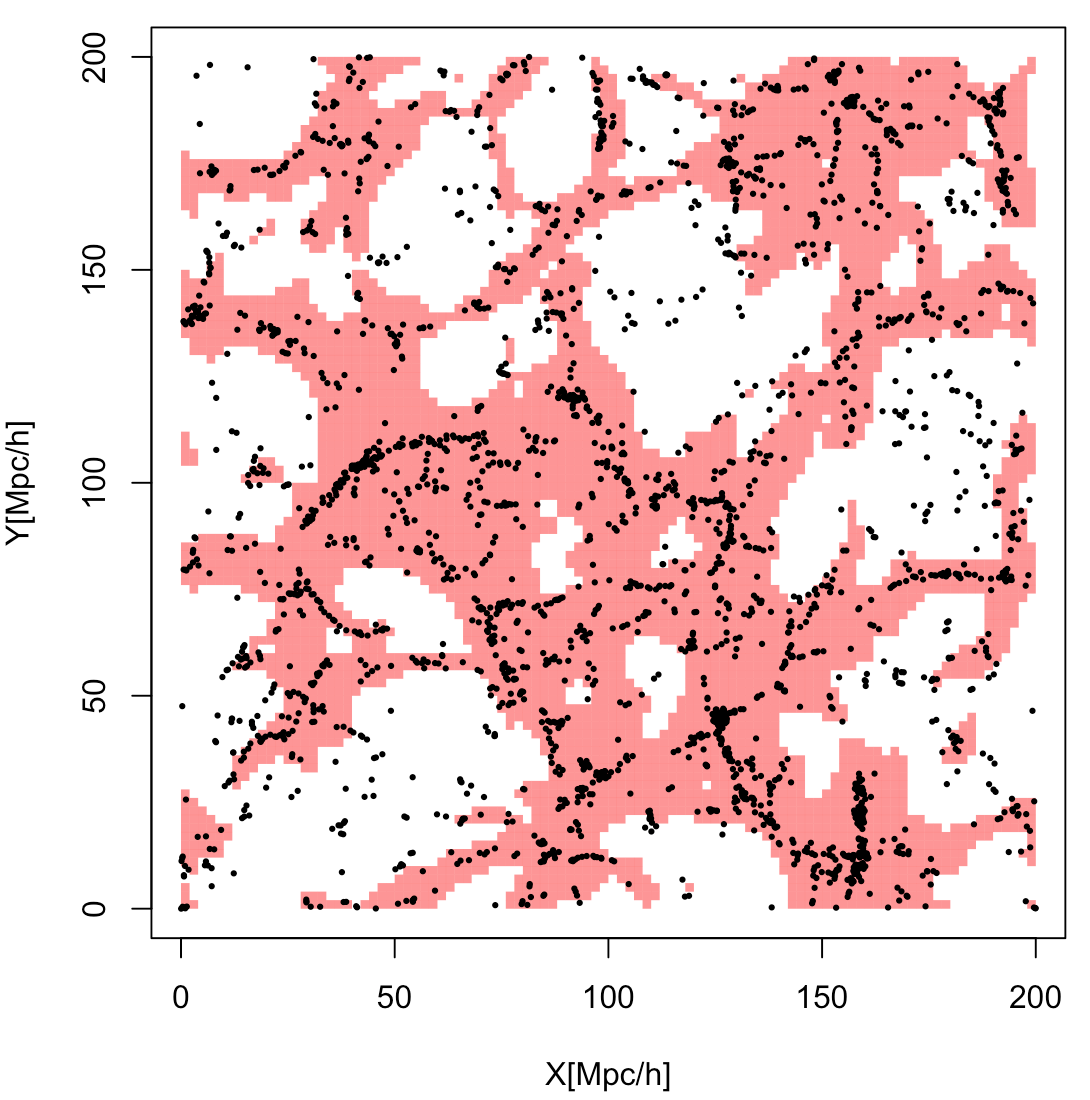}
        \caption{T-web} \label{tweb}
    \end{subfigure}
        \begin{subfigure}{0.26\textwidth}
        \centering
    \includegraphics[width=\textwidth]{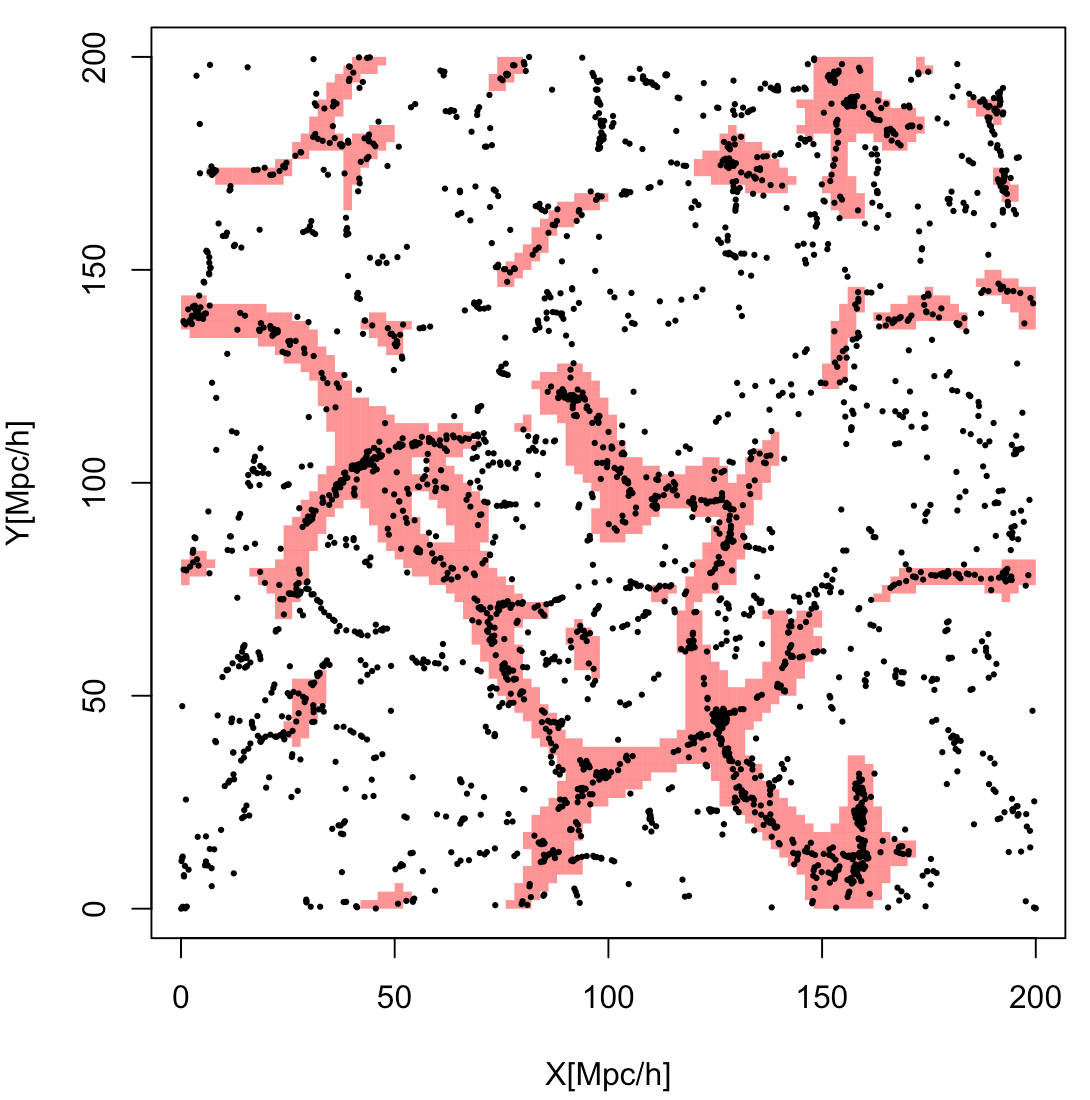}
        \caption{V-web} \label{vweb}
    \end{subfigure}
\caption{Comparison of void-finding methods on the halo catalog of a $2h^{-1}$ Mpc slice of the cosmological simulation from \citet{libeskind2018tracing}. (a) The halos in the slice of the simulation cube considered.  (b) - (l) The various void-finding method results.  The white regions denote voids and red regions denote walls, filaments, and nodes. 
(b) shows all $H_2$ generators in the slice from SCHU, whereas (c) shows only the $H_2$ generators with $p$-value $<0.1$ from SCHU.}\label{comparison}
\end{figure*}

We only compare cosmic voids because filament loops are not defined or detected by other methods. 
Since some of the methods in \citet{libeskind2018tracing} can only find filaments, we only compare SCHU with the nine methods from \citet{libeskind2018tracing} that can find cosmic voids. 
The volume-filling fraction of cosmic voids found by SCHU is $0.302$; the volume-filling fractions of significant cosmic voids with $p$-value $<0.1$ found by SCHU is $0.123$. These volume-filling fractions are relatively small compared to those of the methods listed in Table 2 of \citet{libeskind2018tracing}, which ranges from $0.332$ to $0.903$. Furthermore, Table 2 of the void comparison project of \cite{colberg2008aspen} also shows that the number of voids identified and void volume-filling fractions vary widely from method to method; thus, we primarily focus on qualitative comparisons for the remainder of this section.
A $2h^{-1}$ Mpc-thin slice of the simulation cube was selected to aid in the comparison of the methods, and is displayed in \autoref{GADGET-data}. 
\autoref{comparison} shows cosmic voids detected by different methods within the slice; the white areas are void regions and the red areas represent higher-density cosmic environments.
\autoref{ours1} shows all cosmic voids ($H_2$ generators) in the slice found using SCHU. \autoref{ours2} shows only the cosmic voids with $p$-value $<0.1$, which, as expected, are larger than the non-statistically significant generators of \autoref{ours1}.

The cosmic void regions found by the various methods span a range from visually smooth (larger contiguous regions defined as cosmic voids) to rough (many small regions defined as cosmic voids).
Results of CLASSIC, MSWA, NEXUS+, ORIGAMI, and V-web in Figures~\ref{classic}, \ref{mswa}, \ref{nexus}, \ref{origami}, and \ref{vweb}, respectively,
have similar appearances where the red regions appear tightly constrained to high density regions of matter and the remaining majority of space is defined as cosmic voids. 
Within those five methods, CLASSIC and V-web look smoother with larger contiguous high-density regions, while the other three methods look rougher with thinner, weblike, high-density regions. 
SCHU, DisPerSE, Spineweb, and T-web, Figures~\ref{ours1} and \ref{ours2}, \ref{DisPerSE}, \ref{spineweb}, and \ref{tweb}, respectively, share visual similarities in that the white regions appear tightly constrained to low density regions of matter and the remaining majority of space is defined as matter structures. 
Persistent homology, DisPerSE, and T-web are smoother with large contiguous cosmic void regions, whereas Spineweb is rougher with many small cosmic void regions.
MMF-2 is different from all the other methods as it identifies many disjoint small clusters of matter in regions where there are minimal halos within the slice.

\subsection{Betti functions of the MassiveNuS simulations}

Next we provide a demonstration of the potential usefulness of persistent homology by considering two simulations from the MassiveNuS suite \citep{MassiveNuS}.
These simulations are dark matter-only, and trace the evolution of $1024^3$ particles in a cubic volume of side length $512h^{-1}$ Mpc. The suite consists of 100 simulations with varied values of three cosmological parameters: the neutrino mass sum $\sum m_\nu$, the total matter density $\Omega_m$, and the primordial power spectrum amplitude $A_s$. 
In the following analysis, we study the publicly-available Rockstar \citep{Rockstar} halo catalogs generated from the $z=0$ snapshots of two of these simulations\footnote{\href{http://columbialensing.org}{http://columbialensing.org}}.  The two selected simulations both have $\Omega_m = 0.3$ and $A_s = 2.1 \times 10^{-9}$, but different neutrino mass sums of $\sum m_\nu=0$ eV and $\sum m_\nu = 0.6$ eV, which we denote $S_1$ and $S_2$, respectively. 
In this case, the goal is to use persistent homology to summarize the two simulation datasets and analyze whether the summaries can discriminate the different cosmologies used to produce the data.

When using persistent homology to analyze datasets, the typical summary statistic obtained is a persistence diagram. However, directly comparing persistence diagrams generated from two datasets in such a way that the differences are clearly captured is not straightforward.  The bottleneck distance is an option for comparing two persistence diagrams, but it does not describe the differences across the filtration.  Instead, functional summaries of persistence diagrams can be used.
For example, the \emph{Betti number} for homology dimension $p$, denoted $B_p$, is the rank of $H_p$, which can be interpreted as the number of $p$-dimensional holes.  In persistent homology, $B_p$ can be computed across the varying threshold values $t$ in order to produce a \textit{Betti function}, $B_p(t)$. 
Intuitively, $B_p(t)$ is the number of $H_p$ generators that were born before $t$ and are still alive at $t$.
As noted in the introduction, Betti functions were used in \citet{van2011alpha} and \citet{pranav2016topology} to compare the persistent homology of different dark energy models and cosmic matter distributions, respectively.
Betti functions, or other functional summaries of persistence diagrams (e.g., \citealt{berry2018functional}), are straightforward to use in a comparison.

\begin{figure*}
 \centering
    \begin{subfigure}{0.4\textwidth}
        \centering
    \includegraphics[width=\textwidth]{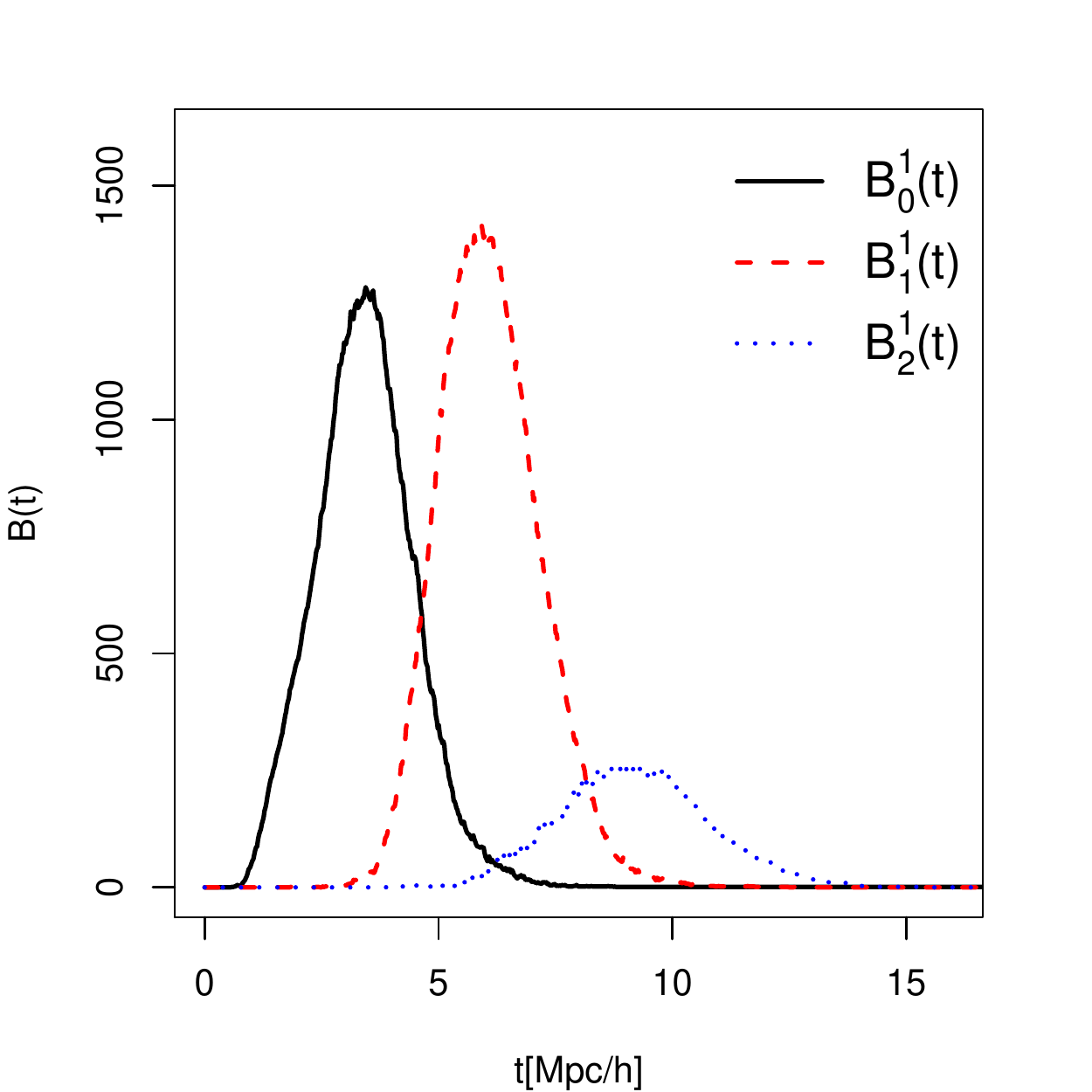}
        \caption{Betti functions of $S_1$}\label{subfig:massive_betti1}
    \end{subfigure}
    \begin{subfigure}{0.4\textwidth}
        \centering
    \includegraphics[width=\textwidth]{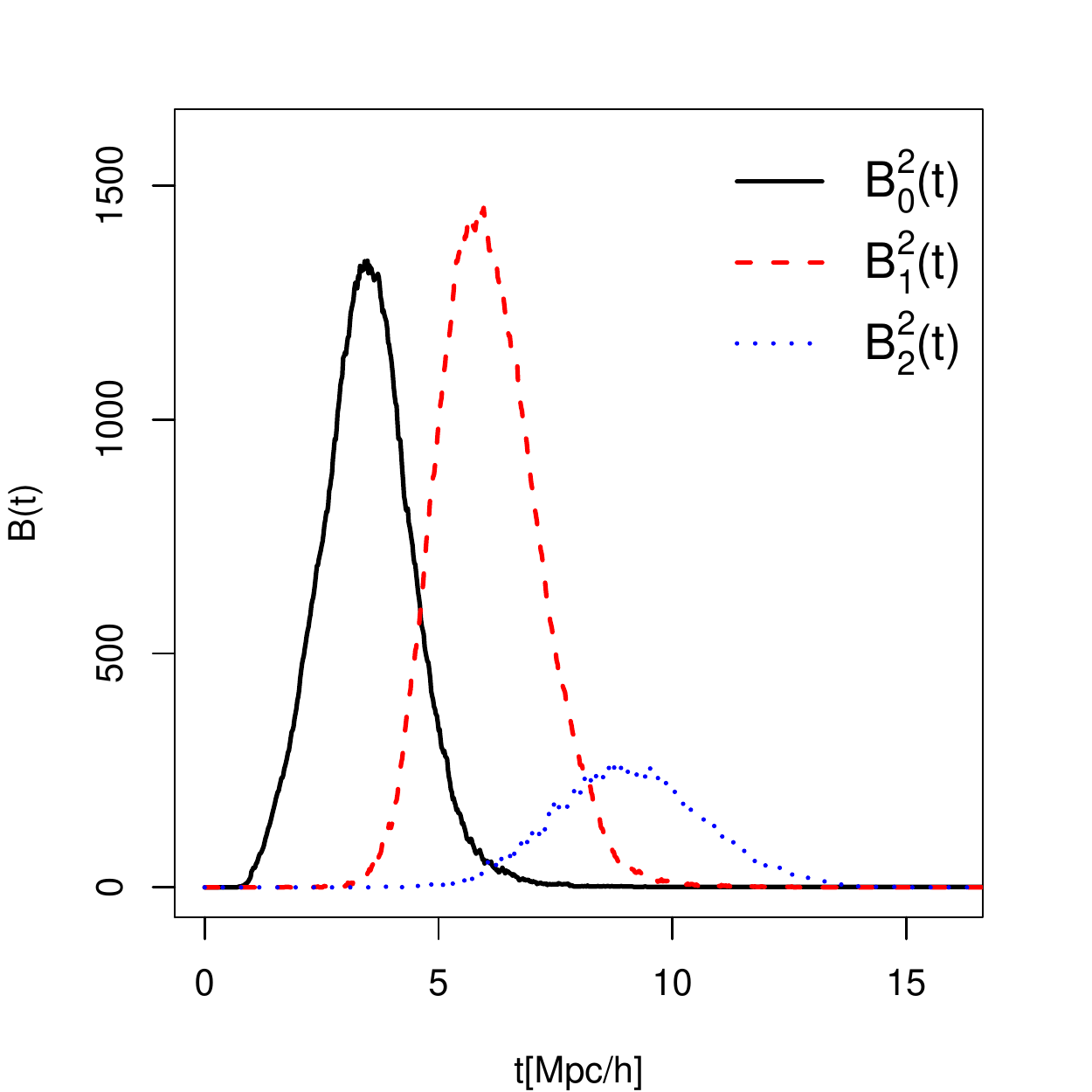}
        \caption{Betti functions of $S_2$}\label{subfig:massive_betti2}
    \end{subfigure}
        \begin{subfigure}{0.4\textwidth}
        \centering
    \includegraphics[width=\textwidth]{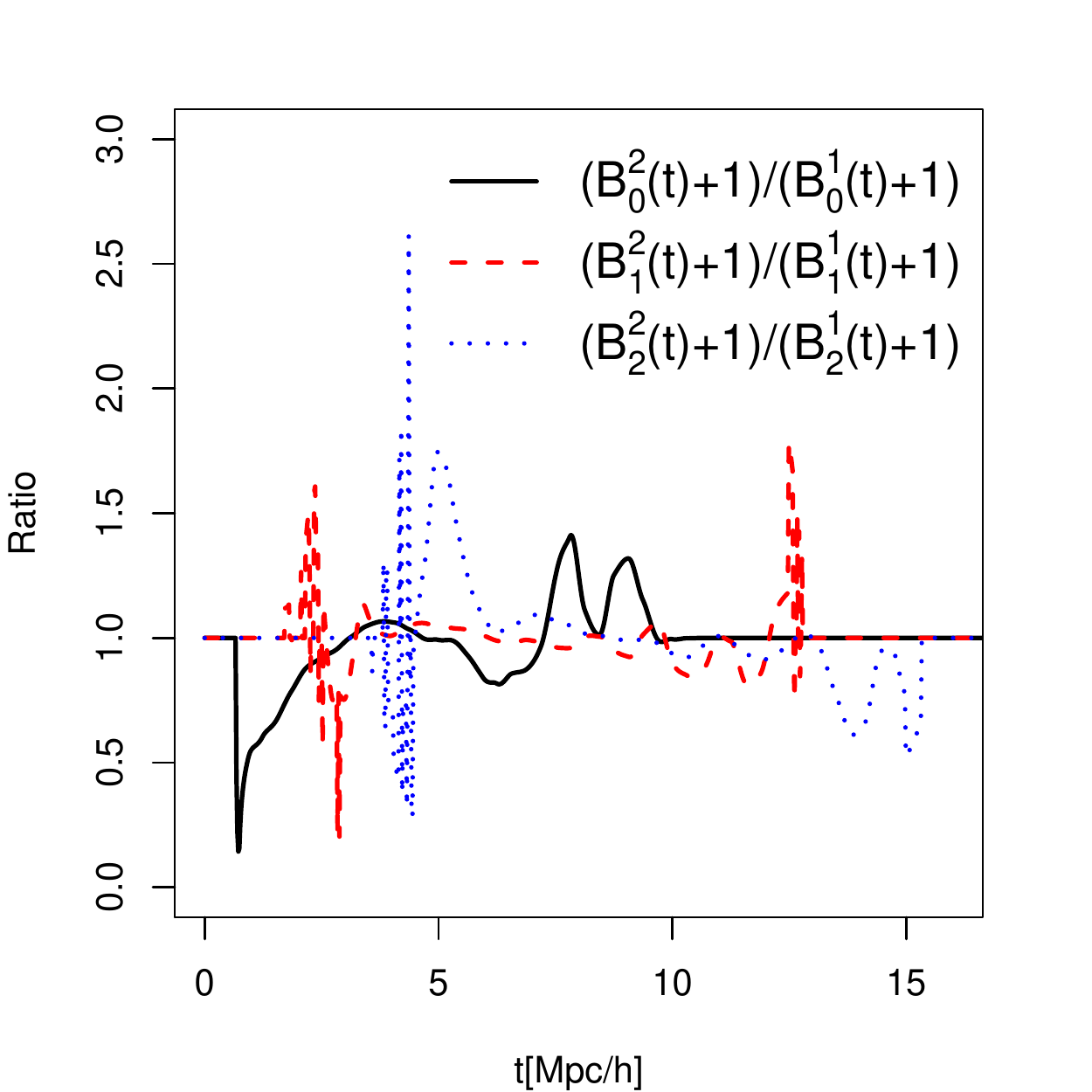}
        \caption{Ratio of Betti functions}\label{subfig:massive_compare}
    \end{subfigure}           \begin{subfigure}{0.4\textwidth}
        \centering
    \includegraphics[width=\textwidth]{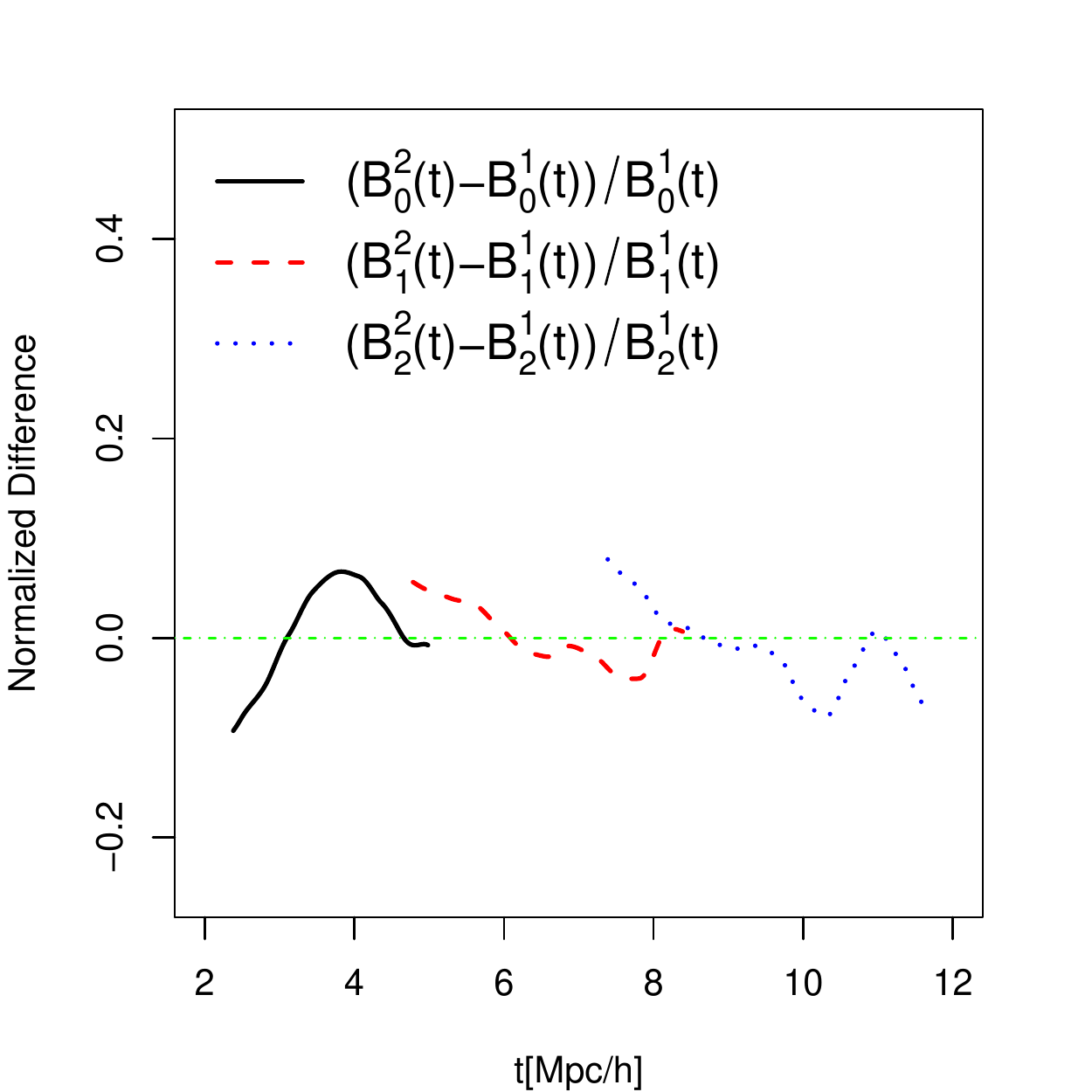}
        \caption{Normalized difference of Betti functions}\label{subfig:massive_compare_diff}
    \end{subfigure}
\caption{Betti functions and ratios of two snapshots from the MassiveNuS simulation suite. The two datasets used are Rockstar halo catalogs generated from the $z=0$ snapshots, both of which have $\Omega_m = 0.3$ and $A_s = 2.1 \times 10^{-9}$, but with different neutrino mass sums of $\sum m_\nu=0$ eV (simulation $S_1$) and $\sum m_\nu = 0.6$ eV (simulation $S_2$). An eighth of the simulation volume is used such that the halos are placed on a grid of size $256 h^{-1}$ Mpc per side with lattice spacing of $1 h^{-1}$ Mpc. A DTM function with $m_0=0.0001$ is used to generate the Betti functions. (a) Betti functions of $S_1$. The black solid, red dashed, and blue dotted lines correspond to $H_0$, $H_1$, and $H_2$. (b) Betti functions of $S_2$. (c) Ratio of smoothed Betti functions of $S_1$ and $S_2$: $\frac{B_p^2(t)+1}{B_p^1(t)+1}$, where $p$ denotes the dimension of the homology group generators. (d) Normalized differences of smoothed Betti functions: $\frac{B_p^2(t) - B_p^1(t)}{B_p^1(t)}$. } \label{massive}
\end{figure*}

\autoref{massive} shows an example of Betti functions. We apply SCHU to an eighth of the simulation volume such that the halos are placed on a grid of size $256h^{-1}$ Mpc per side with grid size of $1h^{-1}$ Mpc. For computing the DTM function, $m_0 =0.0001$ is used, which corresponds to the inclusion of roughly the 40 nearest neighbors.
We denote the 0-, 1-, and 2-dimensional Betti functions of $S_i$ as $B^i_0(t)$, $B^i_1(t)$, and $B^i_2(t)$, respectively, for $i = 1, 2$. 
The resulting Betti functions of $S_1$ and $S_2$ are shown in \autoref{subfig:massive_betti1} and \autoref{subfig:massive_betti2}, respectively. 
The Betti functions of $S_1$ and $S_2$ have similar shapes.
We calculate the ratio for each of the $p$-dimensional Betti functions between $S_1$ and $S_2$, but first carryout mild smoothing to reduce noise fluctuations so that trends in the Betti function ratios can be easily captured. In particular, a local polynomial regression model\footnote{The local polynomial regression model estimates a polynomial model using points in a local neighborhood of the target point with weights defined based on distance from the target point. In this case, the neighborhood is specified to include the nearest $10\%$ of observations.} 
\citep{cleveland1979robust} with an adaptive bandwidth that uses $10\%$ of the nearest observations is considered in order to smooth the Betti functions before computing the ratios displayed in Figures~\ref{subfig:massive_compare} and \ref{subfig:massive_compare_diff}.

The black solid line in \autoref{subfig:massive_compare}, which denotes the ratio of the $B_0^i(t)$, deviates substantially from unity when  $1 < t < 3 h^{-1}$ Mpc and then begins to fluctuate again at $6<t<8 h^{-1}$ Mpc. Both ratios of $B_1(t)$ and $B_2(t)$ display similar behavior at different values of $t$. 
The threshold values where the ratio between the Betti functions of the two simulations is largest correspond to the phases where a particular dimension of homology group generators is just beginning to form. 
Thus, at these points, the ratios are between small counts of generators and can become large and noisy.

However, in the intermediate values of the threshold where the homology group generators of a particular dimension are abundant, there are overall trends between the ratios for both $B_1(t)$ and $B_2(t)$. In \autoref{subfig:massive_compare_diff}, we focus on the normalized difference in the Betti functions between the simulations $\big(B_p^2(t) - B_p^1(t)\big)/B_p^1(t)$ for the threshold values where there are more $p$-dimensional generators present in order to avoid large fluctuations due to low counts. 
The $H_1$ and $H_2$ generator abundances peak in the simulation with massive neutrinos ($S_2$) at a larger magnitude and at lower threshold values than the simulation without massive neutrinos ($S_1$).
While these abundances by threshold are not directly an analog for physical size (the persistence of a generator would be a better proxy for size), the Betti functions do reveal an interesting difference between the simulations.  
The filament loops and voids form earlier in the filtration (i.e., at smaller distances) in the simulation that includes massive neutrinos ($S_2$), suggesting that the presence of massive neutrinos may act to change the density field in such as way that the filament loops and voids form at smaller distances; that is, the neutrinos affect the distribution of the dark matter halos along filaments and on the boundary of voids.
Additional analysis of the dependence of these differences on redshift and the values of ($\sum m_\nu$, $\Omega_m$, $A_s$) is needed to better understand the effect of neutrinos on the density field; this is a topic of future research.
%
Overall, \autoref{subfig:massive_compare_diff} demonstrates that massive neutrinos have a distinct effect on the Betti functions, which could potentially be utilized in larger surveys and simulation datasets to help further constrain cosmological parameters and break degeneracies. 

\section{Conclusions and Discussions}\label{sec:discussion}

In this work, we present a novel method, SCHU, for applying modern statistical methods from topological data analysis, specifically persistent homology, to identify filament loops and cosmic voids in astronomical datasets. While previous works used topological ideas to explore the underlying matter density field in order to study its Betti numbers and topological persistence \citep{pranav2016topology}, 
SCHU strengthens and extends this by assigning $p$-values to individually-identified homology group generators as well as finding a representation of the statistically significant generators in the cosmological volume. 
The addition of this last capability enables simultaneous analysis of void clustering and abundance as well as analysis of the density field's topology via persistence diagrams and Betti numbers. Furthermore, we emphasize the potentially complementary cosmological analysis that can be performed using filament loops, a subset of the cosmic filamentary distribution that SCHU can identify consisting of connected rings of filaments.

In this paper, we (i) introduce the necessary formalism that underlies persistent homology, (ii) detail SCHU for computing persistence diagrams and Betti numbers by embedding the cosmological dataset on a lattice and generating a discrete filtration using the lower-level sets of a DTM function, and (iii) apply SCHU to a controlled Voronoi foam simulation, a subset of the SDSS galaxy survey, a cosmological simulation from the \citet{libeskind2018tracing} cosmic environment identification survey, and a subset of the MassiveNuS simulation suite. We summarize our results as follows:
\begin{itemize}
    \item By using an approximate Voronoi foam model of the cosmic web where the ground truth locations of the voids are known, we demonstrate that SCHU is capable of successfully identifying low-density void regions in cosmological datasets.
    \item Using a subset of SDSS survey data, we demonstrate that SCHU is able to locate statistically significant voids, and these voids are compared to those identified using VIDE \citep{sutter2012public}.
    \item We contrast the results of SCHU as applied to the \citet{libeskind2018tracing} comparison study, finding that while there is still no quantitative consensus between cosmic web environment classifiers, SCHU's results do share similarities with those from various other tools, including Spineweb \citep{aragon2010spine}, DisPerSE \citep{sousbie2011persistent}, and T-web \citep{forero2009dynamical}.
    \item The Betti functions of two simulations from the MassiveNuS suite \citep{MassiveNuS} were analyzed with different values of $\sum m_\nu$, demonstrating that this statistic can potentially be used to help break cosmological parameter degeneracy when applied to larger datasets.
\end{itemize}

The methods presented in this paper lay the groundwork for a detailed follow-up analysis of the large-scale galaxy and matter density fields in state-of-the-art cosmological simulations.
Data products that can be generated from SCHU have potential for constraining cosmological parameters and breaking degeneracy (e.g., $\Omega_m$ and $\sigma_8$) as well as discriminating between various models of dark energy, modified gravity, and massive neutrinos. 
With the imminent, massive influx of cosmological data that will soon be made available through next-generation surveys such as LSST and DESI, it is both important and timely that void-finding methods, such as the one presented here, are applied to realistic mock galaxy catalogs and lightcones in order to predict their model discrimination and parameter constraining power on these future datasets in a statistically rigorous way. 
Simulation suites that span the cosmological parameter space will provide a powerful base for applying SCHU to determine the dependence of void clustering, abundance, and Betti functions on the cosmological parameters. 
Summary statistics of these outputs could also be fed into a machine learning algorithm and trained to predict the underlying set of cosmological parameters. 
Based on our preliminary studies of the Betti functions of the MassiveNuS simulations, some example features of interest include the threshold values where homology group generators are first born and last appear, the threshold values of the peaks in the Betti functions, and the overall normalization of the Betti functions, for each homology group dimension and each simulation.

\section*{Acknowledgements}
The authors thank Jisu Kim, Alessandra Rindalo, and Larry Wasserman for helpful discussions in the early stages of this work.
The authors thank the Yale Center for Research Computing for guidance and use of the research computing infrastructure.
\bibliography{bibliography}

\end{document}